\documentclass[aps,twocolumn,prx]{revtex4-1}
\usepackage{graphicx,amsmath, amsfonts,bm, color}

\usepackage{xcolor}
\usepackage[hypertexnames=false]{hyperref}
\hypersetup{
   colorlinks,
   linkcolor={blue!50!black},%{red!80!black},
   citecolor={blue!50!black},
   urlcolor={blue!80!black}
}
\usepackage{enumitem}
\newcommand{\pa}{\partial}
\newcommand{\pd}[2]{\frac{\partial #1}{\partial #2}}
\renewcommand{\=}{\!=\!}
\newcommand{\1}{^{\mbox{\tiny (1)}}}
\newcommand{\2}{^{\mbox{\tiny (2)}}}
\newcommand{\n}{^{\tiny (n)}}
\newcommand{\eff}{^{\mbox{\tiny eff}}}
\newcommand{\tr}{\operatorname{tr}}

\begin{document}
\title{Frictional sliding without geometrical reflection symmetry}
\author{Michael Aldam$^{1}$}
\thanks{M. Aldam and Y. Bar-Sinai contributed equally to this work.}
\author{Yohai Bar-Sinai$^{1}$}
\thanks{M. Aldam and Y. Bar-Sinai contributed equally to this work.}
\author{Ilya Svetlizky$^{2}$}
\author{Efim A. Brener$^{3}$}
\author{Jay Fineberg$^{2}$}
\author{Eran Bouchbinder$^{1}$}

\affiliation{$^{1}$Chemical Physics Department, Weizmann Institute of Science, Rehovot 7610001, Israel\\
$^{2}$Racah Institute of Physics, The Hebrew University of Jerusalem, Givat Ram, Jerusalem 91904, Israel\\
$^{3}$Peter Gr\"unberg Institut, Forschungszentrum J\"ulich, D-52425 J\"ulich, Germany}

\begin{abstract}
The dynamics of frictional interfaces play an important role in many physical systems spanning a broad range of scales.
It is well-known that frictional interfaces separating two dissimilar materials couple interfacial slip and normal stress variations,
a coupling that has major implications on their stability, failure mechanism and rupture directionality.
In contrast, interfaces separating identical materials are traditionally assumed not to feature such a coupling due to symmetry considerations.
We show, combining theory and experiments, that interfaces which separate bodies made of macroscopically identical materials, but lack geometrical reflection symmetry,
 generically feature such a coupling. We discuss two applications of this novel feature. First, we show that it accounts for a distinct, and previously unexplained,
 experimentally observed weakening effect in frictional cracks. Second, we demonstrate that it can destabilize frictional sliding which is otherwise stable. The emerging framework is expected to find applications in a broad range of systems.
\end{abstract}

\maketitle

\section{Introduction}
\label{sec:intro}

Understanding frictional sliding is a long-standing challenge with important practical and theoretical implications. It is relevant in diverse physical systems spanning a broad range of scales, from the nano-scale to the planetary-scale. A complete analytic treatment of sliding frictional interfaces is generally a formidable task. Two major factors are responsible for the complexity of the problem. First, the friction law, i.e.~the constitutive relation that describes the shear traction at the frictional interface, poses experimental challenges and depends on the slip rate and slip history in a highly nonlinear fashion~\cite{Marone1998, Nakatani2001, Baumberger2006, Putelat2011, Ikari2013, Bar-Sinai2014jgr, DiToro2004, Rice2006, Goldsby2011, Chang2012}. The second factor is the elastodynamics of the sliding bodies, i.e.~the time-dependent long-range stress transfer mechanisms between different points along the interface. It is particularly challenging when the two bodies are made of different materials and in the generic case in which spontaneously-generated interfacial rupture fronts dynamically propagate along the interface~\cite{Weertman1963, Weertman1980, Freund1990, Adams1995, Martins1995a, Martins1995b, Andrews1997, Adams1998, Adams2000, Ranjith2001, Adda-Bedia2003, Kammer201440, Brener2015}.

A significant simplification in relation to the second factor is obtained when the system possesses reflection symmetry across the interface, i.e.~when the two materials are identical, the geometry is symmetric, and the loading configuration is antisymmetric (here and elsewhere we consider macroscopic geometry. Differences in small-scale roughness typically exist and are effectively incorporated into the interfacial constitutive relation). A prototypical example of such a situation is that of two semi-infinite half-spaces made of identical elastic materials, a situation that was extensively studied in the literature (see, for example,~\cite{Madariaga1976, Madariaga1977, Rice1993, Ben-Zion1995, Fukuyama1998, Broberg1999Book, Lapusta2000, Ben-Zion2001, Scholz2002, Rubin2005, Dunham2005, Dunham2007, Ampuero2008}). The main simplification comes from the fact that such a symmetry precludes a coupling between tangential slip and variations in the normal stress. The lack of such symmetry has important implications on the stability of sliding~\cite{Weertman1963, Martins1995a, Martins1995b, Ranjith2001, Adda-Bedia2003,  Rudnicki2006, Dunham2008, Brener2015}, the failure mechanism and rupture directionality~\cite{Comninou1977a, Comninou1977b, Comninou1979, Andrews1997, Adda-Bedia2003, Xia2004, Rudnicki2006, Dunham2008, Ampuero2008b, Brener2015, Shlomai2016, Erickson2016}. Physically, this happens because sliding can enhance (reduce) the normal stress, which in turn can inhibit (facilitate) frictional sliding.

\begin{figure}
 \centering
 \includegraphics[width=\columnwidth]{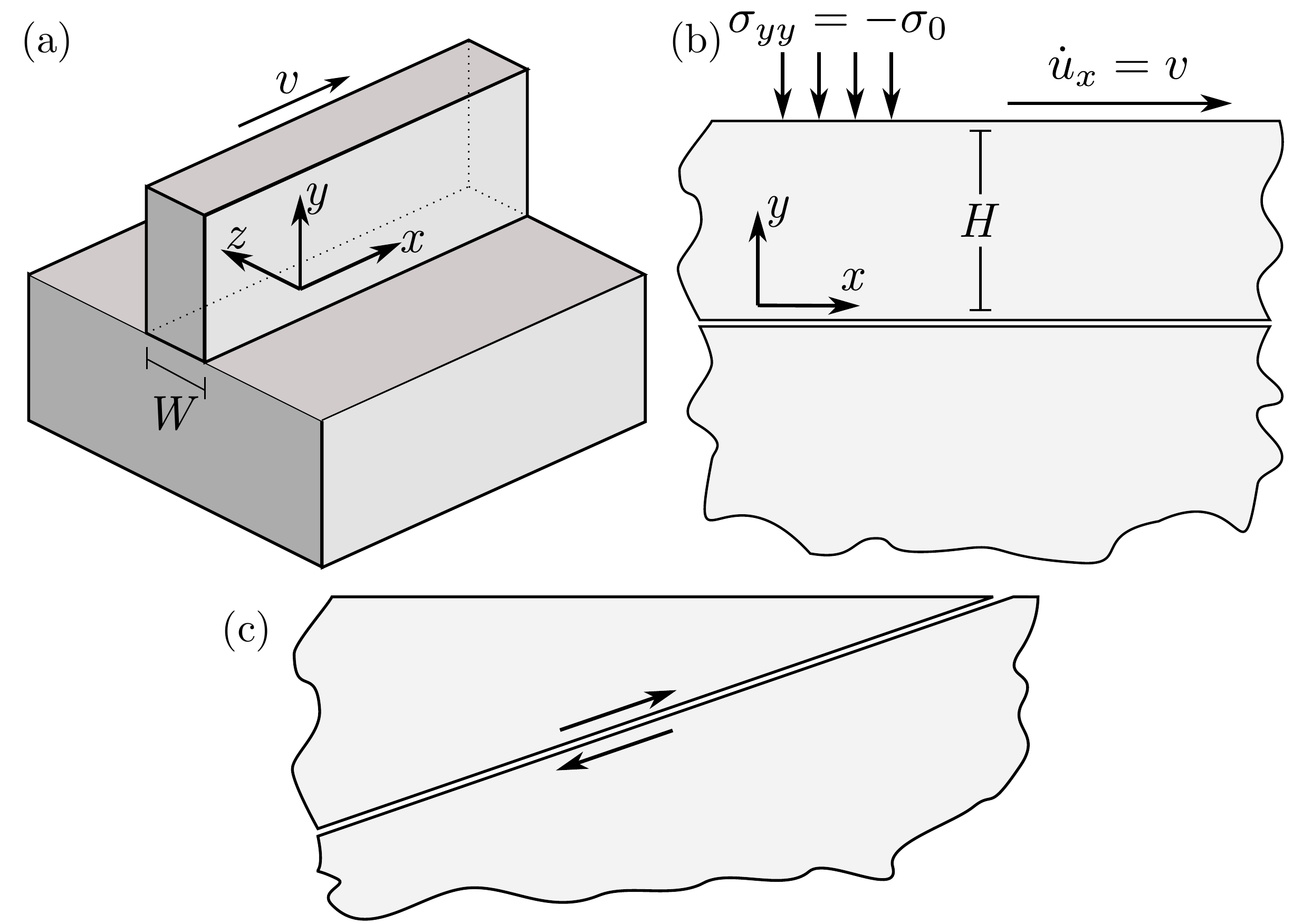}
 \caption{\textbf{Examples of physical systems featuring frictional interfaces separating bodies made of identical materials without geometrical reflection symmetry}: (a) A thin block sliding over a thicker block. (b) A block of finite height $H$ sliding atop a semi-infinite bulk. Sliding occurs in the $x$-direction. (c) An idealized schematic geometry of tectonic subduction motion.}
 \label{fig:geometries}
\end{figure}

The origin of the absence of reflection symmetry is traditionally assumed to be constitutive in nature, i.e.~sliding of dissimilar materials is usually considered. This is known as the bi-material effect. Sliding along such bi-material interfaces has been quite extensively studied in the literature and this material contrast is thought to have important implications for frictional dynamics~\cite{Weertman1963, Comninou1977a, Comninou1977b, Comninou1979, Martins1995a, Martins1995b, Andrews1997, Ranjith2001, Adda-Bedia2003, Xia2004, Rudnicki2006, Dunham2008, Ampuero2008b, Kammer201440, Brener2015, Shlomai2016, Erickson2016}. The purpose of this paper is to explore the possibility of asymmetry of a geometric origin, i.e.~sliding of two bodies made of the same material without geometrical reflection symmetry. Examples of such geometries are depicted in Fig.~\ref{fig:geometries}: sliding of two blocks with different thickness in the direction orthogonal to sliding, an experimental setup that was used in various recent works~\cite{Rubinstein2007, Ben-David2010-fronts, Ben-David2010-ageing, Svetlizky2014, Bayart2016} and will be theoretically addressed below (panel a); sliding of a block of finite height $H$ over a semi-infinite bulk, a simple example to be analyzed in depth in this work (panel b); finally, an idealized sketch of tectonic subduction motion is shown (panel c), a situation in which one lithospheric plate is subducted beneath another one and is responsible for most of the large magnitude earthquakes (``megathrust'') occurring on the Earth's crust~\cite[for example]{Plafker1965, Barrientos1990, Delouis2010, Madariaga2010, Simons2011, Leon-Rios2016}. Obviously, many other sliding geometries which lack reflection symmetry can be conceived. Generally speaking, this situation is expected to be the rule rather than the exception, since no physical system features perfect reflection symmetry.

In this paper we lay out a rather general theoretical framework to address frictional sliding in the absence of geometrical reflection symmetry and support it by extensive experiments. A major outcome is that the effect of geometric asymmetry resembles, sometimes qualitatively and sometimes semi-quantitatively, that of material asymmetry. Consequently, many results obtained for bi-material interfaces are also relevant to interfaces separating bodies made of identical materials with different geometry. As first applications, two main results are obtained within the newly developed framework:
\begin{itemize}[leftmargin=0.5cm]
\item A novel explanation of a sizable weakening effect observed in recent experiments on rupture fronts propagation along frictional interfaces~\cite{Svetlizky2014}. The weakening effect is directly linked to geometric asymmetry and is shown experimentally to disappear in its absence. This result has important implications for the failure dynamics of frictional interfaces.
\item We demonstrate that geometric asymmetry can destabilize frictional sliding which is otherwise stable. We consequently expect geometric asymmetry to play an important role in frictional instabilities.
\end{itemize}
The emerging framework should find additional applications in a broad range of frictional systems.

\section{General framework}
\label{sec:general}

Consider two blocks in frictional contact. At this stage, the discussion remains completely general, allowing the two blocks to be made of different materials, to feature different geometries and to experience general external loadings. We denote the displacement vector fields in the two blocks as ${\bm u}\1({\bm x},t)$ and ${\bm u}\2({\bm x},t)$, where the superscripts correspond to the upper and lower blocks, respectively. Each of these satisfies the momentum balance equation $\nabla\!\cdot\!{\bm \sigma}\!=\!\rho\,\ddot{\bm u}$, where $\rho$ is the mass density of each block. Cauchy's stress tensor ${\bm \sigma}$ is related to the displacement gradient tensor $\nabla {\bm u}$ according to the isotropic Hooke's law $(1+\nu)\mu\left[\nabla {\bm u}\!+\! (\nabla {\bm u})^{\mbox{\tiny T}}\right]\!=\!{\bm \sigma}-\nu({\bm I}\tr{\bm \sigma}-{\bm \sigma})$. Here ${\bm I}$ is the identity tensor, $\nu$ is Poisson's ratio and $\mu$ is the shear modulus of each block. The coordinates are chosen such that the interface lies along the $x$-axis, which is also the direction of sliding, see~Fig.~\ref{fig:geometries}. The direction normal to the interface is the $y$-axis and the interface is the surface $y\!=\!0$. The $z$-axis is in the thickness direction, where $z\=0$ is the center line. While the formulation below and the analysis in Sect.~\ref{sec:stability} are two-dimensional (2D), we shall see that three-dimensional (3D) effects involving the $z$-coordinate play an important role in Sect.~\ref{sec:thin-on-thick}.

Since the bulk equations are linear, one can analyze separately each interfacial Fourier mode, i.e.~write ${\bm u}\n (x,y\!=\!0,t)\={\bm u}\n e^{ik(x-ct)}$, where $k\!>\!0$, $c$ is the complex phase/propagation velocity and $n\=1,2$. The relation between the interfacial displacements and stresses is also linear and can be written as $u_i\n\!=\!M_{ij}\n(c,k)\sigma_{yj}\n$, where the matrix $\bm M\n$ can be obtained from the Green's function of the corresponding medium and $\sigma_{yi}\n$ are the interfacial stresses, i.e.~at $y\=0$. For example, under quasi-static conditions for semi-infinite blocks in 2D, this relation for the lower block (i.e. the block at $y<0$) takes the form~\cite{SM}
\begin{align}
\begin{pmatrix}
  u_x \\ u_y
 \end{pmatrix}
 =\frac{1}{\mu  k}
 \begin{pmatrix}
 1-\nu  & -\frac{i}{2}(1-2 \nu )\\
 \frac{i}{2}(1-2 \nu )& 1-\nu  \\
\end{pmatrix}
\begin{pmatrix}
  \sigma_{xy} \\\sigma_{yy}
 \end{pmatrix}\ .
 \label{eq:M_2D_QS}
 \end{align}

The essence of frictional motion is that the displacement field is discontinuous across the interface. We denote the slip discontinuity at the frictional interface by
\begin{equation}
 \epsilon_i(x)\equiv u\1_i(x,y=0^+)-u\2_i(x,y=0^-)\ .
 \label{eq:epsilon}
\end{equation}
On the interface, $y\!=\!0$, no separation or inter-penetration between the bulks implies $\epsilon_y\!=\!0$ and continuity of $\sigma_{yi}$. Together with the known dynamic response matrices $\bm M\n$, these requirements can be used to calculate the relation between the slip discontinuity and the interfacial stresses of the composite system that consists of both bulks. Following~\cite{Geubelle1995}, this is done by noting that for $y\!=\!0$ we have $\sigma_{yi}\!=\!\left({\bm M}\1\!-\!{\bm M}\2\right)^{-1}_{ij}\epsilon_j$. Thus, the response of the composite system in 2D reads (in Fourier space)
\begin{align}
\begin{split}
\sigma_{xy}&=\mu\1 k\,G_x(c,k)\,\epsilon_x(k) \ ,\\
\sigma_{yy}&=i\mu\1 k\,G_y(c,k)\,\epsilon_x(k) \ ,
\end{split}
\label{eq:G}
\end{align}
where we defined the elastic response functions $G_x(c,k)\!\equiv\!(\mu\1\!k)^{-1}\!\left({\bm M}\1\!-\!{\bm M}\2\right)^{-1}_{xx}$~and~$G_y(c,k)\!\equiv\!-i(\mu\1\!k)^{-1}\!\left({\bm M}\1\!-\!{\bm M}\2\right)^{-1}_{yx}$. That is, the $G_i$'s can be expressed as functions of the response coefficients of both bulks.
Note that the imaginary unit $i$ is included in Eq.~\eqref{eq:G} for convenience. Note also that we use the same notation for a function and its Fourier transform, as they are easily distinguishable by the context or the stated arguments (e.g. $k$ or $x$).

The central player in the analysis to follow is $G_y$, which represents the elastodynamic coupling between tangential slip and normal traction along the interface. In systems with complete reflection symmetry along $y\=0$, this coupling is precluded by symmetry. To see this, note that in this case the off-diagonal elements of ${\bm M}\1$ and ${\bm M}\2$ are identical~\cite{SM}, and thus ${\bm M}\1\!-\!{\bm M}\2$, as well as its inverse, is diagonal. This immediately implies $G_y\=0$. In what follows, we study two important frictional problems in which the sliding bodies are made of {\em identical materials}, i.e.~$\rho\1\=\rho\2\!\equiv\!\rho$, $\mu\1\=\mu\2\!\equiv\!\mu$ and $\nu\1\=\nu\2\!\equiv\!\nu$, yet reflection symmetry relative to the interface is absent due to asymmetry in the {\em geometry} of the bodies, leading to $G_y\!\ne\!0$. These problems highlight the importance of geometrical asymmetry to frictional sliding and its relation to the conventional bi-material effect.

\section{``Thin-on-thick'' systems and the propagation of frictional cracks}
\label{sec:thin-on-thick}

The first problem that we examine, which is directly motivated by recent experimental observations~\cite{Svetlizky2014, Shlomai2016}, is depicted in Fig.~\ref{fig:geometries}a. In this system, a thin block of width $W\!=\!5.5$ mm is pushed along its length (the $x$-axis) on top of a significantly thicker block (here $30$ mm). This ``thin-on-thick'' experimental setup was used in various studies~\cite{Rubinstein2007, Ben-David2010-fronts, Ben-David2010-ageing, Svetlizky2014, Bayart2016}, where a transparent glassy polymer (poly(methyl-methacrylate), PMMA) was used. The transparent material allows a direct real-time visualization and quantification of a fundamental interfacial quantity: the real contact area, $A_r$. The latter is the sum over isolated micro-contacts formed due to the small scale roughness of macroscopic surfaces.
\begin{figure*}
 \centering
 \includegraphics[width=\textwidth]{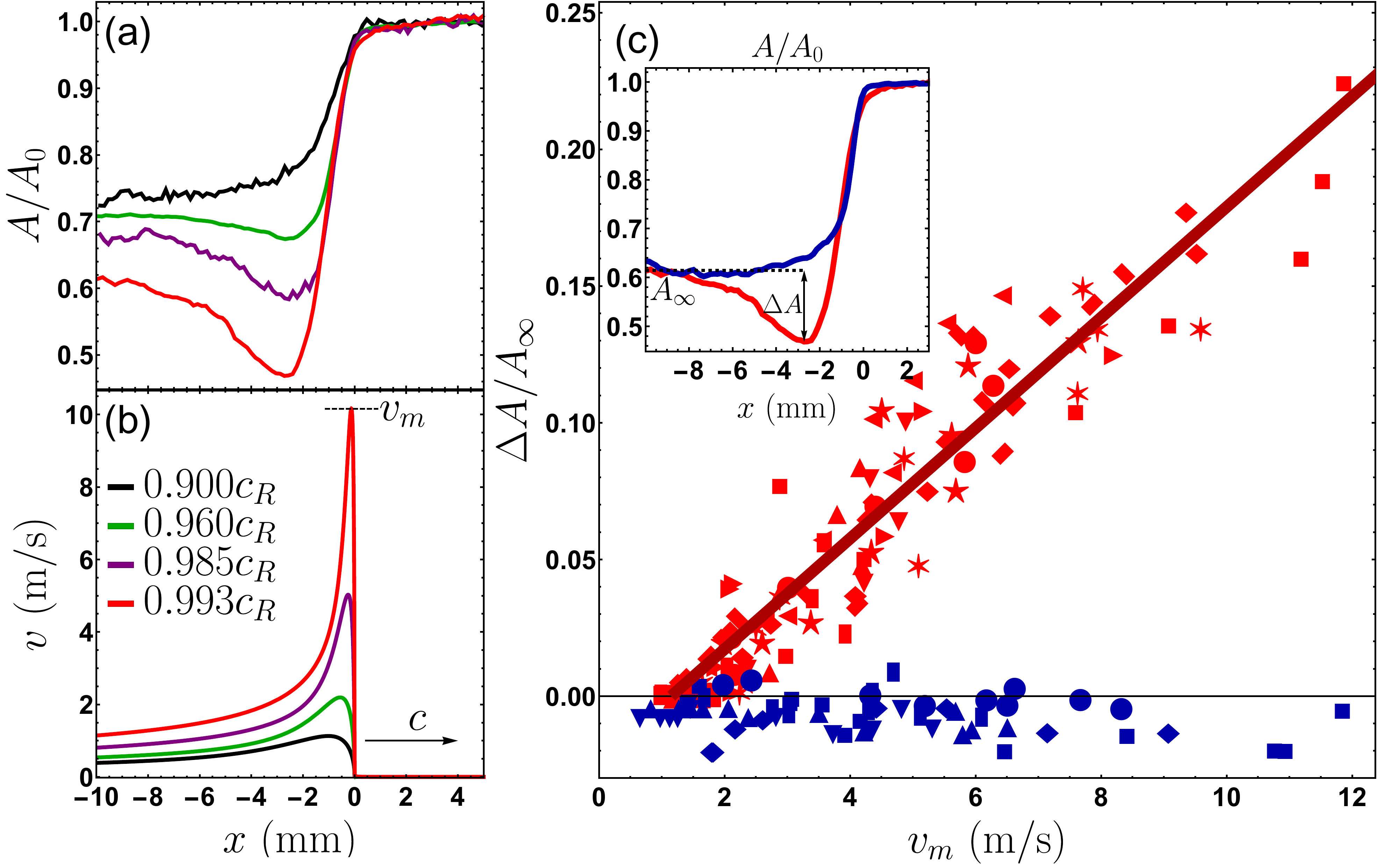}
 \caption{\textbf{Experimental results}. (a) Snapshots of the spatial profile of the contact area $A$ of rupture fronts in the ``thin-on-thick'' setup (see Fig.~\ref{fig:geometries}a and~\cite{Svetlizky2014} for additional details). These fronts propagate to the right at velocities $c$ indicated in the legend of panel b ($0.900c_R\!<\!c\!<\!0.993c_R$, see~\cite{SM}), where $x\!=\!0$ corresponds to the tip of each rupture front. The contact area is normalized by its value $A_0$ before the passage of the front. (b) The slip velocity profiles corresponding to the snapshots in panel a (see~\cite{SM} for details). (c) $\Delta A/A_\infty$, where $\Delta A$ is the magnitude of the real contact area undershoot and $A_\infty$ is the asymptotic value (see inset), vs. the maximal slip velocity $v_{m}$ (see panel b) for both the ``thin-on-thick'' setup (red symbols) and the geometrically symmetric ``thin-on-thin'' setup (blue symbols). Different symbols correspond to different experiments and their size roughly corresponds to the measurement error. The red line is the best linear fit for the red symbols. (inset) The contact area profile for $c\!=\!0.993c_R$ in the ``thin-on-thick'' setup (red line, already appearing in panel a) and in the geometrically symmetric ``thin-on-thin'' setup (blue line).}
 \label{fig:experimental}
\end{figure*}

$A_r$ is typically orders of magnitude smaller than the nominal contact area, $A_n$. Their ratio, $A\!\equiv\!A_r/A_n\!\ll\!1$, plays a critical role in interfacial dynamics~\cite{Bowden1950, Dieterich1994, Nakatani2001, Baumberger2006, Nakatani2006, Rubinstein2007, Nagata2008, Ben-David2010-fronts, Ben-David2010-ageing, Svetlizky2014} because the frictional resistance/stress is proportional to $A$, $\sigma_{xy}\!\propto\!A$, i.e.~the larger the real contact area the larger the frictional resistance. $A$ itself depends on the normal stress and also on the slip history of the interface according to
\begin{equation}
 \sigma_{xy} \propto A\propto \sigma_{yy}(1+\psi)\ ,
 \label{eq:A}
\end{equation}
where $\psi$ is an internal variable characterizing the state of the interface. Frictional sliding leads to reduction of $\psi$, i.e.~to a reduction of the contact area~\cite{Marone1998, Nakatani2001, Nakatani2006, Baumberger2006, Bar-Sinai2013pre, Bar-Sinai2014jgr}. In the absence of sliding, $\psi$ (and hence $A$) grows logarithmically with time, a process known as frictional aging~\cite{Dieterich1972, Dieterich1978, Marone1998, Baumberger2006, Ben-David2010-ageing}.

In~\cite{Svetlizky2014} it was found that sliding is mediated by a succession of crack-like rupture fronts propagating along the frictional interface and that these fronts are surprisingly well described by the classical theory of shear cracks propagating along an interface separating identical materials, Linear Elastic Fracture Mechanics (LEFM). The variation of $A$ along a few of these fronts is shown here in Fig.~\ref{fig:experimental}a. It is seen that the rupture fronts involve a significant overall reduction of the contact area, which weakens the interface (i.e.~reduces $\psi$) and facilitates sliding. We would like to focus our attention on a distinct feature of these curves: As observed in Fig.~\ref{fig:experimental}a, fronts which travel at 90\% of the Rayleigh wave-speed $c_R$, here $c_R\!\simeq\!1237$ m/s (for plane-stress conditions~\cite{SM}), or slower (not shown), feature a monotonic decrease of $A$. However, in fronts propagating even closer to $c_R$, $A$ features a non-monotonic behavior, i.e.~$A$ undershoots the asymptotic value $A_\infty$ (i.e.~$A$ as $x\!\to\!-\infty$) and then rapidly increases, at a rate way too high to be explained by slow frictional aging. This non-monotonic behavior remained unexplained in~\cite{Svetlizky2014}, where it was stated that ``the non-monotonic behavior of $A$ ... suggests interesting dynamics as $c\!\to\!c_R$...''.

In Fig.~\ref{fig:experimental}b we show the spatial profiles of the slip velocity $v$, corresponding to the contact area profiles shown in Fig.~\ref{fig:experimental}a. These profiles were calculated from the experimental data using the simplest cohesive zone model~\cite{Palmer1973, Poliakov2002, Samudrala2002} which is consistent with the measurements of the fracture energy and cohesive zone size (see~\cite{SM} for more details). This model, while generally used to describe identical materials, is motivated by the empirical observation~\cite{SM} that the strain fields are, to first order, quite similar in the thin-on-thin and thin-on-thick setups. This approximation would, of course, have to be modified in cases of strong material contrast, where the fields on both sides of the interface differ strongly~\cite{Shlomai2016}.

We denote the maximum slip velocity in these profiles by $v_m$. Next, in order to quantify the non-monotonic effect,
we define the magnitude of the undershoot $\Delta A$ as the difference between the asymptotic value $A_\infty$ and the
minimum of the profile over the range $-5~\hbox{mm}\!<\!x\!<\!0$, which is the typical spatial range for which $\Delta A\!>\!0$ is observed in the thin-on-thick setup, see Fig.~\ref{fig:experimental}a. $\Delta A/A_\infty$ is plotted vs.~$v_m$ in Fig.~\ref{fig:experimental}c (red symbols), demonstrating that the former is quasi-linear (i.e.~predominantly linear) in the latter. Note that the spread in the data does not allow to identify any systematic deviations from linearity. We stress that the effect is not only qualitatively novel, i.e.~the existence of a non-monotonic contact area behavior $\Delta A/A_\infty\!>\!0$, but it is also quantitatively important. As Fig.~\ref{fig:experimental}c shows, the local {\em reduction} in the real contact area $\Delta A/A_\infty$ can reach nearly 25\%. This is a large quantitative effect, compared to other documented frictional effects, implying the existence of significant local frictional {\em weakening} which can significantly influence interfacial dynamics.

What is the source of this non-monotonicity, why does it scale quasi-linearly with the slip velocity and why does it appear only at sonic propagation velocities? Such behavior has recently been observed in~\cite{Shlomai2016} when investigating the frictional motion of bi-material interfaces in a geometrically symmetric system. There, a very large local reduction of $A$ was observed at sonic propagation velocities, but it entirely disappeared when the upper and lower blocks were made of the same material. We propose that the same happens in our case, only here it is due to geometric asymmetry. That is, we suggest that the non-monotonicity of $A$ stems from the absence of geometrical reflection symmetry of the two blocks, i.e.~from the difference in their thickness. If true, then the fast non-monotonic variation of $A$ is not an intrinsically frictional phenomenon, i.e.~a result of the dynamics of the state of the interface $\psi$, but rather an elastodynamic effect emerging from the coupling between slip and normal stress variations, solely induced by geometrical effects. In terms of Eq.~\eqref{eq:A}, we propose that $\psi$ is monotonic and that the non-monotonicity of $A$ results from a non-monotonic behavior of $\sigma_{yy}$.

Our strategy in testing and exploring this idea is two-fold. First, our idea can be directly tested by a definitive experiment. That is, we expect that when the width of the lower block equals that of the upper one, i.e.~in a ``thin-on-thin'' setup, the non-monotonicity in $A$ disappears altogether even in the limit $c\!\to\!c_R$. We performed this experiment, as in~\cite{Shlomai2016}, and present a representative example (for $c\=0.993c_R$) in the inset of Fig.~\ref{fig:experimental}c (blue line). The curve is indeed monotonic. Moreover, note that the asymptotic value $A_\infty$ is the same as that in the ``thin-on-thick'' setup (cf.~Fig.~\ref{fig:geometries}a, for the same propagation velocity), even though the latter exhibits a large undershoot. In Fig.~\ref{fig:experimental}c (main panel) we added $\Delta A/A_\infty$ of many rupture fronts in the ``thin-on-thin'' setup (blue symbols). $\Delta A/A_\infty$ is indeed very close to zero (small negative values simply correspond to monotonic behavior), i.e.~all of the $A$ profiles in the ``thin-on-thin'' setup are monotonic. This direct experimental evidence provides unquestionable support of our basic idea that the non-monotonic behavior corresponding to the ``thin-on-thick'' setup data in Fig.~\ref{fig:experimental}a emerge from the absence of geometrical reflection symmetry.

Next, our aim is to develop a theoretical understanding of the origin of non-monotonicity. The challenge is to explain both the fact that it emerges at asymptotic propagation velocities ($c\!\to\!c_R$) and the quasi-linear relation between $\Delta A/A_\infty$ and $v_m$. The complete problem, involving a thin block sliding atop a thicker one, is a very complicated 3D elastodynamic problem. We approach the problem by breaking it into two steps. First, we perform a simplified analysis, invoking physically-motivated approximations, which allow us to reduce the mathematical complexity of the problem and gain analytic insight into it.
The major simplification is to consider the corresponding quasi-static problem instead of the full elastodynamic one. The physical rationale for this is clear: the absence of geometrical reflection symmetry should manifest its generic implications also in the framework of static elasticity and hence the simplified analysis is expected to reveal the origin of the non-monotonicity of the real contact area. Then, in the second step, we use the static results in an effective dynamic calculation, to be explained below.

The main outcome of the first step is that the static 3D problem can be approximately mapped onto a 2D problem involving two elastically {\em dissimilar} materials. That is, we show that the \textit{geometric} asymmetry can be approximately mapped onto an effective \textit{constitutive} asymmetry, i.e.~an effective {\em material contrast}. To see how this emerges, we assume that both blocks are infinite in the $y$-direction and that the thicker (lower) block is also infinite in the $z$ direction. That is, the lower block is replaced by a semi-infinite 3D half-space, which allows us to use the well-known interfacial Green's function~\cite{LLElasticity}. More specifically, the 3D real-space Green's function matrix $\hat{\bm M}^{\mbox{\tiny 3D}}(\bm r-\bm r')$~\cite{LLElasticity} allows us to express the interfacial displacements at a point ${\bm r}\=(x,y\=0,z\=0)$ on the symmetry line, ${\bm u}({\bm r})\=(u_x, u_y)$, induced by a point force applied by the upper block at $\bm r'\=(x',y\=0,z')$, ${\bm F}({\bm r'})\=(F_x, F_y)$. Note that the latter is assumed not to contain an out-of-plane component, i.e.~$F_z\=0$, which in principle could emerge from frustrated Poisson expansion at the interface. It is reasonable, though, to neglect it to leading order.
\begin{figure*}
  \centering
  \includegraphics[width=\textwidth]{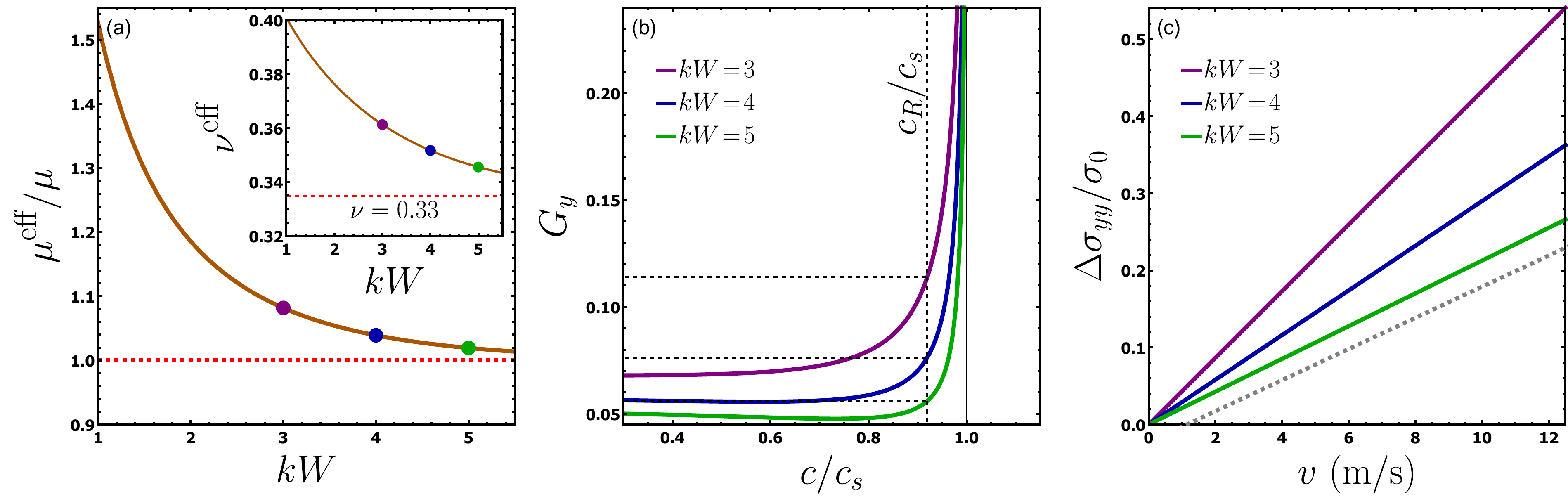}
  \caption{\textbf{Analytical results}. (a) The effective shear modulus $\mu\eff$ of the thicker block, in units of $\mu$, vs.~the dimensionless wavenumber $q\!=\!kW$, cf.~Eq.~\eqref{eq:effective_constants}. (inset) The variation of the effective Poisson's ratio $\nu\eff(q)$ with $q\!=\!kW$. In both we used $\nu\!=\!0.33$, which is relevant for PMMA~\cite{Read1981, Svetlizky2014}. (b) The response function $G_y$, quantifying the {\em effective} bi-material contrast according to $\mu\eff(q)$ and $\nu\eff(q)$ (for the thicker block, the thinner one is represented by plane-stress conditions), corresponding to selected values of $q\!=\!kW$. The corresponding values of the elastic moduli $\mu\eff(q)\!>\!\mu$ and $\nu\eff(q)\!>\!\nu$ are marked in panel a and its inset using the same color code. (c) $\Delta\sigma_{yy}$ given in Eq.~\eqref{eq:dsigma}, normalized by the experimentally applied normal stress $\sigma_0\!=\!4.5$ MPa, vs.~the slip velocity $v$, where the propagation velocity was set to $c\!=\!c_R\!\simeq\!1237$ m/s. The gray dashed line is the red line in Fig.~\ref{fig:experimental}c.}
  \label{fig:analytical}
\end{figure*}

We physically expect shear tractions to be uniform across the thickness $W$, hence they are taken to be constant for $|z|\!\le\!\frac{W}{2}$ (and of course to vanish for $|z|\!>\!\frac{W}{2}$). Thus, we obtain
\begin{align}
\label{eq:Meff}
\begin{pmatrix}
  u_x \\u_y
 \end{pmatrix}
 =\bm M\eff(k)
\begin{pmatrix}
  \sigma_{xy} \\\sigma_{yy}
 \end{pmatrix} \ ,
\end{align}
where the effective 2D response matrix $\bm M\eff(k)$ of the thicker (lower) block is given by the Fourier transform of $\hat{\bm M}^{\mbox{\tiny 3D}}$ over the strip $|z|\!\le\!\frac{W}{2}$,
\begin{equation}
 \bm M\eff(k)\!=\!\int_{-\infty}^{\infty}\!\!\! dx'\int_{-\tfrac{W}{2}}^{\tfrac{W}{2}}\, dz'\,e^{ik(x-x')}\hat{\bm M}^{\mbox{\tiny 3D}}(x-x',z')\ .
\end{equation}
The integration can be carried out, resulting in
\begin{align*}
  \bm M\eff
 \!\simeq\!\frac{1}{\mu k}\!\!
 \begin{pmatrix}
 \!(1\!-\!\nu )B(q) & -\frac{i}{2}(1\!-\!2 \nu)\!\left(1\!-\!e^{-\frac{|q|}{2}}\right)\!\! \\
 \!\frac{i}{2} (1\!-\!2 \nu)\left(1\!-\!e^{-\frac{|q|}{2}}\right) & (1\!-\!\nu )B(q)\!
\end{pmatrix}\!,
\end{align*}
where $q\!\equiv\!kW$ and $B(q)\=\pi^{-1}\!\int_0^q {\cal K}_0(q'/2)dq'$ (${\cal K}_0(z)$ is the modified Bessel function of the second kind of order $0$). The outcome of the analysis, which is presented in full detail in~\cite{SM}, is that $\bm M\eff(k)$ appears to identify with the 2D response matrix of Eq.~\eqref{eq:M_2D_QS}, if one defines the effective elastic moduli of the lower (thicker) block as
\begin{align}
\begin{split}
   \mu\eff(q)&\simeq\frac{\mu }{2 (1-\nu) B(q)-(1-2 \nu) \left(1-e^{-\frac{|q|}{2}}\right)}\ ,\\	
   \nu\eff(q)&\simeq\frac{(1-\nu) B(q)-(1-2 \nu) \left(1-e^{-\frac{|q|}{2}}\right)}{2 (1-\nu) B(q)-(1-2 \nu) \left(1-e^{-\frac{|q|}{2}}\right)} \ .
\end{split}
\label{eq:effective_constants}
\end{align}
These are plotted in Fig.~\ref{fig:analytical}a.

The mapping of the 3D problem onto an effective 2D problem is formally valid as long as the interfacial stresses (and hence displacements) in Eq.~\eqref{eq:Meff} are approximately localized in Fourier $k$-space. Otherwise, Eq.~\eqref{eq:Meff} will not identify with Eq.~\eqref{eq:M_2D_QS} due to the extra $k$-dependence of $\mu\eff(kW)$ and $\nu\eff(kW)$, which is a result of the 3D nature of the original problem. We note in passing that in the limit $q\=kW\!\gg\!1$, $\mu\eff\!\to\!\mu$ and $\nu\eff\!\to\!\nu$, which corresponds to 2D plane-strain conditions~\cite{Timoshenko}. This is expected for small wavelengths, for which the thinner block also appears infinitely thick, and hence is a consistency check on our calculation. The important observation, though, as is clearly seen in Fig.~\ref{fig:analytical}a, is that for the thicker block $\mu\eff(k)\!>\!\mu$ for all experimentally relevant $k$'s~\footnote{Note that at large $kW$, $\mu\eff$ becomes minutely smaller than $\mu$, see~\cite{SM}}. This suggests that the thicker block is effectively stiffer than the thinner one, as hypothesized in~\cite{Radiguet2015276, Kammer2015} where the thicker block was assumed to correspond to plane-strain conditions in numerical simulations. That is, the main physical insight gained from the performed analysis is that geometric asymmetry gives rise to an effective material contrast.

With this physical insight in hand, we aim now at addressing the non-monotonicity of $A$ discussed in panels a and c of Fig.~\ref{fig:experimental}. The 3D static analysis presented above may not yield quantitatively accurate predictions when strongly elastodynamic 2D interfacial rupture fronts are considered. Yet, we believe that the insight embodied in the relations $\mu\eff(k)\!>\!\mu$ and $\nu\eff(k)\!>\!\nu$ is physically robust and hence try to explore their quantitative implications in relation to the experimental observations in the dynamic regime.

To accomplish this, we consider the 2D dynamic transfer function $G_y(c,k)$ in Eq.~\eqref{eq:G} and take it to approximately describe the experimental system when the effective moduli $\mu\eff(k)\!>\!\mu$ and $\nu\eff(k)\!>\!\nu$ are used for the thicker (lower) block and plane-stress conditions~\cite{Timoshenko} are assumed for the thinner (upper) block. Note that it is justified to treat the heights of the two blocks as infinite since the experimental rupture fronts are so fast that they do not interact with the upper and lower boundaries before traversing the whole system. Therefore, Eq.~\eqref{eq:G} can be rewritten as
\begin{equation}
 \Delta \sigma_{yy}(c,k,v) = -c^{-1}\mu\,G_y\left[c,k; \mu\eff(k), \nu\eff(k)\right] v \ ,
 \label{eq:dsigma}
\end{equation}
where we used $v\=\dot\epsilon_x\=-i\,c\,k\,\epsilon_x$ for a constant propagation velocity $c$.

The 2D infinite-system dynamic transfer function $G_y(\cdot)$ in Eq.~\eqref{eq:dsigma} was calculated by Weertman for sliding of dissimilar materials quite some time ago~\cite{Weertman1980}. We reiterate that the basic idea here is to use a known result for dissimilar materials to represent a system composed of identical materials with geometric asymmetry, utilizing the effective moduli derived in Eq.~\eqref{eq:effective_constants}, $\mu\eff(k)\!>\!\mu$ and $\nu\eff(k)\!>\!\nu$. In the presence of any contrast between the shear moduli of the materials, $G_y(c)$ is finite and increases significantly at elastodynamic velocities (in fact, it diverges when $c$ approaches the shear wave-speed $c_s$ of the more compliant material), as shown in Fig.~\ref{fig:analytical}b. Thus, we expect rupture fronts that propagate at near-sonic velocities to be accompanied by a significant reduction in the local normal stress as implied by Eq.~\eqref{eq:dsigma}, reducing locally the real contact area. In turn, this reduces the interfacial strength, which facilitates sliding. This is consistent with the experimental observations of Fig.~\ref{fig:experimental}a, where the non-monotonicity of $A$ becomes substantial at asymptotic propagation velocities ($c\!\to\!c_R$). This normal stress reduction is also remarkably similar to the recent observations of~\cite{Shlomai2016} in bi-material systems, a similarity that further strengthens the analogy between geometric asymmetry and material asymmetry.

The connection between geometric and material asymmetries is yet further strengthened when the directionality of rupture is
considered. The sub-Rayleigh ($c\!<\!c_R$) rupture fronts, shown in Fig.~\ref{fig:experimental}b, propagate from left to right,
in the direction of sliding of the {\em thinner} (upper) block (see also Fig.~\ref{fig:geometries}a). Sub-Rayleigh rupture fronts that are accompanied by normal stress reduction are known to propagate in the direction of sliding of the more {\em compliant} material in a bi-material setup, the so-called ``preferred direction''~\cite{Weertman1980, Andrews1997, Shlomai2016}. This is fully consistent with our result that the thinner (upper) block is effectively softer than the thicker (lower) block (or alternatively, that the thicker block is effectively stiffer than the thinner one).

The quasi-linearity of $\Delta A$ with the (maximal) slip velocity, observed in Fig.~\ref{fig:experimental}c, naturally emerges from Eq.~\eqref{eq:dsigma}. To see this, note that $\Delta A\!\propto\!\Delta \sigma_{yy}$ according to Eq.~\eqref{eq:A} (recall that $\psi$ in that equation is expected to be monotonic) and that $c$ remains close to $c_R$ to within a few percent. In this regime ($c\!\simeq\!c_R$), $G_y$ does not change appreciably as a function of $c$, while the maximal $v$ varies quite substantially (cf.~Fig.~\ref{fig:experimental}a). That means that while $c\!\simeq\!c_R$ is required for the existence of the weakening effect, its variability is mainly determined by $v$. Put together, we obtain $\Delta A\!\propto\! v$. To obtain some estimate of the proportionality factor between $\Delta A$ and $v$ along this line of reasoning, we interpret $\Delta\sigma_{yy}$ in Eq.~\eqref{eq:dsigma} to be a function of $v$ alone, with $c\=c_R$ and $k\!\sim\!{\cal O}(W^{-1})$, where $\mu\eff(k)\!>\!\mu$ (cf.~Fig.~\ref{fig:analytical}a).

The results for $\Delta\sigma_{yy}(v)$ with $kW\=3, 4, 5$, normalized by the experimentally applied normal stress $\sigma_0$, are shown in Fig.~\ref{fig:analytical}c. The slope of the $kW\=5$ line is very close to the slope of the linear fit in Fig.~\ref{fig:experimental}c, which was added to Fig.~\ref{fig:analytical}c for comparison (gray dashed line). Note that the experimental line features a finite $v$ intercept, which is absent in the theoretical one. This is expected since the undershoot, $\Delta A$, is generally susceptible to variations both of $\sigma_{yy}$ and the fracture of contacts (variations of $\psi$ in Eq.~\eqref{eq:A}). For low values of $v$, variations of $\sigma_{yy}$ should be small, and the spatial profile of $A$ is therefore dominated by variations of $\psi$. $A(x)$ should therefore be monotonic in space, similar to the spatial profile in the ``thin-on-thin'' setup, thus rendering any undershoots (i.e. $\Delta A$) to be unmeasurable.

This quantitative agreement should be taken with some caution in light of the various approximations invoked above. Yet, the existence of a characteristic wavenumber $kW\=5$
is not unreasonable as the typical scale of the velocity peaks (see Fig.~\ref{fig:experimental}b), the spatial scale of the undershoot in the contact area (see Fig.~\ref{fig:experimental}a) and $W$ are all in the mm-scale. Furthermore, the {\em relative magnitudes} of the slopes in Fig.~\ref{fig:analytical}c provide a testable prediction for how the slope decreases with increasing $W$. This should be experimentally tested in the future. Finally, as $W$ increases and approaches the width of the lower block, the non-monotonicity is predicted to disappear, as demonstrated experimentally in Fig.~\ref{fig:experimental}c (blue symbols).

The results presented in this section demonstrate that global geometric features of the sliding bodies in a frictional problem, here a difference in their thickness, affect the frictional resistance to sliding and in fact makes it easier for interfacial rupture fronts that mediate sliding to propagate. In fact, the effect of geometric asymmetry is maximal at the extreme rupture velocities that are the norm in frictional sliding. This reduction in frictional dissipation applies to any engineering or tribological system involving identical materials and geometric asymmetry. As such, it implies that the design and friction control of any real-life tribological application must take into account not only the interfacial properties, but also the relative size of the sliding bodies. In the next section we show that the same concept applies to another class of important sliding friction problems, where a different form of geometric asymmetry controls the dynamic response of the system.

\section{Stability of frictional sliding}
\label{sec:stability}

We now focus on a different, yet conceptually related, physical situation in which geometric asymmetry plays a crucial role as well. While in Sect.~\ref{sec:thin-on-thick} geometric asymmetry was associated with a difference in the thickness of the sliding bodies, here its origin is a difference in their height. Moreover, while in Sect.~\ref{sec:thin-on-thick} we addressed the propagation of spatially-localized interfacial cracks, here the focus will be on the stability of homogeneous sliding. Yet, in both cases a geometry-induced coupling between interfacial slip and normal stress variations, encapsulated in the function $G_y$ in~Eq.~\eqref{eq:G}, is the dominant physical player.

We consider an elastic block of a height $H\1$ sliding atop a block of height $H\2\=\eta H\1$ (with a dimensionless positive $\eta$, $0\!<\!\eta\!<\!\infty$),
both made of the same material under plane-strain conditions~\cite{Timoshenko}, as depicted in Fig.~\ref{fig:geometries}b. Note that $\eta\!=\!1$ corresponds to a symmetric system.
The blocks initially slide at a fixed velocity and all of the fields are assumed to reach steady state. A homogeneous compressive normal stress $\sigma_{yy}\1\!=\!-\sigma_0$ is imposed at both $y\!=\!H\1$ and $y\=-H\2$. In addition, a constant velocity ${\dot u}\1_x\!=\!v$ in the positive $x$-direction is imposed at $y\=H\1$ and ${\dot u}\2_x\!=\!0$ at $y\!=\!-H\2$. In this problem, unlike the problem considered in Sect.~\ref{sec:thin-on-thick}, the interfacial dynamics are coupled to the boundaries at $y\=H\1,-H\2$, and hence the heights $H_{1,2}$ are expected to play a central role here.

To fully define the problem, one needs to specify the frictional boundary condition at the interface. Friction is commonly modeled as a linear relation between the interfacial normal stress and the interfacial shear stress (frictional resistance/stress), i.e.
\begin{align}
\sigma_{xy}&=-f(\cdot)\sigma_{yy} \ ,
\label{eq:friction_BC}
\end{align}
where $f(\cdot)$ represents the friction law. Our major goal here is to understand the destabilizing effect associated with geometric asymmetry, i.e.~$\eta\!\ne\!1$,
which to the best of knowledge has not been studied before. Consequently, in order to isolate the geometric effect, we will focus below on situations in which friction is intrinsically stabilizing such that any instability, if exists, is associated with the absence of geometrical reflection symmetry.

To achieve this, we proceed in two steps. First, in Sec.~\ref{sec:simplified},
we present a simplified analysis involving a simple velocity-dependent friction law and strong geometrical asymmetry.
This will allow us to gain much insight into the role of geometric asymmetry in frictional sliding and to clearly identify
the physical origin of instability. Then, in Sec.~\ref{sec:rsf}, we present a significantly generalized analysis for a realistic friction law, including an internal state variable and an interfacial memory length, and for any level of geometric asymmetry. The emerging results strengthen the findings of Sec.~\ref{sec:simplified} and extend them.

\subsection{Simplified analysis: Velocity-dependent friction and large geometric asymmetry}
\label{sec:simplified}

As a primer, we use here a simple friction law where $f(\cdot)$ in Eq.~\eqref{eq:friction_BC} depends only on the interfacial slip velocity $v\!\equiv\!\dot\epsilon_x$, i.e.~$f(v)$. We focus on velocity-strengthening interfaces, $f'(v)\!>\!0$, because in this case sliding is unconditionally stable for symmetric systems~\cite{Rice2001, Perfettini2008, SM} and thus the origin of any emerging instability must be associated with the absence of geometrical reflection symmetry.
Moreover, steady state velocity-strengthening friction has been recently shown to be a generic feature of dry interfaces over some velocity range~\cite{Bar-Sinai2014jgr}. Finally, to simplify the analysis further we consider the case in which the lower block is much higher than the upper one, $\eta\!\gg\!1$. That is, we take the limit $H\2\!\to\!\infty$, such that $H\1\!\equiv\!H$ is the only lengthscale in the problem.

%%%%%%%%%%%%%%%%%%%% Figure %%%%%%%%%%%%%%%%%%%%%%%%%
\begin{figure*}
\centering
 \includegraphics[width=\textwidth]{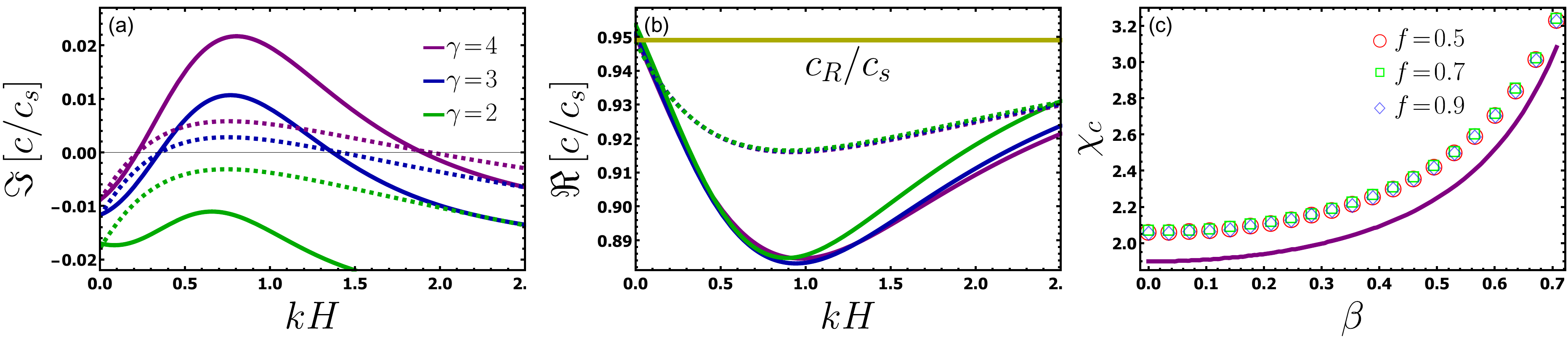}
  \caption{\textbf{Linear stability: Simplified analysis}. Imaginary (a) and real (b) parts of solutions to the linear stability spectrum in Eq.~\eqref{eq:implicit_spectrum}. $\Im(c)\!>\!0$ implies an instability and note that only one solution branch is discussed (other solution branches exist as well, but are not discussed here). The solid lines show numerical solutions to Eq.~\eqref{eq:implicit_spectrum} and the dashed lines show the approximate analytic solutions obtained by a linear expansion around $c\!=\!c_R$. The parameters used are $f\!=\!0.9$ and $\beta\!=\!0.3$, where $\gamma\!\equiv\!\mu/(\sigma_0 c_s f'(v))$ is varied according to the legend. (c) The instability threshold $\chi_c$, i.e.~for $\chi\!\equiv\!\gamma f\!<\!\chi_c$ sliding is stable for all $k$, vs.~$\beta\!\equiv\!c_s/c_d$. The open symbols show direct numerical results and the solid line is the prediction in Eq.~\eqref{eq:criterion}.}
  \label{fig:spectrum}
\end{figure*}
%%%%%%%%%%%%%%%%%%%%%%%%%%%%%%%%%%%%%%%%%%%%%%%%%%%

Under what conditions is homogeneous sliding stable? This question, which is of fundamental importance in a broad range of frictional problems
(see, for example~\cite{Rice1983, Armstrong-Helouvry1994, Olsson1998, Rice2001, Ranjith2001, Nguyen2003, Baumberger2006, Ikari2010, Pomeau2011, Bar-Sinai2013pre, Putelat2015,
Brener2015, BarSinai2012}), is first investigated in the context of the simplified problem defined above. As the interface is characterized by velocity-strengthening friction, $f'(v)\!>\!0$, friction itself tends to {\em stabilize} sliding. Consequently, the only possible destabilizing piece of physics can be the geometric-asymmetry-induced coupling between interfacial slip and normal stress variations, encapsulated in the function $G_y$ (cf.~Eq.~\eqref{eq:G}), which also played a crucial role in Sect.~\ref{sec:thin-on-thick}. Can geometric asymmetry destabilize velocity-strengthening frictional interfaces in much the same way as material asymmetry (the bi-material effect) can~\cite{Brener2015}?

To address the stability question, we perturb Eq.~\eqref{eq:friction_BC} to linear order, obtaining~\cite{SM}
\begin{eqnarray}
 \mu\,G_x(c,k)+i\,\mu\,f\,G_y(c,k)+i\,c\,\sigma_0\,\delta\!f/\delta v=0 \ ,
 \label{eq:implicit_spectrum}
\end{eqnarray}
which is an implicit equation defining the linear stability spectrum $c(k)$. In the simple velocity-dependent friction case considered here, we have $\delta\!f/\delta v\!=\!f'(v)$ (more general interfacial constitutive laws are considered in Sect.~\ref{sec:rsf}). Perturbations with $\Im[c]\!>\!0$ are unstable and will grow exponentially, while perturbations with $\Im[c]\!<\!0$ are stable (remember that $k\!>\!0$). An explicit calculation shows that $G_y$ reads~\cite{SM}
\begin{equation}
G_y\=\frac{c_s^2}{c^2}\left(\frac{2(\alpha_s^2+1)}{1\!+\!\tanh(kH \alpha_d)}-\frac{2(\alpha_s^2+1)}{1\!+\!\tanh(kH\alpha_s)}\right) \ ,
\label{eq:Gy_LSA}
\end{equation}
where $c_s$ and $c_d$ are respectively the shear and dilatational wave-speeds and $\alpha_{s,d}^2\!\equiv\!1-c^2/c_{s,d}^2$ was introduced.

The limit $H\!\to\!\infty$ amounts to a symmetric system, in which case $\eta\!\to\!1$, and indeed $G_y$ vanishes in this limit. We can thus expect the system to be unconditionally stable for $H\!\to\!\infty$.
$G_y$ also vanishes in the limit $H\!\to\!0$. Similarly, $G_x$ takes the form~\cite{SM}
\begin{equation}
G_x\!=\!\frac{c_s^2}{c^2}\!\left(\frac{\left(\alpha _s^2+1\right)^2\alpha_s^{-1}}{1\!+\!\tanh(kH \alpha_s)}\!-\!\frac{4 \alpha _d}{1\!+\!\tanh(kH \alpha_d)}\right) \ .
\label{eq:Gx_LSA}
\end{equation}

Equipped with the results for the dynamic response functions $G_i(c,k)$, the implicit equation for the spectrum, Eq.~\eqref{eq:implicit_spectrum}, can be in principle solved, at least numerically. The equation admits a few solution branches, and in general its analysis is far from trivial. However, since the purpose of the present discussion is not a complete analysis of Eq.~\eqref{eq:implicit_spectrum}, but rather a demonstration of the qualitative effect of the absence of geometrical reflection symmetry, we focus here on a particular branch of solutions which is shown in Fig.~\ref{fig:spectrum}a. It is observed that for a range of parameters, and for a finite range of wavenumbers, the solutions are unstable ($\Im[c]\!>\!0$). This is a direct numerical evidence that geometric asymmetry can destabilize systems which are otherwise stable (remember that $f'(v)\!>\!0$).

It seems natural at this point to ask under what conditions this instability is observed. What are the conditions on the various system parameters such that there will be a range of $k$'s for which $\Im[c(k)]\!>\!0$? As a prelude, we perform a dimensional analysis. Clearly, the only lengthscale in the problem is $H$ and indeed the wavenumber $k$ only appears in the dimensionless combination $kH$. Thus, large (small) $k$ is equivalent to large (small) $H$ and since $G_y$ vanishes in both limits $H\!\to\!0$ and $H\!\to\!\infty$, we expect to find unstable modes only in a finite range $k_{min}\!<\!k\!<\!k_{max}$, if any.

Another dimensionless combination is $\gamma\!\equiv\!\mu/(\sigma_0 c_s f'(v))$, which is the ratio of the elastodynamic quantity $\mu/c_s$ --- proportional to the so-called radiation damping factor for sliding~\citep{Rice1993, Rice2001, Crupi2013, Brener2015} --- and the response of the frictional stress to variations in the sliding velocity. As such, $\gamma$ quantifies the importance of elastodynamics, which tends to destablize sliding when geometrical asymmetry is present, relative to velocity-strengthening friction, which generically stabilizes sliding. We thus expect large $\gamma$ to promote instability, if $G_y\!\ne\!0$. In addition, as $G_y$ is the only possible source of instability in the problem, the appearance of $f G_y$ is associated with destabilization (because $f$ and $G_y$ enter the spectrum in Eq.~\eqref{eq:implicit_spectrum} only through the combination $f G_y$). Finally, the ratio of the two wave-speeds $\beta\!\equiv\!c_s/c_d\=\sqrt{(1-2\nu)/(2-2\nu)}$ is also a dimensionless parameter of the system which depends only on the bulk Poisson's ratio.

To obtain analytic insight into the instability presented in Fig.~\ref{fig:spectrum}a, note that solutions in this instability branch are located near the Rayleigh wave-speed, as shown in Fig.~\ref{fig:spectrum}b (note that here $c_R\!\simeq\!0.95c_s$). Consequently, we expand Eq.~\eqref{eq:implicit_spectrum} to linear order around $c\!=\!c_R+\delta c$, obtaining an explicit expression for $\delta c(kH)$~\cite{SM}. A consequence of this expansion is that the transition between stable and unstable modes occurs for $k$'s which approximately satisfy~\cite{SM}
\begin{equation}
 \gamma f G_y(c_R,k) \approx -c_R/c_s\ .
 \label{eq:criterion}
\end{equation}
This approximate stability criterion explains the existence of an instability and in fact gives reasonable quantitative estimates for its onset.

To see this, note that since $G_y(c_R,k)$ (which is negative, cf.~Eq.~\eqref{eq:Gy_LSA} and~\cite{SM}) vanishes for both $k\!=\!0$ and $k\!=\!\infty$, and attains a global minimum for $k$ of order $H^{-1}$, Eq.~\eqref{eq:criterion} admits solutions only for certain values of the product $\chi\!\equiv\!\gamma f$. When $\chi$ is smaller than a critical value $\chi_c$, no solutions exist and this branch of solutions is stable for all wavenumbers. Note that this criterion has exactly the expected structure: the instability is indeed governed by $G_y$, and large $\gamma$ or $f$ promote instability, which only happens at a finite range of wavenumbers. These predictions are quantitatively verified in Fig.~\ref{fig:spectrum}c. In addition, the real and imaginary parts of the approximate solution for $\delta c(kH)$~\cite{SM} are added to Figs.~\ref{fig:spectrum}a-b (dashed lines), demonstrating reasonable quantitative agreement with the full numerical solution for various parameters.

The results presented in this section demonstrate the destabilizing role that the absence of geometrical reflection symmetry may play in frictional dynamics. In the next section, we significantly extend the analysis to include more realistic friction laws and any geometric contrast.

\subsection{Generalized analysis: State dependence, memory length and arbitrary geometric asymmetry}
\label{sec:rsf}

The analysis presented in the previous section adopted two simplifying assumptions, i.e.~that the frictional response depends
only on the instantaneous slip velocity $v$ and that the lower block is much higher than the upper one, $\eta\!\to\!\infty$.
Frictional interfaces, however, are known to depend also on the state of the interface, not just on the slip velocity,
and obviously the sliding bodies can feature any geometric asymmetry, i.e.~the system can attain any value of $\eta$.
Consequently, our goal here is to relax these simplifying assumptions and to present a significantly generalized
analysis applicable to a broad range of realistic frictional systems.

It is experimentally well-established that the response of frictional interfaces depends, in addition to the slip velocity $v$,
on the state of the interface through the (normalized) real contact area $A(\phi)\!\propto\!\sigma_{yy}(1+\psi(\phi))$~\cite{Marone1998, Nakatani2001, Baumberger2006},
as discussed in relation to Eq.~\eqref{eq:A}. The auxiliary internal state variable $\phi$, which represents the age/maturity of
the contact and is of time dimensions, carries memory of
the history of the interface. This implies that irrespective of the exact functional form of $\psi(\phi)$
(with $d\psi/d\phi\!>\!0$) the frictional response $f(\cdot)$
in Eq.~\eqref{eq:friction_BC} depends on both $v$ and $\phi$, i.e.~we have $f(v,\phi)$.
Since $f$ does not depend solely on the instantaneous sliding velocity, but also on $\phi$, one should distinguish between
\begin{equation}
\pa_v\!f \equiv \frac{\pa f(v,\phi)}{\pa v} \qquad\hbox{and}\qquad d_v\!f\equiv \frac{df(v,\phi_0(v))}{dv}\ ,
\end{equation}
where $\phi_0(v)$ is the steady state value of $\phi$.

It is also well-established that after a rapid variation in $v$, accompanied by an instantaneous frictional response characterized by $\pa_v\!f$,
a new steady state is established over a characteristic slip distance $D$, which can be regarded as an interfacial memory length. This generic behaviour is described by the
following evolution equation for $\phi$~\cite{Rice1983, Marone1998, Baumberger2006}
\begin{equation}
\label{eq:rsf_phi}
\dot\phi = g\Big(\frac{v\,\phi}{D}\Big) \ ,
\end{equation}
with $g(1)\!=\!0$ and $g'(1)\!<\!0$. While several functions $g(\cdot)$ were proposed and extensively studied in the
literature~\cite{Marone1998, Baumberger2006}, the only property that affects the linear stability is $g'(1)$.
Note that if $g(0)\!>\!0$ (corresponding to $v\=0$), the equation describes frictional aging ($\phi$ increases linearly with time under quiescent conditions) and that $g(1)\!=\!0$ corresponds to steady state,
 $\dot\phi\!=\!0$, implying $\phi_0(v)\=D/v$. The latter describes contact rejuvenation, where the typical contact lifetime is inversely proportional to $v$.

The physics incorporated in the distinction between $\pa_v\!f$ and $d_v\!f$, and in the memory length $D$ --- within the
so-called rate-and-state friction constitutive framework --- imply the existence
of two dimensionless parameters that are absent in the simplified analysis of Sect.~\ref{sec:simplified}
\begin{equation}
\Delta\equiv \frac{d_v\!f}{\pa_v\!f}, \qquad\qquad \xi \equiv \frac{D c_s}{H v |g'(1)|} \ .
\end{equation}
Frictional interfaces generically feature $\pa_v\!f\!>\!0$~\cite{Marone1998, Nakatani2001, Baumberger2006}, which is termed the ``direct effect''
(associated with thermally activated rheology~\cite{Baumberger2006, Rice2001}). As in Sect.~\ref{sec:simplified}, we are interested in $d_v\!f\!>\!0$
(i.e.~in steady state velocity-strengthening friction), which implies a positive $\Delta$.
In fact, $\Delta$ varies in the range $0\!<\!\Delta\!<\!1$~\cite{Brener2015}, while $\xi$ can attain any positive value.

%%%%%%%%%%%%%%%%%%%% Figure %%%%%%%%%%%%%%%%%%%%%%%%%
\begin{figure*}
\centering
 \includegraphics[width=\textwidth]{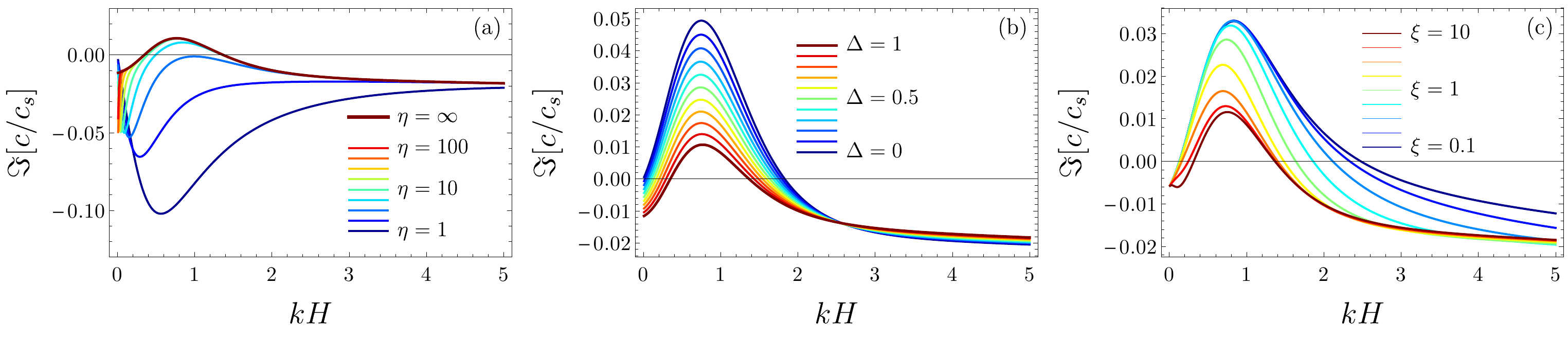}
  \caption{\textbf{Linear stability: Generalized analysis}. $\Im[c/c_s]$ (i.e.~the rate of exponential growth/decay of perturbations, $\Im[c]\!>\!0$ corresponds to instability) vs.~$kH$ for a broad range of physical parameters.
In all panels the parameters are the same as in Fig.~\ref{fig:spectrum} with $\gamma\!=\!3$.
(a) The dependence of $\Im[c(kH)/c_s]$ on $\eta$ for $\Delta\!=\!1$. The curve $\eta\!=\!\infty$ identifies with the blue curve of Fig.~\ref{fig:spectrum}a.
The case $\eta\!=\!1$ corresponds to a symmetric system and is thus stable for all $k$. (b) The dependence of $\Im[c(kH)/c_s]$ on $\Delta$ for $\xi\!=\!1$ and $\eta\!=\!\infty$.
(c) The dependence of $\Im[c(kH)/c_s]$ on $\xi$ for $\Delta\!=\!0.5$ and $\eta\!=\!\infty$.}
\label{fig:spectrum1}
\end{figure*}
%%%%%%%%%%%%%%%%%%%%%%%%%%%%%%%%%%%%%%%%%%%%%%%%%%%

Within this generalized framework, $\delta\!f/\delta v$ of Eq.~\eqref{eq:implicit_spectrum} takes the form~\cite{SM}
\begin{equation}
\label{eq:delta f RSF}
 \frac{\delta\!f}{\delta v}= \pa_v\!f\left(1 + \frac{\Delta - 1}{1-i\,\xi\frac{c}{c_s}kH}\right) \ .
\end{equation}
In the limit $\Delta\!\to\!1$, i.e.~when there is no distinction between $\pa_v\!f$ and $d_v\!f$ ($\pa_v\!f\!\to\!d_v\!f$), and
when $\xi\!\to\!0$, i.e.~when the memory length $D$ becomes vanishingly small, we obtain $\delta\!f/\delta v\!\to\!d_v\!f$.
This recovers the result of Sect.~\ref{sec:simplified} where $d_v\!f$ simply identifies with $f'(v)$.

To understand the effect of $\Delta$ and $\xi$ on frictional stability, we need to solve Eq.~\eqref{eq:implicit_spectrum} using
Eq.~\eqref{eq:delta f RSF}. As we also want to consider arbitrary values of the height ratio $\eta$, we should first derive expressions for the interfacial elastodynamic transfer
function $G_{x,y}$ for any $\eta$. The generalized result takes the form~\cite{SM}
\begin{widetext}
\begin{equation}
 \begin{split}
G_x&=\frac{c_s^2\left(1+\alpha_s^2\right)^2 \Big(\tanh (\eta  kH\alpha_d)+\tanh (kH\alpha_d)\Big)-4 \alpha_d\alpha_s\Big(\tanh (\eta  kH\alpha_s)+\tanh (kH\alpha_s)\Big)}{c^2\,\alpha_s \Big(\tanh (\eta  kH\alpha_d)+\tanh (kH\alpha_d)\Big) \Big(\tanh (\eta  kH\alpha_s)+\tanh (kH\alpha_s)\Big)}\ ,\\
G_y&=\frac{2c_s^2\left(1+\alpha_s^2\right)}{c^2}\frac{\tanh (kH\alpha_s) \tanh ( \eta  kH\alpha_d)-\tanh (kH\alpha_d ) \tanh ( \eta  kH\alpha_s)}{\Big(\tanh (\eta kH\alpha_d)+\tanh(kH\alpha_d)\Big) \Big(\tanh (\eta kH\alpha_s)+\tanh (kH\alpha_s)\Big)}\ ,
 \end{split}
 \label{eq:Gs_general}
\end{equation}
\end{widetext}
Note that Eqs.~\eqref{eq:Gy_LSA}-\eqref{eq:Gx_LSA} are obtained from Eq.~\eqref{eq:Gs_general} by taking the $\eta\!\to\!\infty$
limit, which amounts to setting $\tanh(\eta kH\alpha_i)$ to unity (since both $k$ and $\Re[\alpha_i]$ are positive).
In addition, as expected, $G_y$ vanishes for symmetric systems, i.e.~for $\eta\!=\!1$.

We are now ready to study the effect of the geometric dimensionless parameter $\eta$, and of the constitutive dimensionless
parameters $\Delta$ and $\xi$, on the linear stability of frictional interfaces. That is, we aim at solving the implicit linear
stability spectrum in Eq.~\eqref{eq:implicit_spectrum}, with Eqs.~\eqref{eq:delta f RSF}-\eqref{eq:Gs_general}.
The ultimate goal of such a generalized linear stability analysis is to derive the stability phase-diagram in the
$\gamma$ (here $\pa_v\!f$ replaces $f'(v)$ in the definition of $\gamma$ in Sect.~\ref{sec:simplified}), $f$, $\beta$, $\eta$, $\Delta$ and $\xi$ parameter space, where the stability boundary is a complex hypersurface
in this multi-dimensional space.

As it is obviously impossible to visualize this high-dimensional stability boundary and in order to gain clear physical insight,
we analyze this hypersurface by studying its sections along various parameter directions. A first step was done in Sect.~\ref{sec:simplified},
where the analysis was performed for fixed values of geometric asymmetry $\eta$, frictional resistance $f$
and wave-speed ratio $\beta$, while $\gamma$ varied. As a simple velocity-dependent friction model was adopted there,
we also had $\Delta\=1$. As observed in Fig.~\ref{fig:spectrum}a
and analyzed theoretically in relation to Eq.~\eqref{eq:criterion}, an instability emerges when $\gamma$ becomes sufficiently
large (here somewhere between $\gamma\=2$ and $\gamma\=3$). As $\gamma\!=\!\mu/(\sigma_0 c_s \pa_v\!f)$ quantifies the
importance of elastodynamics relative to instantaneous velocity-strengthening friction, the instability emerges when
elastodynamics becomes more dominant in the presence of large geometric asymmetry, $\eta\=\infty$.

Our next step is to isolate the geometric asymmetry effect embodied in $\eta$. We therefore use the parameters of
Fig.~\ref{fig:spectrum}a-b, together with $\gamma\=3$, and vary $\eta$ over a very broad range, essentially from $\eta\=1$ (corresponding to a symmetric system)
to $\eta\=\infty$. $\Im[c(kH)/c_s]$, obtained by numerically solving Eqs.~\eqref{eq:implicit_spectrum}, \eqref{eq:delta f RSF} and
\eqref{eq:Gs_general}, is shown in Fig.~\ref{fig:spectrum1}a. It is observed that for symmetric systems, $\eta\=1$, sliding is stable for all wave-numbers. As $\eta$ is
increased, $\Im[c(kH)/c_s]$ approaches the x-axis until they first intersect when $\eta\!\simeq\!3.3$ at $kH\!\sim\!{\cal O}(1)$, signaling the onset of instability. This result provides direct evidence for the destabilizing role played by geometric asymmetry in frictional sliding. As $\eta$ is further increased, the system becomes more unstable in the sense of an increased range of unstable wave-numbers and a larger growth rate. Obviously, the result in the $\eta\=\infty$ limit identifies with that of Fig.~\ref{fig:spectrum}a. In fact, the $\eta\=\infty$ analysis well captures the salient features of the instability spectrum
for $\eta$ values moderately above the critical value $\eta\!\simeq\!3.3$.

Next, we would like to understand the effect of $\Delta$, i.e. of a difference between the instantaneous response $\pa_v\!f$ and the steady state response $d_v\!f$, on the sliding
stability in the presence of geometric asymmetry. For that aim, we plot in Fig.~\ref{fig:spectrum1}b $\Im[c(kH)/c_s]$ for various values of $\Delta$, spanning the
whole range $0\!<\!\Delta\!<\!1$, and fixed $\xi\=1$ and $\eta\=\infty$. It is observed that as $d_v\!f$ decreases
relative to $\pa_v\!f$, i.e.~as $\Delta$ decreases, sliding becomes
less stable, resulting in a broader range of unstable wave-numbers and a larger instability growth rate. This result demonstrates the stabilizing role played by steady state velocity-strengthening friction in frictional sliding. We note, though, that the qualitative properties of the
instability spectrum are rather well captured by the $\Delta\=1$ analysis (i.e.~for velocity-dependent friction, where no
distinction is made between $d_v\!f$ and $\pa_v\!f$). We stress that while $\Delta$ affects the properties of instability,
the origin of instability is still geometric asymmetry (i.e.~sufficiently large $\eta$).

Finally, we explore the effect of varying the interfacial memory length $D$, corresponding to varying $\xi$, on frictional stability in the presence of geometric asymmetry.
We plot in Fig.~\ref{fig:spectrum1}c $\Im[c(kH)/c_s]$ for a broad range of $\xi$ values, and fixed $\eta\=\infty$ and
$\Delta\=0.5$. It is observed that increasing $D$ (i.e.~$\xi$) tends to stabilize sliding
(i.e.~shrink the instability range and growth rate) as it makes the real contact area less sensitive to slip velocity perturbations. We also stress here that while $\xi$ affects the range and growth rate of instability, its origin is geometric asymmetry (i.e.~sufficiently large $\eta$).

The results presented in this section provide a rather comprehensive physical picture of the implications of geometric asymmetry on the stability of frictional
sliding, and of the interplay between geometric asymmetry and generic constitutive properties of frictional interfaces, most notably the effect of the state of the interface
and of an interfacial memory length. The results significantly extend those presented in Sect.~\ref{sec:simplified}, yet they show that the simplified analysis properly captured
the destabilizing geometric asymmetry effect. We stress again that additional solutions to Eq.~\eqref{eq:implicit_spectrum}
(with Eqs.~\eqref{eq:delta f RSF}-\eqref{eq:Gs_general}) exist. These additional solution branches, along with a more detailed analysis of the multi-dimensional stability phase-diagram, will be presented in a follow-up report.

The results presented in this section regarding the stability of homogeneous sliding in the presence of geometric asymmetry may have far reaching implications for the dynamics of frictional interfaces in a variety of frictional systems. Under homogeneous loading applied to the top of long enough sliding bodies, as assumed in the analysis, we predict that no homogeneous steady state will be established experimentally under certain conditions that were carefully quantified. Instead, the interface separating geometrically asymmetric bodies will experience inhomogeneous slip related to the most unstable mode identified in the analysis. This will lead to spatiotemporal stick-slip-like motion, accompanied by distinct acoustic signature as in squeaking door hinges.

In frictional systems where the loading configuration promotes inhomogeneous slip, the obtained results may still be relevant. Inhomogeneous slip in slowly driven frictional interfaces typically takes the form of an expanding creep patch. The conditions under which an expanding creep patch spontaneously generates rapid/unstable slip, an important process known as nucleation, may be related to the minimal unstable wavelength in the stability analysis presented in this section for geometrically asymmetric systems. In particular, the minimal unstable wavelength may determine the size at which the expanding creep patch loses stability.

Finally, when rapid slip develops, it is typically mediated by the propagation of rupture modes. Which mode is actually realized in a given experimental system may be affected by the stability analysis presented here. In particular, extended crack-like rupture modes leave behind them a homogeneous sliding state, which may be precluded under certain conditions predicted by our analysis. Instead, localized pulse-like rupture modes may develop. Consequently, the results presented in this section may affect rupture modes selection, a basic open problem in the field of friction. Additional theoretical and
experimental research should be carried out in order to fully explore these potential implications.

\section{Concluding remarks}
\label{sec:conclusions}

In this paper, combining experiments and theory, we showed that frictional interfaces which separate bodies made of identical materials, but lack geometric reflection symmetry about the interface, generically feature coupling between interfacial slip and normal stress variations. This geometric asymmetry effect is shown to account for a sizable, and previously unexplained, normal-stress-induced weakening effect in frictional cracks. New experiments support the theoretical predictions. We then showed that geometric asymmetry can destabilize homogeneous sliding with velocity-strengthening friction which is otherwise stable. These analyses demonstrate that the effect of geometric asymmetry resembles, sometimes qualitatively and sometimes semi-quantitatively, that of material asymmetry (the bi-material effect).

Since no system is perfectly symmetric, we expect the geometrically-induced coupling between interfacial slip and normal stress variations to generically exist in a broad range of man-made and natural frictional systems. Consequently, it should be incorporated into various theoretical approaches, into engineering models and employed
in interpreting experimental observations. The implications in geophysical contexts, such as in subduction zone sliding (cf.~Fig.~\ref{fig:geometries}c), call for further investigation.

\textit{Acknowledgements}
E.B. acknowledges support of the Israel Science Foundation (grant 295/16), the William Z. and Eda Bess Novick Young Scientist Fund, COST Action
MP1303 and of the Harold Perlman Family. J.F. and E.B. acknowledge support of the James S. McDonnell Fund (grant 220020221). J.F. and I.S. acknowledge support of the European Research Council (grant 267256) and the Israel Science Foundation (grants 76/11 and 1523/15).

%\bibliographystyle{apsrev4-1_PRX}
%\bibliography{/home/yohai/Dropbox/mendBib/lib2}

\begin{thebibliography}{82}%
\makeatletter
\providecommand \@ifxundefined [1]{%
 \@ifx{#1\undefined}
}%
\providecommand \@ifnum [1]{%
 \ifnum #1\expandafter \@firstoftwo
 \else \expandafter \@secondoftwo
 \fi
}%
\providecommand \@ifx [1]{%
 \ifx #1\expandafter \@firstoftwo
 \else \expandafter \@secondoftwo
 \fi
}%
\providecommand \natexlab [1]{#1}%
\providecommand \enquote  [1]{``#1''}%
\providecommand \bibnamefont  [1]{#1}%
\providecommand \bibfnamefont [1]{#1}%
\providecommand \citenamefont [1]{#1}%
\providecommand \href@noop [0]{\@secondoftwo}%
\providecommand \href [0]{\begingroup \@sanitize@url \@href}%
\providecommand \@href[1]{\@@startlink{#1}\@@href}%
\providecommand \@@href[1]{\endgroup#1\@@endlink}%
\providecommand \@sanitize@url [0]{\catcode `\\12\catcode `\$12\catcode
  `\&12\catcode `\#12\catcode `\^12\catcode `\_12\catcode `\%12\relax}%
\providecommand \@@startlink[1]{}%
\providecommand \@@endlink[0]{}%
\providecommand \url  [0]{\begingroup\@sanitize@url \@url }%
\providecommand \@url [1]{\endgroup\@href {#1}{\urlprefix }}%
\providecommand \urlprefix  [0]{URL }%
\providecommand \Eprint [0]{\href }%
\providecommand \doibase [0]{http://dx.doi.org/}%
\providecommand \selectlanguage [0]{\@gobble}%
\providecommand \bibinfo  [0]{\@secondoftwo}%
\providecommand \bibfield  [0]{\@secondoftwo}%
\providecommand \translation [1]{[#1]}%
\providecommand \BibitemOpen [0]{}%
\providecommand \bibitemStop [0]{}%
\providecommand \bibitemNoStop [0]{.\EOS\space}%
\providecommand \EOS [0]{\spacefactor3000\relax}%
\providecommand \BibitemShut  [1]{\csname bibitem#1\endcsname}%
\let\auto@bib@innerbib\@empty
%</preamble>
\bibitem [{\citenamefont {Marone}(1998)}]{Marone1998}%
  \BibitemOpen
  \bibfield  {author} {\bibinfo {author} {\bibfnamefont {C.~J.}\ \bibnamefont
  {Marone}},\ }\bibfield  {title} {\emph {\bibinfo {title} {{Laboratory-derived
  friction laws and their application to seismic faulting}},\ }}\href
  {http://arjournals.annualreviews.org/doi/abs/10.1146/annurev.earth.26.1.643}
  {\bibfield  {journal} {\bibinfo  {journal} {Annu. Rev. Earth Planet. Sci.}\
  }\textbf {\bibinfo {volume} {26}},\ \bibinfo {pages} {643} (\bibinfo {year}
  {1998})}\BibitemShut {NoStop}%
\bibitem [{\citenamefont {Nakatani}(2001)}]{Nakatani2001}%
  \BibitemOpen
  \bibfield  {author} {\bibinfo {author} {\bibfnamefont {M.}~\bibnamefont
  {Nakatani}},\ }\bibfield  {title} {\emph {\bibinfo {title} {{Conceptual and
  physical clarification of rate and state friction: Frictional sliding as a
  thermally activated rheology}},\ }}\href
  {http://www.agu.org/pubs/crossref/2001/2000JB900453.shtml} {\bibfield
  {journal} {\bibinfo  {journal} {J. Geophys. Res.}\ }\textbf {\bibinfo
  {volume} {106}},\ \bibinfo {pages} {13347} (\bibinfo {year}
  {2001})}\BibitemShut {NoStop}%
\bibitem [{\citenamefont {Baumberger}\ and\ \citenamefont
  {Caroli}(2006)}]{Baumberger2006}%
  \BibitemOpen
  \bibfield  {author} {\bibinfo {author} {\bibfnamefont {T.}~\bibnamefont
  {Baumberger}}\ and\ \bibinfo {author} {\bibfnamefont {C.}~\bibnamefont
  {Caroli}},\ }\bibfield  {title} {\emph {\bibinfo {title} {{Solid friction
  from stick–slip down to pinning and aging}},\ }}\href
  {http://dx.doi.org/10.1080/00018730600732186}
  {\bibfield  {journal} {\bibinfo  {journal} {Adv. Phys.}\ }\textbf {\bibinfo
  {volume} {55}},\ \bibinfo {pages} {279} (\bibinfo {year} {2006})}\BibitemShut
  {NoStop}%
\bibitem [{\citenamefont {Putelat}\ \emph {et~al.}(2011)\citenamefont
  {Putelat}, \citenamefont {Dawes},\ and\ \citenamefont
  {Willis}}]{Putelat2011}%
  \BibitemOpen
  \bibfield  {author} {\bibinfo {author} {\bibfnamefont {T.}~\bibnamefont
  {Putelat}}, \bibinfo {author} {\bibfnamefont {J.~H.}\ \bibnamefont {Dawes}},
  \ and\ \bibinfo {author} {\bibfnamefont {J.~R.}\ \bibnamefont {Willis}},\
  }\bibfield  {title} {\emph {\bibinfo {title} {{On the microphysical
  foundations of rate-and-state friction}},\ }}\href
  {http://linkinghub.elsevier.com/retrieve/pii/S0022509611000305} {\bibfield
  {journal} {\bibinfo  {journal} {J. Mech. Phys. Solids}\ }\textbf {\bibinfo
  {volume} {59}},\ \bibinfo {pages} {1062} (\bibinfo {year}
  {2011})}\BibitemShut {NoStop}%
\bibitem [{\citenamefont {Ikari}\ \emph {et~al.}(2013)\citenamefont {Ikari},
  \citenamefont {Marone}, \citenamefont {Saffer},\ and\ \citenamefont
  {Kopf}}]{Ikari2013}%
  \BibitemOpen
  \bibfield  {author} {\bibinfo {author} {\bibfnamefont {M.~J.}\ \bibnamefont
  {Ikari}}, \bibinfo {author} {\bibfnamefont {C.~J.}\ \bibnamefont {Marone}},
  \bibinfo {author} {\bibfnamefont {D.~M.}\ \bibnamefont {Saffer}}, \ and\
  \bibinfo {author} {\bibfnamefont {A.~J.}\ \bibnamefont {Kopf}},\ }\bibfield
  {title} {\emph {\bibinfo {title} {{Slip weakening as a mechanism for slow
  earthquakes}},\ }}\href {http://www.nature.com/doifinder/10.1038/ngeo1818}
  {\bibfield  {journal} {\bibinfo  {journal} {Nat. Geosci.}\ }\textbf {\bibinfo
  {volume} {6}},\ \bibinfo {pages} {468} (\bibinfo {year} {2013})}\BibitemShut
  {NoStop}%
\bibitem [{\citenamefont {Bar-Sinai}\ \emph {et~al.}(2014)\citenamefont
  {Bar-Sinai}, \citenamefont {Spatschek}, \citenamefont {Brener},\ and\
  \citenamefont {Bouchbinder}}]{Bar-Sinai2014jgr}%
  \BibitemOpen
  \bibfield  {author} {\bibinfo {author} {\bibfnamefont {Y.}~\bibnamefont
  {Bar-Sinai}}, \bibinfo {author} {\bibfnamefont {R.}~\bibnamefont
  {Spatschek}}, \bibinfo {author} {\bibfnamefont {E.~A.}\ \bibnamefont
  {Brener}}, \ and\ \bibinfo {author} {\bibfnamefont {E.}~\bibnamefont
  {Bouchbinder}},\ }\bibfield  {title} {\emph {\bibinfo {title} {{On the
  velocity-strengthening behavior of dry friction}},\ }}\href
  {http://doi.wiley.com/10.1002/2013JB010586}
  {\bibfield  {journal} {\bibinfo  {journal} {J. Geophys. Res. Solid Earth}\
  }\textbf {\bibinfo {volume} {119}},\ \bibinfo {pages} {1738} (\bibinfo {year}
  {2014})}\BibitemShut {NoStop}%
\bibitem [{\citenamefont {{Di Toro}}\ \emph {et~al.}(2004)\citenamefont {{Di
  Toro}}, \citenamefont {Goldsby},\ and\ \citenamefont {Tullis}}]{DiToro2004}%
  \BibitemOpen
  \bibfield  {author} {\bibinfo {author} {\bibfnamefont {G.}~\bibnamefont {{Di
  Toro}}}, \bibinfo {author} {\bibfnamefont {D.~L.}\ \bibnamefont {Goldsby}}, \
  and\ \bibinfo {author} {\bibfnamefont {T.~E.}\ \bibnamefont {Tullis}},\
  }\bibfield  {title} {\emph {\bibinfo {title} {{Friction falls towards zero in
  quartz rock as slip velocity approaches seismic rates.}}\ }}\href
  {http://www.ncbi.nlm.nih.gov/pubmed/14749829} {\bibfield  {journal} {\bibinfo
   {journal} {Nature}\ }\textbf {\bibinfo {volume} {427}},\ \bibinfo {pages}
  {436} (\bibinfo {year} {2004})}\BibitemShut {NoStop}%
\bibitem [{\citenamefont {Rice}(2006)}]{Rice2006}%
  \BibitemOpen
  \bibfield  {author} {\bibinfo {author} {\bibfnamefont {J.~R.}\ \bibnamefont
  {Rice}},\ }\bibfield  {title} {\emph {\bibinfo {title} {{Heating and
  weakening of faults during earthquake slip}},\ }}\href
  {http://doi.wiley.com/10.1029/2005JB004006} {\bibfield  {journal} {\bibinfo
  {journal} {J. Geophys. Res.}\ }\textbf {\bibinfo {volume} {111}},\ \bibinfo
  {pages} {B05311} (\bibinfo {year} {2006})}\BibitemShut {NoStop}%
\bibitem [{\citenamefont {Goldsby}\ and\ \citenamefont
  {Tullis}(2011)}]{Goldsby2011}%
  \BibitemOpen
  \bibfield  {author} {\bibinfo {author} {\bibfnamefont {D.~L.}\ \bibnamefont
  {Goldsby}}\ and\ \bibinfo {author} {\bibfnamefont {T.~E.}\ \bibnamefont
  {Tullis}},\ }\bibfield  {title} {\emph {\bibinfo {title} {{Flash heating
  leads to low frictional strength of crustal rocks at earthquake slip
  rates}},\ }}\href {http://www.sciencemag.org/cgi/doi/10.1126/science.1207902}
  {\bibfield  {journal} {\bibinfo  {journal} {Science}\ }\textbf {\bibinfo
  {volume} {334}},\ \bibinfo {pages} {216} (\bibinfo {year}
  {2011})}\BibitemShut {NoStop}%
\bibitem [{\citenamefont {Chang}\ \emph {et~al.}(2012)\citenamefont {Chang},
  \citenamefont {Lockner},\ and\ \citenamefont {Reches}}]{Chang2012}%
  \BibitemOpen
  \bibfield  {author} {\bibinfo {author} {\bibfnamefont {J.~C.}\ \bibnamefont
  {Chang}}, \bibinfo {author} {\bibfnamefont {D.~A.}\ \bibnamefont {Lockner}},
  \ and\ \bibinfo {author} {\bibfnamefont {Z.}~\bibnamefont {Reches}},\
  }\bibfield  {title} {\emph {\bibinfo {title} {{Rapid acceleration leads to
  rapid weakening in earthquake-like laboratory experiments.}}\ }}\href
  {http://www.ncbi.nlm.nih.gov/pubmed/23042892} {\bibfield  {journal} {\bibinfo
   {journal} {Science}\ }\textbf {\bibinfo {volume} {338}},\ \bibinfo {pages}
  {101} (\bibinfo {year} {2012})}\BibitemShut {NoStop}%
\bibitem [{\citenamefont {Weertman}(1963)}]{Weertman1963}%
  \BibitemOpen
  \bibfield  {author} {\bibinfo {author} {\bibfnamefont {J.}~\bibnamefont
  {Weertman}},\ }\bibfield  {title} {\emph {\bibinfo {title} {{Dislocations
  moving uniformly on the interface between isotropic media of different
  elastic properties}},\ }}\href
  {http://www.sciencedirect.com/science/article/pii/0022509663900528}
  {\bibfield  {journal} {\bibinfo  {journal} {J. Mech. Phys. Solids}\ }\textbf
  {\bibinfo {volume} {11}},\ \bibinfo {pages} {197} (\bibinfo {year}
  {1963})}\BibitemShut {NoStop}%
\bibitem [{\citenamefont {Weertman}(1980)}]{Weertman1980}%
  \BibitemOpen
  \bibfield  {author} {\bibinfo {author} {\bibfnamefont {J.}~\bibnamefont
  {Weertman}},\ }\bibfield  {title} {\emph {\bibinfo {title} {{Unstable
  slippage across a fault that separates elastic media of different elastic
  constants}},\ }}\href {http://doi.wiley.com/10.1029/JB085iB03p01455}
  {\bibfield  {journal} {\bibinfo  {journal} {J. Geophys. Res.}\ }\textbf
  {\bibinfo {volume} {85}},\ \bibinfo {pages} {1455} (\bibinfo {year}
  {1980})}\BibitemShut {NoStop}%
\bibitem [{\citenamefont {Freund}(1990)}]{Freund1990}%
  \BibitemOpen
  \bibfield  {author} {\bibinfo {author} {\bibfnamefont {L.}~\bibnamefont
  {Freund}},\ }\href
  {http://www.google.co.il/books?hl=en{\&}lr={\&}id=B2sophpCOIYC{\&}pgis=1}
  {\emph {\bibinfo {title} {{Dynamic Fracture Mechanics}}}}\ (\bibinfo
  {publisher} {Cambridge University Press},\ \bibinfo {year} {1990})\ p.\
  \bibinfo {pages} {563}\BibitemShut {NoStop}%
\bibitem [{\citenamefont {Adams}(1995)}]{Adams1995}%
  \BibitemOpen
  \bibfield  {author} {\bibinfo {author} {\bibfnamefont {G.~G.}\ \bibnamefont
  {Adams}},\ }\bibfield  {title} {\emph {\bibinfo {title} {{Self-excited
  oscillations of two elastic half-spaces sliding with a constant coefficient
  of friction}},\ }}\href
  {http://appliedmechanics.asmedigitalcollection.asme.org/article.aspx?articleid=1411672}
  {\bibfield  {journal} {\bibinfo  {journal} {J. Appl. Mech.}\ }\textbf
  {\bibinfo {volume} {62}},\ \bibinfo {pages} {867} (\bibinfo {year}
  {1995})}\BibitemShut {NoStop}%
\bibitem [{\citenamefont {Martins}\ and\ \citenamefont
  {Sim{\~{o}}es}(1995)}]{Martins1995a}%
  \BibitemOpen
  \bibfield  {author} {\bibinfo {author} {\bibfnamefont {J.}~\bibnamefont
  {Martins}}\ and\ \bibinfo {author} {\bibfnamefont {F.~M.~F.}\ \bibnamefont
  {Sim{\~{o}}es}},\ }in\ \href
  {http://link.springer.com/10.1007/978-1-4615-1983-6{\_}14} {\emph {\bibinfo
  {booktitle} {Contact Mechanics}}},\ \bibinfo {editor} {edited by\ \bibinfo
  {editor} {\bibfnamefont {M.}~\bibnamefont {Raous}}, \bibinfo {editor}
  {\bibfnamefont {M.}~\bibnamefont {Jean}}, \ and\ \bibinfo {editor}
  {\bibfnamefont {J.}~\bibnamefont {Moreau}}}\ (\bibinfo  {publisher} {Springer
  US},\ \bibinfo {year} {1995})\ pp.\ \bibinfo {pages} {95--106}\BibitemShut
  {NoStop}%
\bibitem [{\citenamefont {Martins}\ \emph {et~al.}(1995)\citenamefont
  {Martins}, \citenamefont {Guimarães},\ and\ \citenamefont
  {Faria}}]{Martins1995b}%
  \BibitemOpen
  \bibfield  {author} {\bibinfo {author} {\bibfnamefont {J.~A.~C.}\
  \bibnamefont {Martins}}, \bibinfo {author} {\bibfnamefont {J.}~\bibnamefont
  {Guimarães}}, \ and\ \bibinfo {author} {\bibfnamefont {L.~O.}\ \bibnamefont
  {Faria}},\ }\bibfield  {title} {\emph {\bibinfo {title} {{Dynamic surface
  solutions in linear elasticity and viscoelasticity with frictional boundary
  conditions}},\ }}\href
  {http://vibrationacoustics.asmedigitalcollection.asme.org/article.aspx?articleid=1469639}
  {\bibfield  {journal} {\bibinfo  {journal} {J. Vib. Acoust.}\ }\textbf
  {\bibinfo {volume} {117}},\ \bibinfo {pages} {445} (\bibinfo {year}
  {1995})}\BibitemShut {NoStop}%
\bibitem [{\citenamefont {Andrews}\ and\ \citenamefont
  {Ben-Zion}(1997)}]{Andrews1997}%
  \BibitemOpen
  \bibfield  {author} {\bibinfo {author} {\bibfnamefont {D.~J.}\ \bibnamefont
  {Andrews}}\ and\ \bibinfo {author} {\bibfnamefont {Y.}~\bibnamefont
  {Ben-Zion}},\ }\bibfield  {title} {\emph {\bibinfo {title} {{Wrinkle-like
  slip pulse on a fault between different materials}},\ }}\href
  {http://doi.wiley.com/10.1029/96JB02856} {\bibfield  {journal} {\bibinfo
  {journal} {J. Geophys. Res. Solid Earth}\ }\textbf {\bibinfo {volume}
  {102}},\ \bibinfo {pages} {553} (\bibinfo {year} {1997})}\BibitemShut
  {NoStop}%
\bibitem [{\citenamefont {Adams}(1998)}]{Adams1998}%
  \BibitemOpen
  \bibfield  {author} {\bibinfo {author} {\bibfnamefont {G.~G.}\ \bibnamefont
  {Adams}},\ }\bibfield  {title} {\emph {\bibinfo {title} {{Steady sliding of
  two elastic half-spaces with friction reduction due to interface
  stick-slip}},\ }}\href {http://dx.doi.org/10.1115/1.2789077} {\bibfield
  {journal} {\bibinfo  {journal} {J. Appl. Mech.}\ }\textbf {\bibinfo {volume}
  {65}},\ \bibinfo {pages} {470} (\bibinfo {year} {1998})}\BibitemShut
  {NoStop}%
\bibitem [{\citenamefont {Adams}(2001)}]{Adams2000}%
  \BibitemOpen
  \bibfield  {author} {\bibinfo {author} {\bibfnamefont {G.~G.}\ \bibnamefont
  {Adams}},\ }\bibfield  {title} {\emph {\bibinfo {title} {{An intersonic slip
  pulse at a frictional interface between dissimilar materials}},\ }}\href
  {http://appliedmechanics.asmedigitalcollection.asme.org/article.aspx?articleid=1555320}
  {\bibfield  {journal} {\bibinfo  {journal} {J. Appl. Mech.}\ }\textbf
  {\bibinfo {volume} {68}},\ \bibinfo {pages} {81} (\bibinfo {year}
  {2001})}\BibitemShut {NoStop}%
\bibitem [{\citenamefont {Ranjith}\ and\ \citenamefont
  {Rice}(2001)}]{Ranjith2001}%
  \BibitemOpen
  \bibfield  {author} {\bibinfo {author} {\bibfnamefont {K.}~\bibnamefont
  {Ranjith}}\ and\ \bibinfo {author} {\bibfnamefont {J.~R.}\ \bibnamefont
  {Rice}},\ }\bibfield  {title} {\emph {\bibinfo {title} {{Slip dynamics at an
  interface between dissimilar materials}},\ }}\href
  {http://linkinghub.elsevier.com/retrieve/pii/S0022509600000296} {\bibfield
  {journal} {\bibinfo  {journal} {J. Mech. Phys. Solids}\ }\textbf {\bibinfo
  {volume} {49}},\ \bibinfo {pages} {341} (\bibinfo {year} {2001})}\BibitemShut
  {NoStop}%
\bibitem [{\citenamefont {Adda-Bedia}\ and\ \citenamefont {{Ben
  Amar}}(2003)}]{Adda-Bedia2003}%
  \BibitemOpen
  \bibfield  {author} {\bibinfo {author} {\bibfnamefont {M.}~\bibnamefont
  {Adda-Bedia}}\ and\ \bibinfo {author} {\bibfnamefont {M.}~\bibnamefont {{Ben
  Amar}}},\ }\bibfield  {title} {\emph {\bibinfo {title} {{Self-sustained slip
  pulses of finite size between dissimilar materials}},\ }}\href
  {http://linkinghub.elsevier.com/retrieve/pii/S0022509603000681} {\bibfield
  {journal} {\bibinfo  {journal} {J. Mech. Phys. Solids}\ }\textbf {\bibinfo
  {volume} {51}},\ \bibinfo {pages} {1849} (\bibinfo {year}
  {2003})}\BibitemShut {NoStop}%
\bibitem [{\citenamefont {Kammer}\ \emph {et~al.}(2014)\citenamefont {Kammer},
  \citenamefont {Yastrebov}, \citenamefont {Anciaux},\ and\ \citenamefont
  {Molinari}}]{Kammer201440}%
  \BibitemOpen
  \bibfield  {author} {\bibinfo {author} {\bibfnamefont {D.}~\bibnamefont
  {Kammer}}, \bibinfo {author} {\bibfnamefont {V.}~\bibnamefont {Yastrebov}},
  \bibinfo {author} {\bibfnamefont {G.}~\bibnamefont {Anciaux}}, \ and\
  \bibinfo {author} {\bibfnamefont {J.}~\bibnamefont {Molinari}},\ }\bibfield
  {title} {\emph {\bibinfo {title} {The existence of a critical length scale in
  regularised friction},\ }}\href {\doibase
  http://dx.doi.org/10.1016/j.jmps.2013.10.007} {\bibfield  {journal} {\bibinfo
   {journal} {J. Mech. Phys. Solids}\ }\textbf
  {\bibinfo {volume} {63}},\ \bibinfo {pages} {40} (\bibinfo {year}
  {2014})}\BibitemShut {NoStop}%
\bibitem [{\citenamefont {Brener}\ \emph {et~al.}(2016)\citenamefont {Brener},
  \citenamefont {Weikamp}, \citenamefont {Spatschek}, \citenamefont
  {Bar-Sinai},\ and\ \citenamefont {Bouchbinder}}]{Brener2015}%
  \BibitemOpen
  \bibfield  {author} {\bibinfo {author} {\bibfnamefont {E.~A.}\ \bibnamefont
  {Brener}}, \bibinfo {author} {\bibfnamefont {M.}~\bibnamefont {Weikamp}},
  \bibinfo {author} {\bibfnamefont {R.}~\bibnamefont {Spatschek}}, \bibinfo
  {author} {\bibfnamefont {Y.}~\bibnamefont {Bar-Sinai}}, \ and\ \bibinfo
  {author} {\bibfnamefont {E.}~\bibnamefont {Bouchbinder}},\ }\bibfield
  {title} {\emph {\bibinfo {title} {{Dynamic instabilities of frictional
  sliding at a bimaterial interface}},\ }}\href
  {http://linkinghub.elsevier.com/retrieve/pii/S0022509616300370} {\bibfield
  {journal} {\bibinfo  {journal} {J. Mech. Phys. Solids}\ }\textbf {\bibinfo
  {volume} {89}},\ \bibinfo {pages} {149} (\bibinfo {year} {2016})}\BibitemShut
  {NoStop}%
\bibitem [{\citenamefont {Madariaga}(1976)}]{Madariaga1976}%
  \BibitemOpen
  \bibfield  {author} {\bibinfo {author} {\bibfnamefont {R.}~\bibnamefont
  {Madariaga}},\ }\bibfield  {title} {\emph {\bibinfo {title} {{Dynamics of an
  expanding circular fault}},\ }}\href@noop {} {\bibfield  {journal} {\bibinfo
  {journal} {Bull. Seismol. Soc. Am.}\ }\textbf {\bibinfo {volume} {66}},\
  \bibinfo {pages} {639} (\bibinfo {year} {1976})}\BibitemShut {NoStop}%
\bibitem [{\citenamefont {Madariaga}(1977)}]{Madariaga1977}%
  \BibitemOpen
  \bibfield  {author} {\bibinfo {author} {\bibfnamefont {R.}~\bibnamefont
  {Madariaga}},\ }\bibfield  {title} {\emph {\bibinfo {title} {{High-frequency
  radiation from crack (stress drop) models of earthquake faulting}},\ }}\href
  {http://gji.oxfordjournals.org/cgi/doi/10.1111/j.1365-246X.1977.tb04211.x}
  {\bibfield  {journal} {\bibinfo  {journal} {Geophys. J. Int.}\ }\textbf
  {\bibinfo {volume} {51}},\ \bibinfo {pages} {625} (\bibinfo {year}
  {1977})}\BibitemShut {NoStop}%
\bibitem [{\citenamefont {Rice}(1993)}]{Rice1993}%
  \BibitemOpen
  \bibfield  {author} {\bibinfo {author} {\bibfnamefont {J.~R.}\ \bibnamefont
  {Rice}},\ }\bibfield  {title} {\emph {\bibinfo {title} {{Spatio-temporal
  complexity of slip on a fault}},\ }}\href
  {http://doi.wiley.com/10.1029/93JB00191} {\bibfield  {journal} {\bibinfo
  {journal} {J. Geophys. Res.}\ }\textbf {\bibinfo {volume} {98}},\ \bibinfo
  {pages} {9885} (\bibinfo {year} {1993})}\BibitemShut {NoStop}%
\bibitem [{\citenamefont {Ben-Zion}\ and\ \citenamefont
  {Rice}(1995)}]{Ben-Zion1995}%
  \BibitemOpen
  \bibfield  {author} {\bibinfo {author} {\bibfnamefont {Y.}~\bibnamefont
  {Ben-Zion}}\ and\ \bibinfo {author} {\bibfnamefont {J.~R.}\ \bibnamefont
  {Rice}},\ }\bibfield  {title} {\emph {\bibinfo {title} {{Slip patterns and
  earthquake populations along different classes of faults in elastic
  solids}},\ }}\href {http://doi.wiley.com/10.1029/94JB03037} {\bibfield
  {journal} {\bibinfo  {journal} {J. Geophys. Res.}\ }\textbf {\bibinfo
  {volume} {100}},\ \bibinfo {pages} {12959} (\bibinfo {year}
  {1995})}\BibitemShut {NoStop}%
\bibitem [{\citenamefont {Fukuyama}\ and\ \citenamefont
  {Madariaga}(1998)}]{Fukuyama1998}%
  \BibitemOpen
  \bibfield  {author} {\bibinfo {author} {\bibfnamefont {E.}~\bibnamefont
  {Fukuyama}}\ and\ \bibinfo {author} {\bibfnamefont {R.}~\bibnamefont
  {Madariaga}},\ }\bibfield  {title} {\emph {\bibinfo {title} {{Rupture
  dynamics of a planar fault in a 3D elastic medium: rate-and slip-weakening
  friction}},\ }}\href@noop {} {\bibfield  {journal} {\bibinfo  {journal}
  {Bull. Seismol. Soc. Am.}\ }\textbf {\bibinfo {volume} {88}},\ \bibinfo
  {pages} {1} (\bibinfo {year} {1998})}\BibitemShut {NoStop}%
\bibitem [{\citenamefont {Broberg}(1999)}]{Broberg1999Book}%
  \BibitemOpen
  \bibfield  {author} {\bibinfo {author} {\bibfnamefont {K.~B.}\ \bibnamefont
  {Broberg}},\ }\href@noop {} {\emph {\bibinfo {title} {Cracks and Fracture}}}\
  (\bibinfo  {publisher} {Elsevier},\ \bibinfo {year} {1999})\BibitemShut
  {NoStop}%
\bibitem [{\citenamefont {Lapusta}\ \emph {et~al.}(2000)\citenamefont
  {Lapusta}, \citenamefont {Rice}, \citenamefont {Ben-Zion},\ and\
  \citenamefont {Zheng}}]{Lapusta2000}%
  \BibitemOpen
  \bibfield  {author} {\bibinfo {author} {\bibfnamefont {N.}~\bibnamefont
  {Lapusta}}, \bibinfo {author} {\bibfnamefont {J.~R.}\ \bibnamefont {Rice}},
  \bibinfo {author} {\bibfnamefont {Y.}~\bibnamefont {Ben-Zion}}, \ and\
  \bibinfo {author} {\bibfnamefont {G.}~\bibnamefont {Zheng}},\ }\bibfield
  {title} {\emph {\bibinfo {title} {{Elastodynamic analysis for slow tectonic
  loading with spontaneous rupture episodes on faults with rate- and
  state-dependent friction}},\ }}\href
  {http://www.agu.org/pubs/crossref/2000/2000JB900250.shtml} {\bibfield
  {journal} {\bibinfo  {journal} {J. Geophys. Res.}\ }\textbf {\bibinfo
  {volume} {105}},\ \bibinfo {pages} {23765} (\bibinfo {year}
  {2000})}\BibitemShut {NoStop}%
\bibitem [{\citenamefont {Ben-Zion}(2001)}]{Ben-Zion2001}%
  \BibitemOpen
  \bibfield  {author} {\bibinfo {author} {\bibfnamefont {Y.}~\bibnamefont
  {Ben-Zion}},\ }\bibfield  {title} {\emph {\bibinfo {title} {{Dynamic ruptures
  in recent models of earthquake faults}},\ }}\href
  {http://linkinghub.elsevier.com/retrieve/pii/S0022509601000369} {\bibfield
  {journal} {\bibinfo  {journal} {J. Mech. Phys. Solids}\ }\textbf {\bibinfo
  {volume} {49}},\ \bibinfo {pages} {2209} (\bibinfo {year}
  {2001})}\BibitemShut {NoStop}%
\bibitem [{\citenamefont {Scholz}(2002)}]{Scholz2002}%
  \BibitemOpen
  \bibfield  {author} {\bibinfo {author} {\bibfnamefont {C.~H.}\ \bibnamefont
  {Scholz}},\ }\href@noop {} {\emph {\bibinfo {title} {{The Mechanics of
  Earthquakes and Faulting}}}}\ (\bibinfo  {publisher} {Cambridge University
  Press},\ \bibinfo {year} {2002})\BibitemShut {NoStop}%
\bibitem [{\citenamefont {Rubin}\ and\ \citenamefont
  {Ampuero}(2005)}]{Rubin2005}%
  \BibitemOpen
  \bibfield  {author} {\bibinfo {author} {\bibfnamefont {A.~M.}\ \bibnamefont
  {Rubin}}\ and\ \bibinfo {author} {\bibfnamefont {J.-P.}\ \bibnamefont
  {Ampuero}},\ }\bibfield  {title} {\emph {\bibinfo {title} {{Earthquake
  nucleation on (aging) rate and state faults}},\ }}\href
  {http://doi.wiley.com/10.1029/2005JB003686} {\bibfield  {journal} {\bibinfo
  {journal} {J. Geophys. Res.}\ }\textbf {\bibinfo {volume} {110}},\ \bibinfo
  {pages} {B11312} (\bibinfo {year} {2005})}\BibitemShut {NoStop}%
\bibitem [{\citenamefont {Dunham}(2005)}]{Dunham2005}%
  \BibitemOpen
  \bibfield  {author} {\bibinfo {author} {\bibfnamefont {E.~M.}\ \bibnamefont
  {Dunham}},\ }\bibfield  {title} {\emph {\bibinfo {title} {{Dissipative
  interface waves and the transient response of a three-dimensional sliding
  interface with Coulomb friction}},\ }}\href
  {http://linkinghub.elsevier.com/retrieve/pii/S0022509604001309} {\bibfield
  {journal} {\bibinfo  {journal} {J. Mech. Phys. Solids}\ }\textbf {\bibinfo
  {volume} {53}},\ \bibinfo {pages} {327} (\bibinfo {year} {2005})}\BibitemShut
  {NoStop}%
\bibitem [{\citenamefont {Dunham}(2007)}]{Dunham2007}%
  \BibitemOpen
  \bibfield  {author} {\bibinfo {author} {\bibfnamefont {E.~M.}\ \bibnamefont
  {Dunham}},\ }\bibfield  {title} {\emph {\bibinfo {title} {{Conditions
  governing the occurrence of supershear ruptures under slip-weakening
  friction}},\ }}\href {http://doi.wiley.com/10.1029/2006JB004717} {\bibfield
  {journal} {\bibinfo  {journal} {J. Geophys. Res.}\ }\textbf {\bibinfo
  {volume} {112}},\ \bibinfo {pages} {B07302} (\bibinfo {year}
  {2007})}\BibitemShut {NoStop}%
\bibitem [{\citenamefont {Ampuero}\ and\ \citenamefont
  {Rubin}(2008)}]{Ampuero2008}%
  \BibitemOpen
  \bibfield  {author} {\bibinfo {author} {\bibfnamefont {J.-P.}\ \bibnamefont
  {Ampuero}}\ and\ \bibinfo {author} {\bibfnamefont {A.~M.}\ \bibnamefont
  {Rubin}},\ }\bibfield  {title} {\emph {\bibinfo {title} {{Earthquake
  nucleation on rate and state faults – Aging and slip laws}},\ }}\href
  {http://doi.wiley.com/10.1029/2007JB005082} {\bibfield  {journal} {\bibinfo
  {journal} {J. Geophys. Res.}\ }\textbf {\bibinfo {volume} {113}},\ \bibinfo
  {pages} {B01302} (\bibinfo {year} {2008})}\BibitemShut {NoStop}%
\bibitem [{\citenamefont {Rudnicki}\ and\ \citenamefont
  {Rice}(2006)}]{Rudnicki2006}%
  \BibitemOpen
  \bibfield  {author} {\bibinfo {author} {\bibfnamefont {J.~W.}\ \bibnamefont
  {Rudnicki}}\ and\ \bibinfo {author} {\bibfnamefont {J.~R.}\ \bibnamefont
  {Rice}},\ }\bibfield  {title} {\emph {\bibinfo {title} {{Effective normal
  stress alteration due to pore pressure changes induced by dynamic slip
  propagation on a plane between dissimilar materials}},\ }}\href
  {http://doi.wiley.com/10.1029/2006JB004396} {\bibfield  {journal} {\bibinfo
  {journal} {J. Geophys. Res.}\ }\textbf {\bibinfo {volume} {111}},\ \bibinfo
  {pages} {B10308} (\bibinfo {year} {2006})}\BibitemShut {NoStop}%
\bibitem [{\citenamefont {Dunham}\ and\ \citenamefont
  {Rice}(2008)}]{Dunham2008}%
  \BibitemOpen
  \bibfield  {author} {\bibinfo {author} {\bibfnamefont {E.~M.}\ \bibnamefont
  {Dunham}}\ and\ \bibinfo {author} {\bibfnamefont {J.~R.}\ \bibnamefont
  {Rice}},\ }\bibfield  {title} {\emph {\bibinfo {title} {{Earthquake slip
  between dissimilar poroelastic materials}},\ }}\href
  {http://doi.wiley.com/10.1029/2007JB005405} {\bibfield  {journal} {\bibinfo
  {journal} {J. Geophys. Res.}\ }\textbf {\bibinfo {volume} {113}},\ \bibinfo
  {pages} {B09304} (\bibinfo {year} {2008})}\BibitemShut {NoStop}%
\bibitem [{\citenamefont {Comninou}(1977{\natexlab{a}})}]{Comninou1977a}%
  \BibitemOpen
  \bibfield  {author} {\bibinfo {author} {\bibfnamefont {M.}~\bibnamefont
  {Comninou}},\ }\bibfield  {title} {\emph {\bibinfo {title} {{Interface crack
  with friction in the contact zone}},\ }}\href
  {http://appliedmechanics.asmedigitalcollection.asme.org/article.aspx?articleid=1403617}
  {\bibfield  {journal} {\bibinfo  {journal} {J. Appl. Mech.}\ }\textbf
  {\bibinfo {volume} {44}},\ \bibinfo {pages} {780} (\bibinfo {year}
  {1977}{\natexlab{a}})}\BibitemShut {NoStop}%
\bibitem [{\citenamefont {Comninou}(1977{\natexlab{b}})}]{Comninou1977b}%
  \BibitemOpen
  \bibfield  {author} {\bibinfo {author} {\bibfnamefont {M.}~\bibnamefont
  {Comninou}},\ }\bibfield  {title} {\emph {\bibinfo {title} {{The interface
  crack}},\ }}\href
  {http://appliedmechanics.asmedigitalcollection.asme.org/article.aspx?articleid=1403555}
  {\bibfield  {journal} {\bibinfo  {journal} {J. Appl. Mech.}\ }\textbf
  {\bibinfo {volume} {44}},\ \bibinfo {pages} {631} (\bibinfo {year}
  {1977}{\natexlab{b}})}\BibitemShut {NoStop}%
\bibitem [{\citenamefont {Comninou}\ and\ \citenamefont
  {Schmueser}(1979)}]{Comninou1979}%
  \BibitemOpen
  \bibfield  {author} {\bibinfo {author} {\bibfnamefont {M.}~\bibnamefont
  {Comninou}}\ and\ \bibinfo {author} {\bibfnamefont {D.}~\bibnamefont
  {Schmueser}},\ }\bibfield  {title} {\emph {\bibinfo {title} {{The interface
  crack in a combined tension-compression and shear field}},\ }}\href
  {http://appliedmechanics.asmedigitalcollection.asme.org/article.aspx?articleid=1404484}
  {\bibfield  {journal} {\bibinfo  {journal} {J. Appl. Mech.}\ }\textbf
  {\bibinfo {volume} {46}},\ \bibinfo {pages} {345} (\bibinfo {year}
  {1979})}\BibitemShut {NoStop}%
\bibitem [{\citenamefont {Xia}\ \emph {et~al.}(2004)\citenamefont {Xia},
  \citenamefont {Rosakis},\ and\ \citenamefont {Kanamori}}]{Xia2004}%
  \BibitemOpen
  \bibfield  {author} {\bibinfo {author} {\bibfnamefont {K.}~\bibnamefont
  {Xia}}, \bibinfo {author} {\bibfnamefont {A.~J.}\ \bibnamefont {Rosakis}}, \
  and\ \bibinfo {author} {\bibfnamefont {H.}~\bibnamefont {Kanamori}},\
  }\bibfield  {title} {\emph {\bibinfo {title} {{Laboratory earthquakes: the
  sub-Rayleigh-to-supershear rupture transition.}}\ }}\href
  {http://www.ncbi.nlm.nih.gov/pubmed/15031503} {\bibfield  {journal} {\bibinfo
   {journal} {Science}\ }\textbf {\bibinfo {volume} {303}},\ \bibinfo {pages}
  {1859} (\bibinfo {year} {2004})}\BibitemShut {NoStop}%
\bibitem [{\citenamefont {Ampuero}\ and\ \citenamefont
  {Ben-Zion}(2008)}]{Ampuero2008b}%
  \BibitemOpen
  \bibfield  {author} {\bibinfo {author} {\bibfnamefont {J.-P.}\ \bibnamefont
  {Ampuero}}\ and\ \bibinfo {author} {\bibfnamefont {Y.}~\bibnamefont
  {Ben-Zion}},\ }\bibfield  {title} {\emph {\bibinfo {title} {{Cracks, pulses
  and macroscopic asymmetry of dynamic rupture on a bimaterial interface with
  velocity-weakening friction}},\ }}\href
  {http://gji.oxfordjournals.org/cgi/doi/10.1111/j.1365-246X.2008.03736.x}
  {\bibfield  {journal} {\bibinfo  {journal} {Geophys. J. Int.}\ }\textbf
  {\bibinfo {volume} {173}},\ \bibinfo {pages} {674} (\bibinfo {year}
  {2008})}\BibitemShut {NoStop}%
\bibitem [{\citenamefont {Shlomai}\ and\ \citenamefont
  {Fineberg}(2016)}]{Shlomai2016}%
  \BibitemOpen
  \bibfield  {author} {\bibinfo {author} {\bibfnamefont {H.}~\bibnamefont
  {Shlomai}}\ and\ \bibinfo {author} {\bibfnamefont {J.}~\bibnamefont
  {Fineberg}},\ }\bibfield  {title} {\emph {\bibinfo {title} {{The structure of
  slip-pulses and supershear ruptures driving slip in bimaterial friction}},\
  }}\href {\doibase doi:10.1038/ncomms11787} {\bibfield  {journal} {\bibinfo
  {journal} {Nat. Comm.}\ }\textbf {\bibinfo {volume} {7}},\ \bibinfo {pages}
  {11787} (\bibinfo {year} {2016})}\BibitemShut {NoStop}%
\bibitem [{\citenamefont {Erickson}\ and\ \citenamefont
  {Day}(2016)}]{Erickson2016}%
  \BibitemOpen
  \bibfield  {author} {\bibinfo {author} {\bibfnamefont {B.~A.}\ \bibnamefont
  {Erickson}}\ and\ \bibinfo {author} {\bibfnamefont {S.~M.}\ \bibnamefont
  {Day}},\ }\bibfield  {title} {\emph {\bibinfo {title} {Bimaterial effects in
  an earthquake cycle model using rate-and-state friction},\ }}\href {\doibase
  10.1002/2015JB012470} {\bibfield  {journal} {\bibinfo  {journal} {Journal of
  Geophysical Research: Solid Earth}\ }\textbf {\bibinfo {volume} {121}},\
  \bibinfo {pages} {2480} (\bibinfo {year} {2016})}\BibitemShut {NoStop}%
\bibitem [{\citenamefont {Rubinstein}\ \emph {et~al.}(2007)\citenamefont
  {Rubinstein}, \citenamefont {Cohen},\ and\ \citenamefont
  {Fineberg}}]{Rubinstein2007}%
  \BibitemOpen
  \bibfield  {author} {\bibinfo {author} {\bibfnamefont {S.}~\bibnamefont
  {Rubinstein}}, \bibinfo {author} {\bibfnamefont {G.}~\bibnamefont {Cohen}}, \
  and\ \bibinfo {author} {\bibfnamefont {J.}~\bibnamefont {Fineberg}},\
  }\bibfield  {title} {\emph {\bibinfo {title} {{Dynamics of precursors to
  frictional sliding}},\ }}\href
  {http://link.aps.org/doi/10.1103/PhysRevLett.98.226103} {\bibfield  {journal}
  {\bibinfo  {journal} {Phys. Rev. Lett.}\ }\textbf {\bibinfo {volume} {98}},\
  \bibinfo {pages} {226103} (\bibinfo {year} {2007})}\BibitemShut {NoStop}%
\bibitem [{\citenamefont {Ben-David}\ \emph
  {et~al.}(2010{\natexlab{a}})\citenamefont {Ben-David}, \citenamefont
  {Cohen},\ and\ \citenamefont {Fineberg}}]{Ben-David2010-fronts}%
  \BibitemOpen
  \bibfield  {author} {\bibinfo {author} {\bibfnamefont {O.}~\bibnamefont
  {Ben-David}}, \bibinfo {author} {\bibfnamefont {G.}~\bibnamefont {Cohen}}, \
  and\ \bibinfo {author} {\bibfnamefont {J.}~\bibnamefont {Fineberg}},\
  }\bibfield  {title} {\emph {\bibinfo {title} {{The dynamics of the onset of
  frictional slip}},\ }}\href
  {http://www.sciencemag.org/cgi/doi/10.1126/science.1194777} {\bibfield
  {journal} {\bibinfo  {journal} {Science}\ }\textbf {\bibinfo {volume}
  {330}},\ \bibinfo {pages} {211} (\bibinfo {year}
  {2010}{\natexlab{a}})}\BibitemShut {NoStop}%
\bibitem [{\citenamefont {Ben-David}\ \emph
  {et~al.}(2010{\natexlab{b}})\citenamefont {Ben-David}, \citenamefont
  {Rubinstein},\ and\ \citenamefont {Fineberg}}]{Ben-David2010-ageing}%
  \BibitemOpen
  \bibfield  {author} {\bibinfo {author} {\bibfnamefont {O.}~\bibnamefont
  {Ben-David}}, \bibinfo {author} {\bibfnamefont {S.~M.}\ \bibnamefont
  {Rubinstein}}, \ and\ \bibinfo {author} {\bibfnamefont {J.}~\bibnamefont
  {Fineberg}},\ }\bibfield  {title} {\emph {\bibinfo {title} {{Slip-stick and
  the evolution of frictional strength.}}\ }}\href
  {http://www.ncbi.nlm.nih.gov/pubmed/20054393} {\bibfield  {journal} {\bibinfo
   {journal} {Nature}\ }\textbf {\bibinfo {volume} {463}},\ \bibinfo {pages}
  {76} (\bibinfo {year} {2010}{\natexlab{b}})}\BibitemShut {NoStop}%
\bibitem [{\citenamefont {Svetlizky}\ and\ \citenamefont
  {Fineberg}(2014)}]{Svetlizky2014}%
  \BibitemOpen
  \bibfield  {author} {\bibinfo {author} {\bibfnamefont {I.}~\bibnamefont
  {Svetlizky}}\ and\ \bibinfo {author} {\bibfnamefont {J.}~\bibnamefont
  {Fineberg}},\ }\bibfield  {title} {\emph {\bibinfo {title} {{Classical shear
  cracks drive the onset of dry frictional motion.}}\ }}\href
  {http://www.ncbi.nlm.nih.gov/pubmed/24805344} {\bibfield  {journal} {\bibinfo
   {journal} {Nature}\ }\textbf {\bibinfo {volume} {509}},\ \bibinfo {pages}
  {205} (\bibinfo {year} {2014})}\BibitemShut {NoStop}%
\bibitem [{\citenamefont {Bayart}\ \emph {et~al.}(2016)\citenamefont {Bayart},
  \citenamefont {Svetlizky},\ and\ \citenamefont {Fineberg}}]{Bayart2016}%
  \BibitemOpen
  \bibfield  {author} {\bibinfo {author} {\bibfnamefont {E.}~\bibnamefont
  {Bayart}}, \bibinfo {author} {\bibfnamefont {I.}~\bibnamefont {Svetlizky}}, \
  and\ \bibinfo {author} {\bibfnamefont {J.}~\bibnamefont {Fineberg}},\
  }\bibfield  {title} {\emph {\bibinfo {title} {{Slippery but Tough: The Rapid
  Fracture of Lubricated Frictional Interfaces}},\ }}\href {\doibase
  10.1103/PhysRevLett.116.194301} {\bibfield  {journal} {\bibinfo  {journal}
  {Phys. Rev. Lett.}\ }\textbf {\bibinfo {volume} {116}},\ \bibinfo {pages}
  {194301} (\bibinfo {year} {2016})}\BibitemShut {NoStop}%
\bibitem [{\citenamefont {Plafker}(1965)}]{Plafker1965}%
  \BibitemOpen
  \bibfield  {author} {\bibinfo {author} {\bibfnamefont {G.}~\bibnamefont
  {Plafker}},\ }\bibfield  {title} {\emph {\bibinfo {title} {{Tectonic
  Deformation Associated with the 1964 Alaska Earthquake: The earthquake of 27
  March 1964 resulted in observable crustal deformation of unprecedented areal
  extent}},\ }}\href
  {http://www.sciencemag.org/cgi/doi/10.1126/science.148.3678.1675} {\bibfield
  {journal} {\bibinfo  {journal} {Science}\ }\textbf {\bibinfo {volume}
  {148}},\ \bibinfo {pages} {1675} (\bibinfo {year} {1965})}\BibitemShut
  {NoStop}%
\bibitem [{\citenamefont {Barrientos}\ and\ \citenamefont
  {Ward}(1990)}]{Barrientos1990}%
  \BibitemOpen
  \bibfield  {author} {\bibinfo {author} {\bibfnamefont {S.~E.}\ \bibnamefont
  {Barrientos}}\ and\ \bibinfo {author} {\bibfnamefont {S.~N.}\ \bibnamefont
  {Ward}},\ }\bibfield  {title} {\emph {\bibinfo {title} {{The 1960 Chile
  earthquake: inversion for slip distribution from surface deformation}},\
  }}\href
  {http://gji.oxfordjournals.org/cgi/doi/10.1111/j.1365-246X.1990.tb05673.x}
  {\bibfield  {journal} {\bibinfo  {journal} {Geophys. J. Int.}\ }\textbf
  {\bibinfo {volume} {103}},\ \bibinfo {pages} {589} (\bibinfo {year}
  {1990})}\BibitemShut {NoStop}%
\bibitem [{\citenamefont {Delouis}\ \emph {et~al.}(2010)\citenamefont
  {Delouis}, \citenamefont {Nocquet},\ and\ \citenamefont
  {Vall{\'{e}}e}}]{Delouis2010}%
  \BibitemOpen
  \bibfield  {author} {\bibinfo {author} {\bibfnamefont {B.}~\bibnamefont
  {Delouis}}, \bibinfo {author} {\bibfnamefont {J.-M.}\ \bibnamefont
  {Nocquet}}, \ and\ \bibinfo {author} {\bibfnamefont {M.}~\bibnamefont
  {Vall{\'{e}}e}},\ }\bibfield  {title} {\emph {\bibinfo {title} {{Slip
  distribution of the February 27, 2010 Mw = 8.8 Maule Earthquake, central
  Chile, from static and high-rate GPS, InSAR, and broadband teleseismic
  data}},\ }}\href {http://doi.wiley.com/10.1029/2010GL043899} {\bibfield
  {journal} {\bibinfo  {journal} {Geophys. Res. Lett.}\ }\textbf {\bibinfo
  {volume} {37}} (\bibinfo {year} {2010})}\BibitemShut {NoStop}%
\bibitem [{\citenamefont {Madariaga}\ \emph {et~al.}(2010)\citenamefont
  {Madariaga}, \citenamefont {Metois}, \citenamefont {Vigny},\ and\
  \citenamefont {Campos}}]{Madariaga2010}%
  \BibitemOpen
  \bibfield  {author} {\bibinfo {author} {\bibfnamefont {R.}~\bibnamefont
  {Madariaga}}, \bibinfo {author} {\bibfnamefont {M.}~\bibnamefont {Metois}},
  \bibinfo {author} {\bibfnamefont {C.}~\bibnamefont {Vigny}}, \ and\ \bibinfo
  {author} {\bibfnamefont {J.}~\bibnamefont {Campos}},\ }\bibfield  {title}
  {\emph {\bibinfo {title} {{Central Chile Finally Breaks}},\ }}\href
  {http://www.sciencemag.org/cgi/doi/10.1126/science.1189197} {\bibfield
  {journal} {\bibinfo  {journal} {Science}\ }\textbf {\bibinfo {volume}
  {328}},\ \bibinfo {pages} {181} (\bibinfo {year} {2010})}\BibitemShut
  {NoStop}%
\bibitem [{\citenamefont {Simons}\ \emph {et~al.}(2011)\citenamefont {Simons},
  \citenamefont {Minson}, \citenamefont {Sladen}, \citenamefont {Ortega},
  \citenamefont {Jiang}, \citenamefont {Owen}, \citenamefont {Meng},
  \citenamefont {Ampuero}, \citenamefont {Wei}, \citenamefont {Chu},
  \citenamefont {Helmberger}, \citenamefont {Kanamori}, \citenamefont
  {Hetland}, \citenamefont {Moore},\ and\ \citenamefont {Webb}}]{Simons2011}%
  \BibitemOpen
  \bibfield  {author} {\bibinfo {author} {\bibfnamefont {M.}~\bibnamefont
  {Simons}}, \bibinfo {author} {\bibfnamefont {S.~E.}\ \bibnamefont {Minson}},
  \bibinfo {author} {\bibfnamefont {A.}~\bibnamefont {Sladen}}, \bibinfo
  {author} {\bibfnamefont {F.}~\bibnamefont {Ortega}}, \bibinfo {author}
  {\bibfnamefont {J.}~\bibnamefont {Jiang}}, \bibinfo {author} {\bibfnamefont
  {S.~E.}\ \bibnamefont {Owen}}, \bibinfo {author} {\bibfnamefont
  {L.}~\bibnamefont {Meng}}, \bibinfo {author} {\bibfnamefont {J.-P.}\
  \bibnamefont {Ampuero}}, \bibinfo {author} {\bibfnamefont {S.}~\bibnamefont
  {Wei}}, \bibinfo {author} {\bibfnamefont {R.}~\bibnamefont {Chu}}, \bibinfo
  {author} {\bibfnamefont {D.~V.}\ \bibnamefont {Helmberger}}, \bibinfo
  {author} {\bibfnamefont {H.}~\bibnamefont {Kanamori}}, \bibinfo {author}
  {\bibfnamefont {E.}~\bibnamefont {Hetland}}, \bibinfo {author} {\bibfnamefont
  {A.~W.}\ \bibnamefont {Moore}}, \ and\ \bibinfo {author} {\bibfnamefont
  {F.~H.}\ \bibnamefont {Webb}},\ }\bibfield  {title} {\emph {\bibinfo {title}
  {{The 2011 Magnitude 9.0 Tohoku-Oki Earthquake: Mosaicking the Megathrust
  from Seconds to Centuries}},\ }}\href
  {http://www.sciencemag.org/cgi/doi/10.1126/science.1206731} {\bibfield
  {journal} {\bibinfo  {journal} {Science}\ }\textbf {\bibinfo {volume}
  {332}},\ \bibinfo {pages} {1421} (\bibinfo {year} {2011})}\BibitemShut
  {NoStop}%
\bibitem [{\citenamefont {Le{\'{o}}n-R{\'{i}}os}\ \emph
  {et~al.}(2016)\citenamefont {Le{\'{o}}n-R{\'{i}}os}, \citenamefont {Ruiz},
  \citenamefont {Maksymowicz}, \citenamefont {Leyton}, \citenamefont
  {Fuenzalida},\ and\ \citenamefont {Madariaga}}]{Leon-Rios2016}%
  \BibitemOpen
  \bibfield  {author} {\bibinfo {author} {\bibfnamefont {S.}~\bibnamefont
  {Le{\'{o}}n-R{\'{i}}os}}, \bibinfo {author} {\bibfnamefont {S.}~\bibnamefont
  {Ruiz}}, \bibinfo {author} {\bibfnamefont {A.}~\bibnamefont {Maksymowicz}},
  \bibinfo {author} {\bibfnamefont {F.}~\bibnamefont {Leyton}}, \bibinfo
  {author} {\bibfnamefont {A.}~\bibnamefont {Fuenzalida}}, \ and\ \bibinfo
  {author} {\bibfnamefont {R.}~\bibnamefont {Madariaga}},\ }\bibfield  {title}
  {\emph {\bibinfo {title} {{Diversity of the 2014 Iquique's foreshocks and
  aftershocks: clues about the complex rupture process of a Mw 8.1
  earthquake}},\ }}\href {\doibase 10.1007/s10950-016-9568-6} {\bibfield
  {journal} {\bibinfo  {journal} {J. Seismol.}\ }\textbf {\bibinfo {volume}
  {1911}},\ \bibinfo {pages} {1} (\bibinfo {year} {2016})}\BibitemShut
  {NoStop}%
%\bibitem [{\citenamefont {Le{\'{o}}n-R{\'{i}}os}\ \emph
%  {et~al.}(2016)\citenamefont {Le{\'{o}}n-R{\'{i}}os}, \citenamefont {Ruiz},
%  \citenamefont {Maksymowicz}, \citenamefont {Leyton}, \citenamefont
%  {Fuenzalida},\ and\ \citenamefont {Madariaga}}]{Leon-Rios2016}%
%  \BibitemOpen
%  \bibfield  {author} {\bibinfo {author} {\bibfnamefont {S.}~\bibnamefont
%  {Le{\'{o}}n-R{\'{i}}os}}, \bibinfo {author} {\bibfnamefont {S.}~\bibnamefont
%  {Ruiz}}, \bibinfo {author} {\bibfnamefont {A.}~\bibnamefont {Maksymowicz}},
%  \bibinfo {author} {\bibfnamefont {F.}~\bibnamefont {Leyton}}, \bibinfo
%  {author} {\bibfnamefont {A.}~\bibnamefont {Fuenzalida}}, \ and\ \bibinfo
%  {author} {\bibfnamefont {R.}~\bibnamefont {Madariaga}},\ }\bibfield  {title}
%  {\emph {\bibinfo {title} {{Diversity of the 2014 Iquique's foreshocks and
%  aftershocks: clues about the complex rupture process of a Mw 8.1
%  earthquake}},\ }}\href {http://link.springer.com/10.1007/s10950-016-9568-6}
%  {\bibfield  {journal} {\bibinfo  {journal} {J. Seismol.}\ } (\bibinfo {year}
%  {2016})}\BibitemShut {NoStop}%
\bibitem [{SM()}]{SM}%
  \BibitemOpen
  \href@noop {} {\emph {\bibinfo {title} {{Supplementary
  Material}}}}\BibitemShut {NoStop}%
\bibitem [{\citenamefont {Geubelle}(1995)}]{Geubelle1995}%
  \BibitemOpen
  \bibfield  {author} {\bibinfo {author} {\bibfnamefont {P.}~\bibnamefont
  {Geubelle}},\ }\bibfield  {title} {\emph {\bibinfo {title} {{A spectral
  method for three-dimensional elastodynamic fracture problems}},\ }}\href
  {http://linkinghub.elsevier.com/retrieve/pii/002250969500043I} {\bibfield
  {journal} {\bibinfo  {journal} {J. Mech. Phys. Solids}\ }\textbf {\bibinfo
  {volume} {43}},\ \bibinfo {pages} {1791} (\bibinfo {year}
  {1995})}\BibitemShut {NoStop}%
\bibitem [{\citenamefont {Bowden}\ and\ \citenamefont
  {Tabor}(1950)}]{Bowden1950}%
  \BibitemOpen
  \bibfield  {author} {\bibinfo {author} {\bibfnamefont {F.}~\bibnamefont
  {Bowden}}\ and\ \bibinfo {author} {\bibfnamefont {D.}~\bibnamefont {Tabor}},\
  }\href@noop {} {\emph {\bibinfo {title} {{The friction and lubrication of
  solids}}}}\ (\bibinfo  {publisher} {Clarendon Press},\ \bibinfo {year}
  {1950})\BibitemShut {NoStop}%
\bibitem [{\citenamefont {Dieterich}\ and\ \citenamefont
  {Kilgore}(1994)}]{Dieterich1994}%
  \BibitemOpen
  \bibfield  {author} {\bibinfo {author} {\bibfnamefont {J.~H.}\ \bibnamefont
  {Dieterich}}\ and\ \bibinfo {author} {\bibfnamefont {B.~D.}\ \bibnamefont
  {Kilgore}},\ }\bibfield  {title} {\emph {\bibinfo {title} {{Direct
  observation of frictional contacts: New insights for state-dependent
  properties}},\ }}\href {http://www.springerlink.com/index/10.1007/BF00874332}
  {\bibfield  {journal} {\bibinfo  {journal} {Pure Appl. Geophys.}\ }\textbf
  {\bibinfo {volume} {143}},\ \bibinfo {pages} {283} (\bibinfo {year}
  {1994})}\BibitemShut {NoStop}%
\bibitem [{\citenamefont {Nakatani}\ and\ \citenamefont
  {Scholz}(2006)}]{Nakatani2006}%
  \BibitemOpen
  \bibfield  {author} {\bibinfo {author} {\bibfnamefont {M.}~\bibnamefont
  {Nakatani}}\ and\ \bibinfo {author} {\bibfnamefont {C.~H.}\ \bibnamefont
  {Scholz}},\ }\bibfield  {title} {\emph {\bibinfo {title} {{Intrinsic and
  apparent short-time limits for fault healing: Theory, observations, and
  implications for velocity-dependent friction}},\ }}\href
  {http://www.agu.org/pubs/crossref/2006/2005JB004096.shtml} {\bibfield
  {journal} {\bibinfo  {journal} {J. Geophys. Res.}\ }\textbf {\bibinfo
  {volume} {111}},\ \bibinfo {pages} {B12208} (\bibinfo {year}
  {2006})}\BibitemShut {NoStop}%
\bibitem [{\citenamefont {Nagata}\ \emph {et~al.}(2008)\citenamefont {Nagata},
  \citenamefont {Nakatani},\ and\ \citenamefont {Yoshida}}]{Nagata2008}%
  \BibitemOpen
  \bibfield  {author} {\bibinfo {author} {\bibfnamefont {K.}~\bibnamefont
  {Nagata}}, \bibinfo {author} {\bibfnamefont {M.}~\bibnamefont {Nakatani}}, \
  and\ \bibinfo {author} {\bibfnamefont {S.}~\bibnamefont {Yoshida}},\
  }\bibfield  {title} {\emph {\bibinfo {title} {{Monitoring frictional strength
  with acoustic wave transmission}},\ }}\href
  {http://doi.wiley.com/10.1029/2007GL033146} {\bibfield  {journal} {\bibinfo
  {journal} {Geophys. Res. Lett.}\ }\textbf {\bibinfo {volume} {35}},\ \bibinfo
  {pages} {L06310} (\bibinfo {year} {2008})}\BibitemShut {NoStop}%
\bibitem [{\citenamefont {Bar-Sinai}\ \emph {et~al.}(2013)\citenamefont
  {Bar-Sinai}, \citenamefont {Spatschek}, \citenamefont {Brener},\ and\
  \citenamefont {Bouchbinder}}]{Bar-Sinai2013pre}%
  \BibitemOpen
  \bibfield  {author} {\bibinfo {author} {\bibfnamefont {Y.}~\bibnamefont
  {Bar-Sinai}}, \bibinfo {author} {\bibfnamefont {R.}~\bibnamefont
  {Spatschek}}, \bibinfo {author} {\bibfnamefont {E.~A.}\ \bibnamefont
  {Brener}}, \ and\ \bibinfo {author} {\bibfnamefont {E.}~\bibnamefont
  {Bouchbinder}},\ }\bibfield  {title} {\emph {\bibinfo {title} {{Instabilities
  at frictional interfaces: Creep patches, nucleation, and rupture fronts}},\
  }}\href {http://link.aps.org/doi/10.1103/PhysRevE.88.060403} {\bibfield
  {journal} {\bibinfo  {journal} {Phys. Rev. E}\ }\textbf {\bibinfo {volume}
  {88}},\ \bibinfo {pages} {060403(R)} (\bibinfo {year} {2013})}\BibitemShut
  {NoStop}%
\bibitem [{\citenamefont {Dieterich}(1972)}]{Dieterich1972}%
  \BibitemOpen
  \bibfield  {author} {\bibinfo {author} {\bibfnamefont {J.~H.}\ \bibnamefont
  {Dieterich}},\ }\bibfield  {title} {\emph {\bibinfo {title} {{Time-Dependent
  Friction in Rocks}},\ }}\href@noop {} {\bibfield  {journal} {\bibinfo
  {journal} {J. Geophys. Res.}\ }\textbf {\bibinfo {volume} {77}},\ \bibinfo
  {pages} {3690} (\bibinfo {year} {1972})}\BibitemShut {NoStop}%
\bibitem [{\citenamefont {Dieterich}(1978)}]{Dieterich1978}%
  \BibitemOpen
  \bibfield  {author} {\bibinfo {author} {\bibfnamefont {J.~H.}\ \bibnamefont
  {Dieterich}},\ }\bibfield  {title} {\emph {\bibinfo {title} {{Time-dependent
  friction and the mechanics of stick-slip}},\ }}\href
  {http://www.springerlink.com/index/10.1007/BF00876539} {\bibfield  {journal}
  {\bibinfo  {journal} {Pure Appl. Geophys.}\ }\textbf {\bibinfo {volume}
  {116}},\ \bibinfo {pages} {790} (\bibinfo {year} {1978})}\BibitemShut
  {NoStop}%
\bibitem [{\citenamefont {Palmer}\ and\ \citenamefont
  {Rice}(1973)}]{Palmer1973}%
  \BibitemOpen
  \bibfield  {author} {\bibinfo {author} {\bibfnamefont {A.~C.}\ \bibnamefont
  {Palmer}}\ and\ \bibinfo {author} {\bibfnamefont {J.~R.}\ \bibnamefont
  {Rice}},\ }\bibfield  {title} {\emph {\bibinfo {title} {{The growth of slip
  surfaces in the progressive failure of over-consolidated clay}},\ }}\href
  {http://rspa.royalsocietypublishing.org/cgi/doi/10.1098/rspa.1973.0040}
  {\bibfield  {journal} {\bibinfo  {journal} {Proc. R. Soc. A Math. Phys. Eng.
  Sci.}\ }\textbf {\bibinfo {volume} {332}},\ \bibinfo {pages} {527} (\bibinfo
  {year} {1973})}\BibitemShut {NoStop}%
\bibitem [{\citenamefont {Poliakov}\ \emph {et~al.}(2002)\citenamefont
  {Poliakov}, \citenamefont {Dmowska},\ and\ \citenamefont
  {Rice}}]{Poliakov2002}%
  \BibitemOpen
  \bibfield  {author} {\bibinfo {author} {\bibfnamefont {A.~N.~B.}\
  \bibnamefont {Poliakov}}, \bibinfo {author} {\bibfnamefont {R.}~\bibnamefont
  {Dmowska}}, \ and\ \bibinfo {author} {\bibfnamefont {J.~R.}\ \bibnamefont
  {Rice}},\ }\bibfield  {title} {\emph {\bibinfo {title} {{Dynamic shear
  rupture interactions with fault bends and off-axis secondary faulting}},\
  }}\href {http://doi.wiley.com/10.1029/2001JB000572} {\bibfield  {journal}
  {\bibinfo  {journal} {J. Geophys. Res. Solid Earth}\ }\textbf {\bibinfo
  {volume} {107}},\ \bibinfo {pages} {ESE 6} (\bibinfo {year}
  {2002})}\BibitemShut {NoStop}%
\bibitem [{\citenamefont {Samudrala}\ \emph {et~al.}(2002)\citenamefont
  {Samudrala}, \citenamefont {Huang},\ and\ \citenamefont
  {Rosakis}}]{Samudrala2002}%
  \BibitemOpen
  \bibfield  {author} {\bibinfo {author} {\bibfnamefont {O.}~\bibnamefont
  {Samudrala}}, \bibinfo {author} {\bibfnamefont {Y.}~\bibnamefont {Huang}}, \
  and\ \bibinfo {author} {\bibfnamefont {A.~J.}\ \bibnamefont {Rosakis}},\
  }\bibfield  {title} {\emph {\bibinfo {title} {{Subsonic and intersonic shear
  rupture of weak planes with a velocity weakening cohesive zone}},\ }}\href
  {http://doi.wiley.com/10.1029/2001JB000460} {\bibfield  {journal} {\bibinfo
  {journal} {J. Geophys. Res.}\ }\textbf {\bibinfo {volume} {107}},\ \bibinfo
  {pages} {2170} (\bibinfo {year} {2002})}\BibitemShut {NoStop}%
\bibitem [{\citenamefont {Landau}\ and\ \citenamefont
  {Lifshitz}(1986)}]{LLElasticity}%
  \BibitemOpen
  \bibfield  {author} {\bibinfo {author} {\bibfnamefont {L.~D.}\ \bibnamefont
  {Landau}}\ and\ \bibinfo {author} {\bibfnamefont {E.~M.}\ \bibnamefont
  {Lifshitz}},\ }\href@noop {} {\emph {\bibinfo {title} {{Theory of elasticity,
  Vol. 7}}}},\ \bibinfo {edition} {3rd}\ ed.\ (\bibinfo  {publisher} {Pergamon
  Press},\ \bibinfo {year} {1986})\BibitemShut {NoStop}%
\bibitem [{\citenamefont {Read}\ and\ \citenamefont {Duncan}(1981)}]{Read1981}%
  \BibitemOpen
  \bibfield  {author} {\bibinfo {author} {\bibfnamefont {B.}~\bibnamefont
  {Read}}\ and\ \bibinfo {author} {\bibfnamefont {J.}~\bibnamefont {Duncan}},\
  }\bibfield  {title} {\emph {\bibinfo {title} {{Measurement of dynamic
  properties of polymeric glasses for different modes of deformation}},\
  }}\href {http://linkinghub.elsevier.com/retrieve/pii/0142941881900313}
  {\bibfield  {journal} {\bibinfo  {journal} {Polym. Test.}\ }\textbf {\bibinfo
  {volume} {2}},\ \bibinfo {pages} {135} (\bibinfo {year} {1981})}\BibitemShut
  {NoStop}%
\bibitem [{\citenamefont {Timoshenko}\ and\ \citenamefont
  {Goodier}(1951)}]{Timoshenko}%
  \BibitemOpen
  \bibfield  {author} {\bibinfo {author} {\bibfnamefont {S.}~\bibnamefont
  {Timoshenko}}\ and\ \bibinfo {author} {\bibfnamefont {J.~N.}\ \bibnamefont
  {Goodier}},\ }\href@noop {} {\emph {\bibinfo {title} {{Theory of
  Elasticity}}}}\ (\bibinfo  {publisher} {McGraw-Hill},\ \bibinfo {year}
  {1951})\BibitemShut {NoStop}%
\bibitem [{Note1()}]{Note1}%
  \BibitemOpen
  \bibinfo {note} {Note that at large $kW$, $\mu ^{\unhbox \voidb@x \hbox
  {\relax \protect \fontsize {5}{6}\protect \selectfont eff}}$ becomes minutely
  smaller than $\mu $, see~\cite {SM}}\BibitemShut {NoStop}%
\bibitem [{\citenamefont {Radiguet}\ \emph {et~al.}(2015)\citenamefont
  {Radiguet}, \citenamefont {Kammer},\ and\ \citenamefont
  {Molinari}}]{Radiguet2015276}%
  \BibitemOpen
  \bibfield  {author} {\bibinfo {author} {\bibfnamefont {M.}~\bibnamefont
  {Radiguet}}, \bibinfo {author} {\bibfnamefont {D.}~\bibnamefont {Kammer}}, \
  and\ \bibinfo {author} {\bibfnamefont {J.}~\bibnamefont {Molinari}},\
  }\bibfield  {title} {\emph {\bibinfo {title} {The role of viscoelasticity on
  heterogeneous stress fields at frictional interfaces},\ }}\href {\doibase
  http://dx.doi.org/10.1016/j.mechmat.2014.03.009} {\bibfield  {journal}
  {\bibinfo  {journal} {Mechanics of Materials}\ }\textbf {\bibinfo {volume}
  {80, Part B}},\ \bibinfo {pages} {276 } (\bibinfo {year} {2015})},\ \bibinfo
  {note} {materials and Interfaces}\BibitemShut {NoStop}%
\bibitem [{\citenamefont {Kammer}\ \emph {et~al.}(2015)\citenamefont {Kammer},
  \citenamefont {Radiguet}, \citenamefont {Ampuero},\ and\ \citenamefont
  {Molinari}}]{Kammer2015}%
  \BibitemOpen
  \bibfield  {author} {\bibinfo {author} {\bibfnamefont {D.~S.}\ \bibnamefont
  {Kammer}}, \bibinfo {author} {\bibfnamefont {M.}~\bibnamefont {Radiguet}},
  \bibinfo {author} {\bibfnamefont {J.-P.}\ \bibnamefont {Ampuero}}, \ and\
  \bibinfo {author} {\bibfnamefont {J.-F.}\ \bibnamefont {Molinari}},\
  }\bibfield  {title} {\emph {\bibinfo {title} {Linear elastic fracture
  mechanics predicts the propagation distance of frictional slip},\ }}\href
  {\doibase 10.1007/s11249-014-0451-8} {\bibfield  {journal} {\bibinfo
  {journal} {Tribology Letters}\ }\textbf {\bibinfo {volume} {57}},\ \bibinfo
  {pages} {1} (\bibinfo {year} {2015})}\BibitemShut {NoStop}%
\bibitem [{\citenamefont {Rice}\ \emph {et~al.}(2001)\citenamefont {Rice},
  \citenamefont {Lapusta},\ and\ \citenamefont {Ranjith}}]{Rice2001}%
  \BibitemOpen
  \bibfield  {author} {\bibinfo {author} {\bibfnamefont {J.~R.}\ \bibnamefont
  {Rice}}, \bibinfo {author} {\bibfnamefont {N.}~\bibnamefont {Lapusta}}, \
  and\ \bibinfo {author} {\bibfnamefont {K.}~\bibnamefont {Ranjith}},\
  }\bibfield  {title} {\emph {\bibinfo {title} {{Rate and state dependent
  friction and the stability of sliding between elastically deformable
  solids}},\ }}\href
  {http://linkinghub.elsevier.com/retrieve/pii/S0022509601000424} {\bibfield
  {journal} {\bibinfo  {journal} {J. Mech. Phys. Solids}\ }\textbf {\bibinfo
  {volume} {49}},\ \bibinfo {pages} {1865} (\bibinfo {year}
  {2001})}\BibitemShut {NoStop}%
\bibitem [{\citenamefont {Perfettini}\ and\ \citenamefont
  {Ampuero}(2008)}]{Perfettini2008}%
  \BibitemOpen
  \bibfield  {author} {\bibinfo {author} {\bibfnamefont {H.}~\bibnamefont
  {Perfettini}}\ and\ \bibinfo {author} {\bibfnamefont {J.-P.}\ \bibnamefont
  {Ampuero}},\ }\bibfield  {title} {\emph {\bibinfo {title} {{Dynamics of a
  velocity strengthening fault region: Implications for slow earthquakes and
  postseismic slip}},\ }}\href {http://doi.wiley.com/10.1029/2007JB005398}
  {\bibfield  {journal} {\bibinfo  {journal} {J. Geophys. Res.}\ }\textbf
  {\bibinfo {volume} {113}},\ \bibinfo {pages} {B09411} (\bibinfo {year}
  {2008})}\BibitemShut {NoStop}%
\bibitem [{\citenamefont {Rice}\ and\ \citenamefont {Ruina}(1983)}]{Rice1983}%
  \BibitemOpen
  \bibfield  {author} {\bibinfo {author} {\bibfnamefont {J.~R.}\ \bibnamefont
  {Rice}}\ and\ \bibinfo {author} {\bibfnamefont {A.}~\bibnamefont {Ruina}},\
  }\bibfield  {title} {\emph {\bibinfo {title} {{Stability of steady frictional
  slipping}},\ }}\href {http://link.aip.org/link/?JAMCAV/50/343/1} {\bibfield
  {journal} {\bibinfo  {journal} {J. Appl. Mech.}\ }\textbf {\bibinfo {volume}
  {50}},\ \bibinfo {pages} {343} (\bibinfo {year} {1983})}\BibitemShut
  {NoStop}%
\bibitem [{\citenamefont {Armstrong-H{\'{e}}louvry}\ \emph
  {et~al.}(1994)\citenamefont {Armstrong-H{\'{e}}louvry}, \citenamefont
  {Dupont},\ and\ \citenamefont {{De Wit}}}]{Armstrong-Helouvry1994}%
  \BibitemOpen
  \bibfield  {author} {\bibinfo {author} {\bibfnamefont {B.}~\bibnamefont
  {Armstrong-H{\'{e}}louvry}}, \bibinfo {author} {\bibfnamefont
  {P.}~\bibnamefont {Dupont}}, \ and\ \bibinfo {author} {\bibfnamefont {C.~C.}\
  \bibnamefont {{De Wit}}},\ }\bibfield  {title} {\emph {\bibinfo {title} {{A
  survey of models, analysis tools and compensation methods for the control of
  machines with friction}},\ }}\href
  {http://www.sciencedirect.com/science/article/pii/0005109894902097} {\bibfield
  {journal} {\bibinfo  {journal} {Automatica}\ }\textbf {\bibinfo {volume}
  {30}},\ \bibinfo {pages} {1083} (\bibinfo {year} {1994})}\BibitemShut
  {NoStop}%
\bibitem [{\citenamefont {Olsson}\ \emph {et~al.}(1998)\citenamefont {Olsson},
  \citenamefont {{\AA}str{\"{o}}m}, \citenamefont {{De Wit}}, \citenamefont
  {G{\"{a}}fvert},\ and\ \citenamefont {Lischinsky}}]{Olsson1998}%
  \BibitemOpen
  \bibfield  {author} {\bibinfo {author} {\bibfnamefont {H.}~\bibnamefont
  {Olsson}}, \bibinfo {author} {\bibfnamefont {K.}~\bibnamefont
  {{\AA}str{\"{o}}m}}, \bibinfo {author} {\bibfnamefont {C.~C.}\ \bibnamefont
  {{De Wit}}}, \bibinfo {author} {\bibfnamefont {M.}~\bibnamefont
  {G{\"{a}}fvert}}, \ and\ \bibinfo {author} {\bibfnamefont {P.}~\bibnamefont
  {Lischinsky}},\ }\bibfield  {title} {\emph {\bibinfo {title} {{Friction
  models and friction compensation}},\ }}\href
  {http://linkinghub.elsevier.com/retrieve/pii/S094735809870113X} {\bibfield
  {journal} {\bibinfo  {journal} {Eur. J. Control}\ }\textbf {\bibinfo {volume}
  {4}},\ \bibinfo {pages} {176} (\bibinfo {year} {1998})}\BibitemShut {NoStop}%
\bibitem [{\citenamefont {Nguyen}(2003)}]{Nguyen2003}%
  \BibitemOpen
  \bibfield  {author} {\bibinfo {author} {\bibfnamefont {Q.~S.}\ \bibnamefont
  {Nguyen}},\ }\bibfield  {title} {\emph {\bibinfo {title} {{Instability and
  friction}},\ }}\href@noop {} {\bibfield  {journal} {\bibinfo  {journal}
  {Comptes Rendus - Mec.}\ }\textbf {\bibinfo {volume} {331}},\ \bibinfo
  {pages} {99} (\bibinfo {year} {2003})}\BibitemShut {NoStop}%
\bibitem [{\citenamefont {Ikari}\ \emph {et~al.}(2010)\citenamefont {Ikari},
  \citenamefont {Marone},\ and\ \citenamefont {Saffer}}]{Ikari2010}%
  \BibitemOpen
  \bibfield  {author} {\bibinfo {author} {\bibfnamefont {M.~J.}\ \bibnamefont
  {Ikari}}, \bibinfo {author} {\bibfnamefont {C.~J.}\ \bibnamefont {Marone}}, \
  and\ \bibinfo {author} {\bibfnamefont {D.~M.}\ \bibnamefont {Saffer}},\
  }\bibfield  {title} {\emph {\bibinfo {title} {{On the relation between fault
  strength and frictional stability}},\ }}\href
  {http://geology.gsapubs.org/cgi/doi/10.1130/G31416.1} {\bibfield  {journal}
  {\bibinfo  {journal} {Geology}\ }\textbf {\bibinfo {volume} {39}},\ \bibinfo
  {pages} {83} (\bibinfo {year} {2010})}\BibitemShut {NoStop}%
\bibitem [{\citenamefont {Pomeau}\ and\ \citenamefont
  {Berre}(2011)}]{Pomeau2011}%
  \BibitemOpen
  \bibfield  {author} {\bibinfo {author} {\bibfnamefont {Y.}~\bibnamefont
  {Pomeau}}\ and\ \bibinfo {author} {\bibfnamefont {M.~L.}\ \bibnamefont
  {Berre}},\ }\bibfield  {title} {\emph {\bibinfo {title} {{Critical speed-up
  vs critical slow-down: a new kind of relaxation oscillation with application
  to stick-slip phenomena}},\ }}\href {http://arxiv.org/abs/1107.3331}
  {\bibfield  {journal} {\bibinfo  {journal} {arXiv:1107.3331}\ } (\bibinfo
  {year} {2011})}\BibitemShut {NoStop}%
\bibitem [{\citenamefont {Putelat}\ and\ \citenamefont
  {Dawes}(2015)}]{Putelat2015}%
  \BibitemOpen
  \bibfield  {author} {\bibinfo {author} {\bibfnamefont {T.}~\bibnamefont
  {Putelat}}\ and\ \bibinfo {author} {\bibfnamefont {J.~H.}\ \bibnamefont
  {Dawes}},\ }\bibfield  {title} {\emph {\bibinfo {title} {{Steady and
  transient sliding under rate-and-state friction}},\ }}\href
  {http://linkinghub.elsevier.com/retrieve/pii/S0022509615000319} {\bibfield
  {journal} {\bibinfo  {journal} {J. Mech. Phys. Solids}\ } (\bibinfo {year}
  {2015})}\BibitemShut {NoStop}%
\bibitem [{\citenamefont {{Bar Sinai}}\ \emph {et~al.}(2012)\citenamefont {{Bar
  Sinai}}, \citenamefont {Brener},\ and\ \citenamefont
  {Bouchbinder}}]{BarSinai2012}%
  \BibitemOpen
  \bibfield  {author} {\bibinfo {author} {\bibfnamefont {Y.}~\bibnamefont {{Bar
  Sinai}}}, \bibinfo {author} {\bibfnamefont {E.~A.}\ \bibnamefont {Brener}}, \
  and\ \bibinfo {author} {\bibfnamefont {E.}~\bibnamefont {Bouchbinder}},\
  }\bibfield  {title} {\emph {\bibinfo {title} {{Slow rupture of frictional
  interfaces}},\ }}\href
  {http://doi.wiley.com/10.1029/2011GL050554} {\bibfield  {journal} {\bibinfo
  {journal} {Geophys. Res. Lett.}\ }\textbf {\bibinfo {volume} {39}},\ \bibinfo
  {pages} {L03308} (\bibinfo {year} {2012})}\BibitemShut {NoStop}%
\bibitem [{\citenamefont {Crupi}\ and\ \citenamefont
  {Bizzarri}(2013)}]{Crupi2013}%
  \BibitemOpen
  \bibfield  {author} {\bibinfo {author} {\bibfnamefont {P.}~\bibnamefont
  {Crupi}}\ and\ \bibinfo {author} {\bibfnamefont {A.}~\bibnamefont
  {Bizzarri}},\ }\bibfield  {title} {\emph {\bibinfo {title} {{The role of
  radiation damping in the modeling of repeated earthquake events}},\ }}\href
  {http://www.annalsofgeophysics.eu/index.php/annals/article/view/6200}
  {\bibfield  {journal} {\bibinfo  {journal} {Ann. Geophys.}\ }\textbf
  {\bibinfo {volume} {56}},\ \bibinfo {pages} {R0111} (\bibinfo {year}
  {2013})}\BibitemShut {NoStop}%
\end{thebibliography}

\begin{thebibliography}{10}%
	\makeatletter
	\providecommand \@ifxundefined [1]{%
		\@ifx{#1\undefined}
	}%
	\providecommand \@ifnum [1]{%
		\ifnum #1\expandafter \@firstoftwo
		\else \expandafter \@secondoftwo
		\fi
	}%
	\providecommand \@ifx [1]{%
		\ifx #1\expandafter \@firstoftwo
		\else \expandafter \@secondoftwo
		\fi
	}%
	\providecommand \natexlab [1]{#1}%
	\providecommand \enquote  [1]{``#1''}%
	\providecommand \bibnamefont  [1]{#1}%
	\providecommand \bibfnamefont [1]{#1}%
	\providecommand \citenamefont [1]{#1}%
	\providecommand \href@noop [0]{\@secondoftwo}%
	\providecommand \href [0]{\begingroup \@sanitize@url \@href}%
	\providecommand \@href[1]{\@@startlink{#1}\@@href}%
	\providecommand \@@href[1]{\endgroup#1\@@endlink}%
	\providecommand \@sanitize@url [0]{\catcode `\\12\catcode `\$12\catcode
		`\&12\catcode `\#12\catcode `\^12\catcode `\_12\catcode `\%12\relax}%
	\providecommand \@@startlink[1]{}%
	\providecommand \@@endlink[0]{}%
	\providecommand \url  [0]{\begingroup\@sanitize@url \@url }%
	\providecommand \@url [1]{\endgroup\@href {#1}{\urlprefix }}%
	\providecommand \urlprefix  [0]{URL }%
	\providecommand \Eprint [0]{\href }%
	\providecommand \doibase [0]{http://dx.doi.org/}%
	\providecommand \selectlanguage [0]{\@gobble}%
	\providecommand \bibinfo  [0]{\@secondoftwo}%
	\providecommand \bibfield  [0]{\@secondoftwo}%
	\providecommand \translation [1]{[#1]}%
	\providecommand \BibitemOpen [0]{}%
	\providecommand \bibitemStop [0]{}%
	\providecommand \bibitemNoStop [0]{.\EOS\space}%
	\providecommand \EOS [0]{\spacefactor3000\relax}%
	\providecommand \BibitemShut  [1]{\csname bibitem#1\endcsname}%
	\let\auto@bib@innerbib\@empty
	%</preamble>
	\bibitem [{\citenamefont {Landau}\ and\ \citenamefont
		{Lifshitz}(1986)}]{S-LLElasticity}%
	\BibitemOpen
	\bibfield  {author} {\bibinfo {author} {\bibfnamefont {L.~D.}\ \bibnamefont
			{Landau}}\ and\ \bibinfo {author} {\bibfnamefont {E.~M.}\ \bibnamefont
			{Lifshitz}},\ }\href@noop {} {\emph {\bibinfo {title} {{Theory of elasticity,
					Vol. 7}}}},\ \bibinfo {edition} {3rd}\ ed.\ (\bibinfo  {publisher} {Pergamon
		Press},\ \bibinfo {year} {1986})\BibitemShut {NoStop}%
	\bibitem [{\citenamefont {Timoshenko}\ and\ \citenamefont
		{Goodier}(1951)}]{S-Timoshenko}%
	\BibitemOpen
	\bibfield  {author} {\bibinfo {author} {\bibfnamefont {S.}~\bibnamefont
			{Timoshenko}}\ and\ \bibinfo {author} {\bibfnamefont {J.~N.}\ \bibnamefont
			{Goodier}},\ }\href
	{http://books.google.co.il/books/about/Theory{\_}of{\_}Elasticity{\_}7.html?id=tpY-VkwCkAIC{\&}redir{\_}esc=y}
	{\enquote {\bibinfo {title} {{Theory of Elasticity 7}},}\ } (\bibinfo {year}
	{1951})\BibitemShut {NoStop}%
	\bibitem [{\citenamefont {Gradshteyn}\ and\ \citenamefont
		{Ryzhik}(2007)}]{S-Gradshteyn}%
	\BibitemOpen
	\bibfield  {author} {\bibinfo {author} {\bibfnamefont {I.}~\bibnamefont
			{Gradshteyn}}\ and\ \bibinfo {author} {\bibfnamefont {I.}~\bibnamefont
			{Ryzhik}},\ }\href@noop {} {\emph {\bibinfo {title} {{Table of Integrals,
					Series, and Products}}}},\ \bibinfo {edition} {7th}\ ed.\ (\bibinfo {year}
	{2007})\BibitemShut {NoStop}%
	\bibitem [{\citenamefont {Svetlizky}\ \emph {et~al.}(2016)\citenamefont
		{Svetlizky}, \citenamefont {{Pino Mu{\~{n}}oz}}, \citenamefont {Radiguet},
		\citenamefont {Kammer}, \citenamefont {Molinari},\ and\ \citenamefont
		{Fineberg}}]{S-Svetlizky2016}%
	\BibitemOpen
	\bibfield  {author} {\bibinfo {author} {\bibfnamefont {I.}~\bibnamefont
			{Svetlizky}}, \bibinfo {author} {\bibfnamefont {D.}~\bibnamefont {{Pino
					Mu{\~{n}}oz}}}, \bibinfo {author} {\bibfnamefont {M.}~\bibnamefont
			{Radiguet}}, \bibinfo {author} {\bibfnamefont {D.~S.}\ \bibnamefont
			{Kammer}}, \bibinfo {author} {\bibfnamefont {J.-F.}\ \bibnamefont
			{Molinari}}, \ and\ \bibinfo {author} {\bibfnamefont {J.}~\bibnamefont
			{Fineberg}},\ }\href {\doibase 10.1073/pnas.1517545113} {\bibfield  {journal}
		{\bibinfo  {journal} {Proc. Natl. Acad. Sci.}\ }\textbf {\bibinfo {volume}
			{113}},\ \bibinfo {pages} {542} (\bibinfo {year} {2016})}\BibitemShut
	{NoStop}%
	\bibitem [{\citenamefont {Svetlizky}\ and\ \citenamefont
		{Fineberg}(2014)}]{S-Svetlizky2014}%
	\BibitemOpen
	\bibfield  {author} {\bibinfo {author} {\bibfnamefont {I.}~\bibnamefont
			{Svetlizky}}\ and\ \bibinfo {author} {\bibfnamefont {J.}~\bibnamefont
			{Fineberg}},\ }\href {\doibase 10.1038/nature13202} {\bibfield  {journal}
		{\bibinfo  {journal} {Nature}\ }\textbf {\bibinfo {volume} {509}},\ \bibinfo
		{pages} {205} (\bibinfo {year} {2014})}\BibitemShut {NoStop}%
	\bibitem [{\citenamefont {Ajovalasit}\ \emph {et~al.}(2013)\citenamefont
		{Ajovalasit}, \citenamefont {Fragapane},\ and\ \citenamefont
		{Zuccarello}}]{S-Ajovalasit2013}%
	\BibitemOpen
	\bibfield  {author} {\bibinfo {author} {\bibfnamefont {A.}~\bibnamefont
			{Ajovalasit}}, \bibinfo {author} {\bibfnamefont {S.}~\bibnamefont
			{Fragapane}}, \ and\ \bibinfo {author} {\bibfnamefont {B.}~\bibnamefont
			{Zuccarello}},\ }\href {\doibase 10.1111/str.12043} {\bibfield  {journal}
		{\bibinfo  {journal} {Strain}\ }\textbf {\bibinfo {volume} {49}},\ \bibinfo
		{pages} {366} (\bibinfo {year} {2013})}\BibitemShut {NoStop}%
	\bibitem [{\citenamefont {Broberg}(1999)}]{S-Broberg1999Book}%
	\BibitemOpen
	\bibfield  {author} {\bibinfo {author} {\bibfnamefont {K.~B.}\ \bibnamefont
			{Broberg}},\ }\href@noop {} {\emph {\bibinfo {title} {Cracks Fract.}}}\
	(\bibinfo  {publisher} {Elsevier},\ \bibinfo {year} {1999})\BibitemShut
	{NoStop}%
	\bibitem [{\citenamefont {Shlomai}\ and\ \citenamefont
		{Fineberg}(2016)}]{S-Shlomai2016}%
	\BibitemOpen
	\bibfield  {author} {\bibinfo {author} {\bibfnamefont {H.}~\bibnamefont
			{Shlomai}}\ and\ \bibinfo {author} {\bibfnamefont {J.}~\bibnamefont
			{Fineberg}},\ }\href {\doibase 10.1038/ncomms11787} {\bibfield  {journal}
		{\bibinfo  {journal} {Nat. Commun.}\ }\textbf {\bibinfo {volume} {7}},\
		\bibinfo {pages} {11787} (\bibinfo {year} {2016})}\BibitemShut {NoStop}%
	\bibitem [{\citenamefont {Poliakov}\ \emph {et~al.}(2002)\citenamefont
		{Poliakov}, \citenamefont {Dmowska},\ and\ \citenamefont
		{Rice}}]{S-Poliakov2002}%
	\BibitemOpen
	\bibfield  {author} {\bibinfo {author} {\bibfnamefont {A.~N.~B.}\
			\bibnamefont {Poliakov}}, \bibinfo {author} {\bibfnamefont {R.}~\bibnamefont
			{Dmowska}}, \ and\ \bibinfo {author} {\bibfnamefont {J.~R.}\ \bibnamefont
			{Rice}},\ }\href {\doibase 10.1029/2001JB000572} {\bibfield  {journal}
		{\bibinfo  {journal} {J. Geophys. Res. Solid Earth}\ }\textbf {\bibinfo
			{volume} {107}},\ \bibinfo {pages} {ESE 6} (\bibinfo {year}
		{2002})}\BibitemShut {NoStop}%
	\bibitem [{\citenamefont {Samudrala}\ \emph {et~al.}(2002)\citenamefont
		{Samudrala}, \citenamefont {Huang},\ and\ \citenamefont
		{Rosakis}}]{S-Samudrala2002}%
	\BibitemOpen
	\bibfield  {author} {\bibinfo {author} {\bibfnamefont {O.}~\bibnamefont
			{Samudrala}}, \bibinfo {author} {\bibfnamefont {Y.}~\bibnamefont {Huang}}, \
		and\ \bibinfo {author} {\bibfnamefont {A.~J.}\ \bibnamefont {Rosakis}},\
	}\href {\doibase 10.1029/2001JB000460} {\bibfield  {journal} {\bibinfo
		{journal} {J. Geophys. Res.}\ }\textbf {\bibinfo {volume} {107}},\ \bibinfo
	{pages} {2170} (\bibinfo {year} {2002})}\BibitemShut {NoStop}%
\end{thebibliography}
%

\onecolumngrid
\newpage
\begin{center}
	\textbf{\large Supplemental Materials for:\\``On the spatial distribution of thermal energy in equilibrium''}
\end{center}
%\twocolumngrid

%%%%%%%%%%%%%%%%%%%%%%%%%%%%%%%%%%%%%%%%%%%%%%%%%%%%%%%%%%%%%%%%%%%%%%%%%%%%%%%%%
%%%%%%%%%%%%%%%%%%%%%% these lines of code handle the concatenation %%%%%%%%%%%%%
%%%%%%%%%%%%%%%%%%%%%%%%%%%%%%%%%%%%%%%%%%%%%%%%%%%%%%%%%%%%%%%%%%%%%%%%%%%%%%%%%
\setcounter{equation}{0}
\setcounter{figure}{0}
\setcounter{section}{0}
\setcounter{table}{0}
\setcounter{page}{1}
\makeatletter
\renewcommand{\theequation}{S\arabic{equation}}
\renewcommand{\thefigure}{S\arabic{figure}}
\renewcommand{\thesection}{S-\Roman{section}}
\renewcommand*{\thepage}{S\arabic{page}}
\renewcommand{\bibnumfmt}[1]{[S#1]}
\renewcommand{\citenumfont}[1]{S#1}
%%%%%%%%%%%%%%%%%%%%%%%%%%%%%%%%%%%%%%%%%%%%%%%%%%%%%%%%%%%%%%%%%%%%%%%%%%%%%%%%%
%%%%%%%%%%%%%%%%%%%%%% these lines of code handle the concatenation %%%%%%%%%%%%%
%%%%%%%%%%%%%%%%%%%%%%%%%%%%%%%%%%%%%%%%%%%%%%%%%%%%%%%%%%%%%%%%%%%%%%%%%%%%%%%%%

\newcommand{\K}{\mathcal{K}}

\section{The response functions \texorpdfstring{$M_{ij}$}{} in plane-strain elasticity}
\label{sec:M}

The purpose of this section is to explicitly calculate the relation between the interfacial stress $\sigma_{yi}$ (which is a vector) and the interfacial displacement $\bm u$ for a two-dimensional (2D) elastic body that occupies the region $-\infty\!<\!x\!<\!\infty$ and $0\!\le\!y\!\le\!H$. The bottom boundary at $y\=0$ is a frictional interface and plane-strain conditions are assumed. Thus, the equations of motion are those of linear elasticity~\cite{S-LLElasticity}, i.e.
\begin{align}
\label{eq:LE}
\nabla \cdot \bm{\sigma}&=\rho\frac{\partial ^2 {\bm u}}{\partial t^2}\ ,
&
\left(
\begin{array}{c}
\sigma _{xx} \\
\sigma _{yy} \\
\sigma _{xy} \\
\end{array}
\right)&=\frac{2 \mu}{1-2 \nu } \left(
\begin{array}{ccc}
1-\nu & \nu & 0 \\
\nu& 1-\nu& 0 \\
0 & 0 & 1-2 \nu \\
\end{array}
\right) \left(
\begin{array}{c}
\varepsilon _{xx} \\
\varepsilon _{yy} \\
\varepsilon _{xy} \\
\end{array}
\right)\ ,
\end{align}
where $\varepsilon_{ij}\!\equiv\!\frac{1}{2}\left(\partial_i u_{j}+\partial_j u_{i}\right)$ is the infinitesimal strain tensor (not to be confused with the slip displacement discontinuity vector $\epsilon_i$), $\bm \sigma$ is Cauchy's stress tensor and $\mu$, $\nu$  and $\rho$ are the shear modulus, Poisson's ratio and mass density, respectively.

At the top boundary $y\=H$ the material is loaded by imposing a horizontal velocity $v$ and a compressive normal stress $\sigma_{yy}\=-\sigma_0$ (with $\sigma_0\!>\!0$). The homogeneous solution $\bm u_h$ consistent with these boundary conditions reads
\begin{equation}
\bm u_h(y,t)\equiv \Big(\frac{f\sigma_0}{\mu}y+v t\ ,\ -\frac{1-2\nu}{2(1-\nu)}\frac{\sigma_0}{\mu} y\Big)\ .
\end{equation}
Since the equations of motion are linear, one can decompose a general solution to a sum of the steady solution of homogeneous sliding and a deviation from it, and write $\bm u(x,y,t)\=\bm u_h(y,t)+\delta\bm u(x,y,t)$. The boundary conditions (BC) at $y\=H$ are
\begin{equation}
\pa_t(\delta u_x)=0 \qquad \mbox{ and } \qquad \delta\sigma_{yy}=0 \ .
\label{eq:SBC}
\end{equation}
In what follows, we calculate the response function of the field $\delta\bm u$. For easier readability we omit henceforth the notation $\delta\bm u$ and denote it simply by $\bm u$.

Consider now a single Fourier mode, i.e. assume that all fields depend on $x$ and $t$ as $\propto e^{ik(x-ct)}$, for which Eq.~\eqref{eq:LE} admits a solution of the form
\begin{equation}
\label{eq:full_sol}
{\bm u}
=\left(\begin{array}{llll}
A_1 \alpha_{s} e^{-k\alpha_{s}y}&+A_2 \alpha_{s} e^{k\alpha_{s}y}&+A_3 e^{-k{\alpha_{d}}y}&+A_4 e^{k{\alpha_{d}}y}\\
iA_1 e^{-k\alpha_{s}y}&-iA_2 e^{k\alpha_{s}y}&+iA_3{\alpha_{d}} e^{-k{\alpha_{d}}y} &-iA_4 {\alpha_{d}} e^{k{\alpha_{d}}y}
\end{array}\right)e^{ik(x-ct)}\ ,
\end{equation}
where $c_s\!=\!\sqrt{\frac{\mu}{\rho}}$ and $c_d\!=\!\sqrt{\frac{2-2 \nu}{1-2\nu}}c_s$ are the shear and dilatational wavespeeds and we defined $\alpha_{i}\equiv\sqrt{1-c^2/c_i^2} $ where $i\in\{s,d\}$. $A_i$ are 4 unknown amplitudes which are determined by employing 4 boundary conditions. These are the 2 conditions at $y\=H$, which are given in Eq.~\eqref{eq:SBC}, and 2 conditions at $y\=0$, given by the interfacial stresses $\sigma_{yi}\,e^{ik(x-ct)}$ (which may arise from the frictional contact with another body or any other force-generating loading conditions). After calculating the amplitudes, one can express the relation between the interfacial displacements $u_i$ and the interfacial stresses $\sigma_{yi}$ in the form $u_i\=M_{ij}(c,k)\sigma_{yj}$, where
\begin{equation}
\label{eq:general_M}
\bm{M}=\frac{1}{\mu k\left(T_d \left(\alpha _s^2+1\right)^2-4 \alpha _d \alpha _s T_s\right)}
\left(
\begin{array}{cc}
T_d T_s \alpha _s(1-\alpha_s^2)&
i \left(T_d \left(\alpha _s^2+1\right)-2 T_s \alpha _d \alpha _s\right) \\
-i \left(T_d \left(\alpha _s^2+1\right)-2 T_s \alpha _d \alpha _s\right) &
\alpha _d(1-\alpha_s^2) \\
\end{array}
\right)
\ ,
\end{equation}
and $T_i\equiv\tanh(k H \alpha_i)$. Note that in case that the body under consideration occupies the region in space $-H\!\le\!y\!\le\!0$ (with $H\!>\!0$) the analysis remains valid, but $H$ should be replaced by $-H$. This simply amounts to changing the sign of the diagonal entries of $\bm M$ in Eq.~\eqref{eq:general_M}.

In the main text, a specific example of Eq.~\eqref{eq:general_M} for a semi-infinite half-space $y\!<\!0$ under quasi-static (QS) conditions was considered in Eq.~(1). This is achieved from Eq.~\eqref{eq:general_M} in two steps. First, a semi-infinite half-space $y\!<\!0$ corresponds to $H\!\to\!-\infty$, which implies $T_i\!\to\!-1$, for which we obtain
\begin{equation}
\bm M =-\frac{1}{\mu k\Big(\left(\alpha _s^2+1\right)^2-4 \alpha _d \alpha _s\Big)}\left(
\begin{array}{cc}
\alpha _s \left(1-\alpha _s^2\right) & -i \left(1+\alpha _s^2-2 \alpha _d \alpha _s\right) \\
i \left(1+\alpha _s^2-2 \alpha _d \alpha _s\right) & \alpha _d \left(1-\alpha _s^2\right) \\
\end{array}
\right)\ .
\label{eq:m2s}
\end{equation}
Second, the QS limit is obtained by taking $c\!\to\!0$. Note that since $\alpha_i\!\to\!1$ in this limit, all entries of the matrix in Eq.~\eqref{eq:m2s} vanish, but the prefactor diverges. Their product approaches a finite limit, yielding
\begin{equation}
\label{eq:M_QS_inf_H}
\bm{M}=\frac{1}{\mu k}\left(
\begin{array}{cc}
1-\nu  & -\frac{i}{2} (1-2 \nu ) \\
\frac{i}{2} (1-2 \nu ) & 1-\nu  \\
\end{array}
\right)\ .
\end{equation}
This expression identifies with Eq. (1) in the manuscript.

\section{The response of a composite system}

Next we aim at calculating the response of a general composite system, composed of two bodies made of different materials with different $H$'s in frictional contact. Both bodies are assumed to be infinite in the $x$-direction. The upper material, denoted by the superscript (1), is assumed to occupy the region $0\!<\!y\!<H\1$ and the lower one, denoted by the superscript (2), the region $-H\2\!<\!y\!<\!0$ (with positive $H^{(i)}$). Since $\alpha_{s,d}$ and $T_{s,d}$ (and $\mu$) may be different for the different materials, they are labeled with a superscript,
\begin{align}
\alpha_{j}^{(i)}&\equiv\left(1-\frac{c}{c_j^{(i)}}\right)^{1/2}\ , &
T_{j}^{(i)}&\equiv \tanh\Big(k H^{(i)} \alpha_{j}^{(i)}\Big)\ , &
i&\in\{1,2\},\ j\in\{s,d\}\ .
\end{align}

Following the previous section, the displacements at the frictional interface $y\=0$ are given as
\begin{align}
\begin{pmatrix}u_x\1\\u_{y}\1\\\end{pmatrix}&=\bm M\1 \begin{pmatrix}\sigma_{xy}\1\\\sigma_{yy}\1\\\end{pmatrix}\ ,
&
\begin{pmatrix}u_x\2\\u_{y}\2\\\end{pmatrix}&=\bm M\2 \begin{pmatrix}\sigma_{xy}\2\\\sigma_{yy}\2\\\end{pmatrix}\ .
\end{align}
Since both $\sigma_{xy}$ and $\sigma_{yy}$ are continuous at $y\=0$, the displacement discontinuity $\epsilon_i$ (cf.~Eq.~\eqref{eq:epsilon} in the main text) can be written as
\begin{equation}
\begin{pmatrix}\epsilon_x \\ \epsilon_{y} \end{pmatrix}=\Big(\bm M\1 - \bm M\2 \Big)\begin{pmatrix}\sigma_{xy}\\ \sigma_{yy} \\\end{pmatrix}\ .
\end{equation}
Since we demand $\epsilon_y\=0$, the relation between $\sigma_{yi}$ and $\epsilon_x$ takes the form
\begin{align}
\begin{pmatrix}\sigma_{xy}\\ \sigma_{yy} \\\end{pmatrix}&=\bm G \begin{pmatrix}\epsilon_x \\ 0 \end{pmatrix}\ ,
\qquad \qquad
\mbox{ where } \qquad
\bm G\equiv \Big(\bm M\1 - \bm M\2 \Big)^{-1} \ .
\label{eq:defG}
\end{align}
We thus write
\begin{align}\label{eq:sig_G}
\sigma_{xy}&=\mu\1 k G_x \epsilon_x
&
\sigma_{yy}&= i \mu\1 k G_y \epsilon_x\ ,
\end{align}
with the definitions $G_x \equiv G_{xx}/(\mu\1 k)$ and $G_y \equiv G_{yx}/(i\mu\1 k)$. Note that in symmetric systems $\bm M\2$ is obtained from $\bm M\1$ simply by taking $H\to-H$. Therefore, the diagonal terms of $\bm M\1$ and $\bm M\2$ have opposite signs and the off-diagonal terms are identical, or in other words, $\bm M\1-\bm M\2$ is diagonal. Thus, $\bm G$ is also diagonal, i.e.~$G_y\=0$, and no coupling exists between tangential motion and normal traction in this case (i.e.~for symmetric systems).

In the case addressed in Sec.~\ref{sec:stability} of the main text we have $\mu\1=\mu\2\equiv\mu$ and $\nu\1=\nu\2\equiv\nu$ and therefore also $\alpha\1_{s,d}=\alpha\2_{s,d}\equiv\alpha_{s,d}$. Plugging these in Eqs.~\eqref{eq:general_M} and \eqref{eq:defG}, one obtains
% \begin{equation}\begin{split}
% \bm G=&\frac{k \mu  c_s^2}{c^2 (T\1_d+T\2_d) (T\1_s+T\2_s)}\times\\
% &\left(
% \begin{array}{cc}
%  \frac{\left(\alpha _s^2+1\right)^2 (T\1_d+T\2_d)}{\alpha _s}-4 \alpha _d (T\1_s+T\2_s) &
%   -2 i \left(\alpha _s^2+1\right) (T\2_d T\1_s-T\1_d T\2_s) \\
%  2 i \left(\alpha _s^2+1\right) (T\2_d T\1_s-T\1_d T\2_s) &
%  \frac{T\1_d T\2_d \left(\alpha _s^2+1\right)^2 (T\1_s+T\2_s)}{\alpha _d}-4 T\1_s T\2_s \alpha _s (T\1_d+T\2_d)\\
% \end{array}
% \right)\ ,
% \end{split}\end{equation}
% such that
\begin{align}\label{eq:G_gen}
G_x&=\frac{c_s^2}{c^2}\left(\frac{\left(\alpha _s^2+1\right)^2}{\alpha _s \left(T\1_s+T\2_s\right)}-\frac{4 \alpha _d}{T\1_d+T\2_d}\right)\ ,&
G_y&=2\left(\alpha _s^2+1\right)\frac{c_s^2}{c^2}\left(\frac{T\2_d}{T\1_d+T\2_d}-\frac{T\2_s}{T\1_s+T\2_s}\right)\ .
\end{align}

\section{Linear stability analysis of homogeneous sliding}

In this section we provide more details regarding the linear stability analysis discussed in Sec.~\ref{sec:stability} of the main text. The basic constitutive relation for the frictional stress reads
\begin{equation}
\label{eq:fric_law}
\sigma_{xy}+f(v,\phi)\sigma_{yy} =0 \ .
\end{equation}
Taking the variation of this relation with respect to sliding velocity perturbations $\delta v$ relative to steady state sliding at $v$, one readily obtains to linear order
\begin{equation}
\frac{\delta\sigma_{xy}}{\delta v}+f\frac{\delta\sigma_{yy}}{\delta v}-\sigma_0 \frac{\delta\!f}{\delta v}=0 \ ,
\end{equation}
where we used the fact that to zeroth order we have $\sigma_{yy}\=-\sigma_0$. Using Eqs.~\eqref{eq:sig_G} and the relation $\delta{v}\=-i c k \delta\epsilon _x$ we obtain
\begin{equation}
\mu\Big(G_x(c,k)+i f G_y(c,k)\Big)+ic\,\sigma_0\frac{\delta\!f }{\delta v}=0 \ ,
\end{equation}
where $f$ is evaluated at the steady state sliding velocity $v$. This is Eq.~\eqref{eq:implicit_spectrum} of the main text.

In the simple case of velocity-dependent friction, where $f\!=\!f(v)$ (i.e.~no state dependence), we simply have $\frac{\delta\!f}{\delta v}\!=\!f'(v)$. In the general case, we have $f\!=\!f(v,\phi)$ with $\dot{\phi}\!=\!g(\tfrac{v\phi}{D})$. In steady state we have $\phi\!=\!D/v$ such that $g(1)\!=\!0$, and in addition we expect $g'(1)\!<\!0$. The perturbation of $f$ takes the form
\begin{equation}
\delta f=\left(\pd{f}{v}+\pd{f}{\phi}\frac{\delta\phi}{\delta{v}}\right)\delta{v}\ .
\label{eq:fint}
\end{equation}
Avoiding direct reference to $\pa_\phi f$, we use the fact that in steady state $\phi\!=\!D/v$ and thus $d_v\!f=\partial_v\!f-\frac{D}{v^2}\partial_\phi f$. We then rewrite Eq.~\eqref{eq:fint} as
\begin{equation}
\frac{\delta f}{\delta v}=\left(1+(1-\Delta)\frac{v^2}{D}\frac{\delta\phi}{\delta{v}}\right)\pa_v f \qquad\hbox{with}\qquad
\Delta \equiv \frac{d_v\!f}{\partial_v\!f}\ .
\label{eq:fint2}
\end{equation}
In order to obtain $\delta\phi/\delta{v}$, we perturb $\dot{\phi}\!=\!g\left(\frac{v\phi}{D}\right)$ by setting $v\!=\!v+\delta{v}$ and $\phi\!=\!D/v+\delta\phi$, which to leading order yields
\begin{equation}
\delta\dot\phi=-ick\delta\phi=g'(1) \left(\frac{v}{D}\delta \phi+\frac{\delta v}{v}\right) \ .
\end{equation}
This is a linear equation that can be solved for $\delta\phi/\delta v$. Plugging the solution into Eq.~\eqref{eq:fint2}, we finally obtain
\begin{equation}
\label{eq:delta f RSF_SM}
\frac{\delta\!f}{\delta v}= \pa_v\!f\left(1 + \frac{\Delta - 1}{1-i\,\xi\frac{c}{c_s}kH}\right) \qquad\qquad\hbox{with}\qquad\qquad \xi \equiv \frac{D c_s}{H v |g'(1)|} \ .
\end{equation}
This is Eq.~\eqref{eq:delta f RSF} in the manuscript.

\subsection{Simplified analysis: $\eta=\infty$ and $\Delta=1$}
\label{subsec:inf_eta}

In Sect.~\ref{sec:simplified} of the main text we examine the stability of the steady state sliding at a velocity $v$ for which the frictional stress takes the form  $\sigma_{xy}\=f(v)\sigma_0$ (i.e. no state-dependence, $\Delta=1$). As detailed above, in this case the equation that defines the stability spectrum $c(k)$ is
\begin{equation}
\mu\Big(G_x(c,k)+i f(v) G_y(c,k)\Big)+ic\,\sigma_0\, f'(v)=0\ .
\label{eq:implicit_spectrum_S}
\end{equation}
We also assume the bottom layer is infinitely deep, which means $\eta\!\to\!\infty$. As a result, $T\2_{s,d}$ of Eq.~\eqref{eq:G_gen} should be replaced by $-1$.

As noted in the main text, Eq.~\eqref{eq:implicit_spectrum_S} admits multiple solution branches, a few of them might be stable or unstable, depending on the system parameters and on $k$. Here we focus on a solution branch which is located near the Rayleigh wavespeed $c\=c_R$ in the complex $c$-plane, cf.~Fig.~\ref{fig:spectrum}b in the manuscript.
% , which is the unique positive real solution to the equation
% \begin{equation}
% 4 \alpha_s(c)\,\alpha_d(c)=\left(1+\alpha_s(c)^2\right)^2\ .
% \end{equation}
First, we non-dimensionalize the equations by defining $z\=c/c_s$, $q\=kH$ and using the definitions introduced in the main text $\gamma\!\equiv\!\mu/(\sigma_0 c_s f'(v))$ and $\beta\=c_s/c_d$. With these, we obtain
\begin{align}
% \alpha_s&=\sqrt{1-z^2} \ ,&
G_x&=z^{-2}\left(\frac{\left(1+\alpha_s^2\right)^2}{\alpha_s \left(\tanh \left(q \alpha_s\right)+1\right)}-\frac{4 \alpha_d}{\tanh \left(q \alpha_d\right)+1}\right)\ ,
\label{eq:Gx}\\[3mm]
% \alpha_d&=\sqrt{1-\beta^2z^2} \ ,&
G_y&=2z^{-2}\left(\alpha _s^2+1\right)\left(\frac{1}{\tanh \left(q \alpha_d\right)+1}-\frac{1}{\tanh \left(q \alpha_s\right)+1}\right)\ ,
\label{eq:Gy}
\end{align}
where $\alpha_s\=\sqrt{1-z^2}$ and $\alpha_d\=\sqrt{1-\beta^2z^2}$, and the implicit spectrum equation reads
\begin{align}
\gamma(G_x +i f G_y)+iz=0\ .
\label{eq:S_z}
\end{align}

We now expand Eq.~\eqref{eq:S_z} to linear order in $\delta z$, where $z\=z_R+\delta z$ and $z_R\!\equiv\!c_R/c_s$ is the dimensionless Rayleigh wavespeed. The solution for $\delta z$ reads
\begin{align}
\delta z \simeq -\frac{\displaystyle z_R-i \gamma  \Big(G_x(z_R,q)+i f G_y(z_R,q)\Big)}{\displaystyle 1-i \gamma  \Big(G_x'(z_R,q)+i f G_y'(z_R,q)\Big)}\ ,
\label{eq:dz}
\end{align}
where $G_j'$ denotes $\pa G_j/\pa z$ evaluated at $z\=z_R$. This solution is plotted in Fig.~\ref{fig:spectrum}a-b of the main text. Since $z_R\!\lesssim\!1$, the functions $G_x(z_R,q)$ and $G_y(z_R,q)$ are real, as well as their derivatives. Thus, the real and imaginary parts of $\delta z$ in Eq.~\eqref{eq:dz} are
\begin{align}
\Re[\delta{z}]&=-\frac{\left(\gamma  f G_y'+1\right) \left(\gamma  f G_y+z_R\right)+\gamma ^2 G_x G_x'}{\left(\gamma  f G_y'+1\right)^2+\gamma ^2 \left(G_x'\right)^2}
\ ,&
\Im[\delta z]&=-\gamma G'_x\frac{\Big(z_R+\gamma  f G_y\Big)+\frac{G_x}{G'_x} \left(1+\gamma  f G_y'\right)}{\left(\gamma  f G_y'+1\right)^2+\gamma ^2 \left(G_x'\right)^2} \ .
\label{eq:fffff}
\end{align}

\begin{figure}
	\centering
	\includegraphics[width=0.4\textwidth]{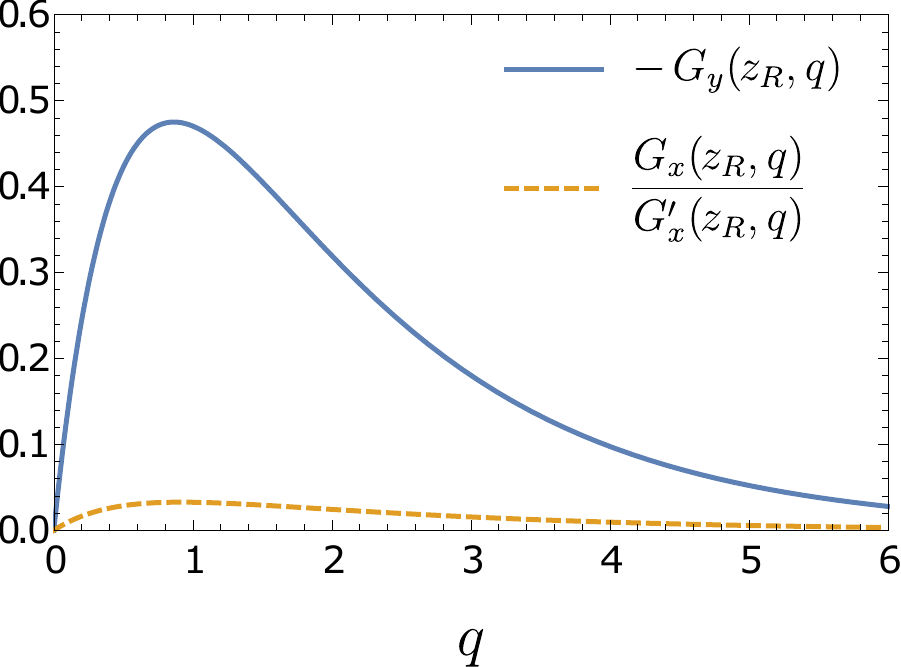}
	\caption{The combinations of the response functions $G_x,G_y$ and $G_x'$ which appear in Eq.~\eqref{eq:fffff}. $\beta=0.3$ was used.}
	\label{fig:compare_G}
\end{figure}

The assumption underlying the expansion to leading order in $\delta z$ is $|\delta z|\!\ll\!z_R$, where $z_R$ is smaller than, but close to, unity. $\Im[\delta z]$ is small by construction as we are interested in understanding the instability threshold, determined by a zero crossing of $\Im[\delta z]$. To assess the smallness of $\Re[\delta z]$, note that $\alpha_s(z_R)\=\sqrt{1-z_R^2}\!\ll\!1$ due to the proximity of $z_R$ to unity (while $\alpha_d(z_R)$ remains of order unity because of the factor $\beta$). Furthermore, note that $G_x(z_R,q)$ contains a term proportional to $\alpha_s^{-1}(z_R)$, cf.~Eq.~\eqref{eq:Gx}. Since $\pa_z (\alpha_s(z)^{-1})=z/\alpha_s(z)^3$, we have  $G_x(z_R,q)/G'_x(z_R,q)\!\sim\!\alpha_s^{2}\!\ll\!1$ and therefore the second term in the numerator of $\Im[\delta z]$ is negligible with respect to the first. This is demonstrated explicitly in Fig.~\ref{fig:compare_G}. Thus, the criterion for the critical wavelnumber at which $\Im[\delta z]$ changes sign is approximately $\gamma f G_y(z_R,q)\!\approx\!-z_R$, which is Eq.~\eqref{eq:criterion} of the main text. Finally, we note that the fact that $G_x(z_R,q)/G'_x(z_R,q)\!\ll\!1$ is self-consistent with our working assumption that $\Re[\delta z]$ is small near the threshold. Near threshold, i.e.~when $\gamma  f G_y+z_R\approx 0$, we have %the real part of $\Re[\delta z]$ reads
\begin{equation}
\Re[\delta z]\approx -\frac{\gamma ^2 G_x/G_x'}{(G_x')^{-2}\left(\gamma  f G_y'+1\right)^2+\gamma^2}\ll1\ .
\end{equation}

\section{Response function of the ``thin-on-thick'' geometry}

In this section we calculate the response function of the ``thin-on-thick'' geometry discussed in Sec.~\ref{sec:thin-on-thick} of the main text and depicted in Fig.~\ref{fig:geometries}a. The thinner (upper) block is assumed to be under plane-stress conditions, for which the elastic response is known to be identical to that of plane-strain, cf.~Eq.~\eqref{eq:M_QS_inf_H}, but with renormalized elastic constants $\nu^{\mbox{\tiny plane-stress}}\=\nu/(1+\nu)$ and $\mu^{\mbox{\tiny plane-stress}}\=\mu$~\cite{S-Timoshenko}.

To calculate the response of the thicker (lower) block, we model it as a semi-infinite half-space. Although in the main text (and in the experimental setup) the thicker block is the lower one, in order to conform to the notations of Sec.~\ref{sec:M} we calculate here the response matrix of a semi-infinite half-space that occupies the upper half-space $y\!>\!0$ and at the end transform the result to be valid for the lower half-space $y\!<\!0$. As discussed in Sec.~\ref{sec:M}, the response matrix of a material that occupies the lower half-space $y\!<\!0$ is obtained by inverting the signs of the diagonal elements of the matrix.

The surface Green function of a half-space, which relates the displacement field ${\bm u}(x,y\=0,z)$ to a point force at the origin ${\bm F} \delta(r)$, reads~\cite[Eq.~(8.19)]{S-LLElasticity}
\begin{align}
\begin{pmatrix}
u_x \\u_y\\u_z
\end{pmatrix}
=-\frac{1}{4\pi \mu r}
\begin{pmatrix}
2 (1-\nu )+2\nu\frac{x^2}{r^2} &  -(1-2 \nu)\frac{x}{r} & 2\nu\frac{x z}{r^2} \\
(1-2\nu) \frac{x}{r} & 2 (1-\nu ) & (1-2 \nu )\frac{z}{r} \\
2\nu \frac{x z}{r^2} & -(1-2\nu)\frac{z}{r} &2 (1-\nu)+2\nu \frac{z^2}{r^2} \\
\end{pmatrix}
\begin{pmatrix}
F_x \\F_y\\F_z
\end{pmatrix}\ ,
\end{align}
where $r\equiv \sqrt{x^2+z^2}$. As explained in the main text, the following properties are used:
\begin{enumerate}[label=(\alph*)]
	\item We examine the response along the symmetry line of the interface, i.e.~$y\=z\=0$.
	\item We set $F_z\=0$, i.e. no out-of-plane forces emerge, consistent with the plane-stress assumption of the thinner block.
	\item $F_x,F_y$ are constant along the width $-\frac{W}{2}\!\le\!z\!\le\!\frac{W}{2}$ and consequently take the form
	$$F_i(x,y=0,z)\=e^{ikx}H\left(\tfrac{W}{2}-z\right)H\left(\tfrac{W}{2}+z\right)\sigma_{iy} dx\,dz \ ,$$
	where $H$ is Heaviside's step function.
\end{enumerate}
Since $u_z$ vanishes on the symmetry line $z\=0$ due to reflection symmetry through the $x\!-\!y$ plane, we will only be interested in the $x,y$ components of the displacement field. Points (a)+(b) imply
\begin{align}
\begin{pmatrix}
u_x \\u_y
\end{pmatrix}
=-\frac{1}{4\pi \mu r}
\begin{pmatrix}
2 (1-\nu ) +2 \nu \frac{ x^2}{r^2}& -(1-2 \nu)\frac{x }{r} \\
(1-2 \nu )\frac{x }{r} & 2 (1-\nu ) \\
\end{pmatrix}
\begin{pmatrix}
F_x \\F_y
\end{pmatrix}\ .
\label{eq:fo}
\end{align}
Using point (c), we obtain the interfacial displacements by integrating over the contact region
\begin{align}
\nonumber
\begin{pmatrix}
u_x \\u_y
\end{pmatrix}
&=-\frac{1}{4\pi \mu}\int_{-\infty}^\infty\!\!\! dx' \int_{-\tfrac{W}{2}}^{\tfrac{W}{2}}dz' \ \frac{e^{ikx'}}{\sqrt{(x-x')^2+z'^2}}
\begin{pmatrix}
2 (1-\nu )+2 \nu \frac{(x-x')^2}{(x-x')^2+z'^2} & -(1-2 \nu)\frac{x-x'}{\sqrt{(x-x')^2+z'^2}} \\
(1-2 \nu )\frac{x-x'}{\sqrt{(x-x')^2+z'^2}} & 2 (1-\nu ) \\
\end{pmatrix}
\begin{pmatrix}
\sigma_{xy} \\\sigma_{yy}
\end{pmatrix}
\\
&=-\frac{e^{ikx}}{4\pi\mu}\int_{-\infty}^\infty\!\!\! dX \int_{-\tfrac{W}{2}}^{\tfrac{W}{2}}dz'
\ e^{ikX}
\begin{pmatrix}
\frac{2 (1-\nu )}{(X^2+z'^2)^{1/2}}+\frac{2 \nu  X^2}{(X^2+z'^2)^{3/2}} & (1-2 \nu)\frac{X }{X^2+z'^2} \\
-(1-2 \nu)\frac{X }{X^2+z'^2} & \frac{2 (1-\nu )}{((x-x')^2+(z-z')^2)^{1/2}} \\
\end{pmatrix}
\begin{pmatrix}
\sigma_{xy} \\\sigma_{yy}
\end{pmatrix} \ .
\end{align}
where we defined $X\!\equiv\!x'\!-\!x$. The result of this integration is, by definition, the interfacial response matrix $\bm M$ (note that the $e^{ikx}$ prefactor is not included in $\bm M$). That is, if we define
\begin{align}
I_\alpha&=\int_{-\infty}^\infty\!\!\! dX \int_{-\tfrac{W}{2}}^{\tfrac{W}{2}}dz'
\frac{X^{\alpha-1}}{(X^2+z'^2)^{\alpha/2}}e^{ikX} \ ,
\qquad
\alpha\in\{1,2,3\}\ ,
\label{eq:Ij}
\end{align}
then $\bm M$ is given by
\begin{align}
\bm M=-\frac{1}{4\pi\mu}
\begin{pmatrix}
2 (1-\nu )I_1 +2\nu I_3& (1-2 \nu)I_2 \\
-(1-2 \nu)I_2 & 2 (1-\nu )I_1 \\
\end{pmatrix}\ .
\label{eq:Mi}
\end{align}
% where we defined
% \begin{align*}
%  I_1&\equiv\int_{-\infty}^\infty\!\!\! dX\!\! \int_{-\tfrac{W}{2}}^{\tfrac{W}{2}}\!dz'
%   \frac{e^{ikX}}{(X^2+z'^2)^{1/2}}\, , &
%   I_2&\equiv\int_{-\infty}^\infty\!\!\! dX\!\! \int_{-\tfrac{W}{2}}^{\tfrac{W}{2}}\!dz'
%   \frac{e^{ikX}X}{(X^2+z'^2)^{1}}\, , &
%   I_3&\equiv\int_{-\infty}^\infty\!\!\! dX\!\! \int_{-\tfrac{W}{2}}^{\tfrac{W}{2}}\!dz'
%   \frac{e^{ikX}X^2}{(X^2+z'^2)^{3/2}} \ ,
% \end{align*}
% or in a more compact form,

\begin{figure}
	\centering
	\includegraphics[width=.55\textwidth]{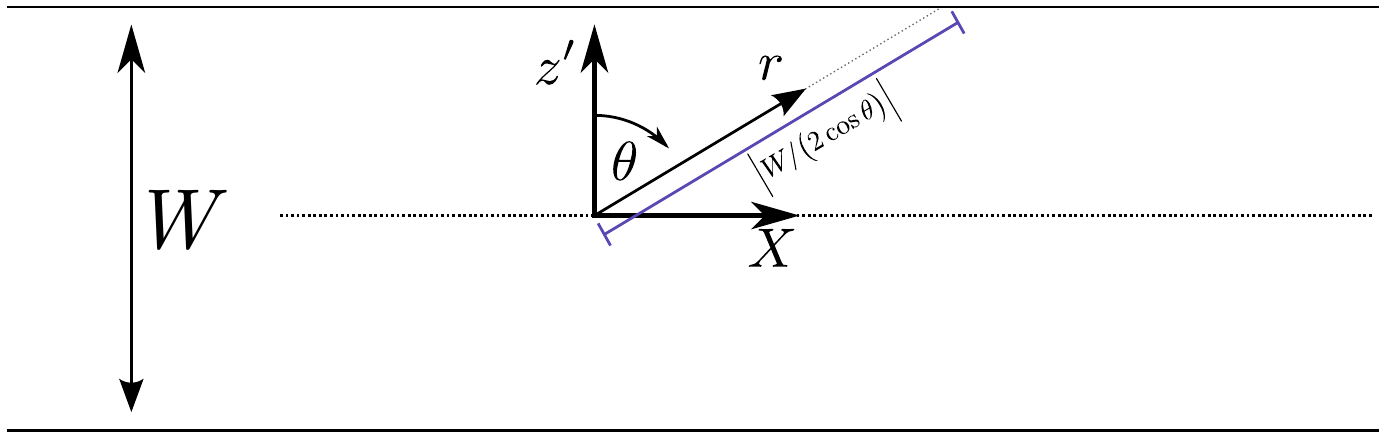}
	\caption{A sketch of the integration domain of Eq.~\eqref{eq:Ij} and the polar coordinates of Eqs.~\eqref{eq:deb}-\eqref{eq:fin}.}
	\label{fig:coor}
\end{figure}

The integration is easier in polar coordinates and it is more convenient to measure the polar angle from the $z$-axis, cf.~Fig.~\ref{fig:coor}. Then, the integrals of Eq.~\eqref{eq:Ij} take the form
\begin{align}\label{eq:deb}
\begin{split}
I_\alpha&%=\int_{-\infty}^\infty\!\!\! dX \int_{-\tfrac{W}{2}}^{\tfrac{W}{2}}dz' \frac{X^{\alpha-1}}{(X^2+z'^2)^{\alpha/2}}e^{ikX}
%=\int_{-\pi}^{\pi}d\theta \int_{0}^{|W/2\cos\theta|}r\,dr\,\frac{(r\sin\theta)^{\alpha-1}}{r^\alpha} e^{i k r \sin\theta}\\
=\int_{-\pi}^{\pi}d\theta \int_{0}^{|W/2\cos\theta|}\!dr \,e^{i k r \sin\theta}(\sin\theta)^{\alpha-1}
=\int_{-\pi}^{\pi} \frac{d\theta}{ik}\left[e^{i \frac{kW}{2} \frac{\sin\theta}{|\cos\theta|}}-1\right](\sin\theta)^{\alpha-2}\ .
\end{split}
\end{align}
The integrand is symmetric with respect to the reflection $z\!\to\!-z$ and thus the integral over $\theta$ can be performed on the domain $-\frac{\pi}{2}\!\le\!\theta\!\le\!\frac{\pi}{2}$. Over this domain the cosine function does not change sign and we have
\begin{align}
I_\alpha=\frac{2W}{iq}\int_{-\pi/2}^{\pi/2} \left[e^{i \frac{q}{2}\tan\theta}-1\right](\sin\theta)^{\alpha-2}d\theta\ ,
\end{align}
where $q\!\equiv\!kW$ was introduced. Employing the change of variables $u=\tan\theta$ we obtain
\begin{align}
I_\alpha=\frac{2W}{iq}\int_{-\infty}^{\infty} \frac{\left(e^{i\frac{qu}{2}}-1\right)u^{\alpha-2}}{(1+u^2)^{\alpha/2}}du\ .
\label{eq:fin}
\end{align}
Using some straightforward manipulations, the explicit integration can be performed using~\cite[Eq.~(3.771.2)]{S-Gradshteyn}, yielding finally
\begin{align}
I_1(q)&=\frac{2 W}{q}\!\int_0^q {\cal K}_0\left(\frac{q'}{2}\right)dq' \equiv \frac{2\pi W}{q}B(q) \ ,&%= \frac{W}{2}\, \mathcal{G}_{1,3}^{2,1}\left(\left(\frac{q}{4}\right)^2\biggl|^{\frac{1}{2}}_{0,0,-\frac{1}{2}}\right)\ .
I_2&=\frac{2 \pi W}{q}i\left(1-e^{-\frac{|q|}{2}}\right)\ ,&
I_3&=2 W \K_0\left(\frac{\left| q\right| }{2}\right)\equiv \frac{2 \pi W}{q}C(q)\ ,
\label{eq:B(q)C(q)}
\end{align}
where $\K_0(z)$ is the modified Bessel function of the second kind of order 0.

Using the expressions for $I_1,I_2$ and $I_3$ in Eq.~\eqref{eq:Mi}, we obtain finally
\begin{align}
\bm M =-\frac{1}{\mu k}
\begin{pmatrix}
(1-\nu )B(q)+\nu C(q) & \frac{i}{2}(1-2 \nu)\left(1-e^{-\frac{|q|}{2}}\right) \\[2mm]
-\frac{i}{2} (1-2 \nu)\left(1-e^{-\frac{|q|}{2}}\right) & (1-\nu )B(q)
\end{pmatrix}\ .
\end{align}
%with the definition
%\begin{align}
% C(q) = \frac{k}{2\pi}I_3=\frac{q}{\pi}\K_0\left(\frac{\left| q\right| }{2}\right)\ .
%\label{eq:bcq}
%\end{align}
This analysis was performed for the upper half-space $y\!>\!0$. As discussed above, the response matrix of the lower half-space, $y\!<\!0$, is obtained by inverting the sings of the diagonal elements, yielding
\begin{equation}
\bm M =\frac{1}{\mu k}
\begin{pmatrix}
(1-\nu )B(q)+\nu C(q) & -\frac{i}{2}(1-2 \nu)\left(1-e^{-\frac{|q|}{2}}\right) \\[2mm]
\frac{i}{2} (1-2 \nu)\left(1-e^{-\frac{|q|}{2}}\right) & (1-\nu )B(q)
\end{pmatrix}\ .
\label{eq:M_exact}
\end{equation}

\begin{figure}
	\centering
	\includegraphics[width=0.4\textwidth]{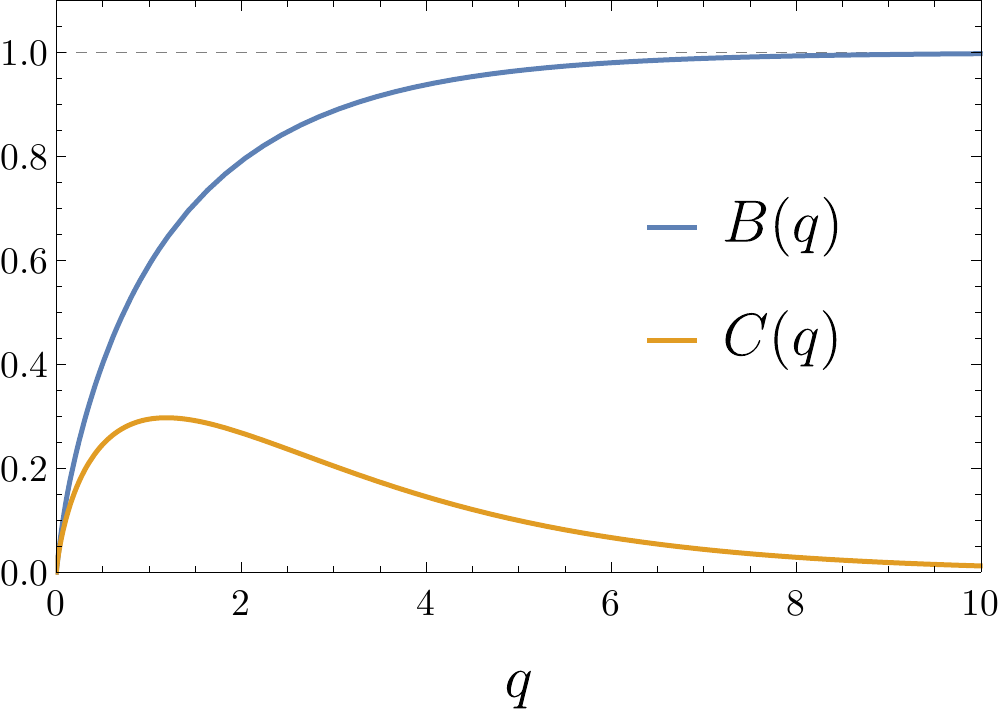}
	\caption{The functions $B(q)$ and $C(q)$, cf.~Eq.~\eqref{eq:B(q)C(q)}, as a function of $q$. It is observed that $C(q)$ is negligible with respect to $B(q)$ except for very small $q$.}
	\label{fig:bcq}
\end{figure}

\subsection{Mapping to an effective 2D material}
How does the response matrix of Eq.~\eqref{eq:M_exact} relate to that of an infinite 2D plane-strain material, i.e.~Eq.~\eqref{eq:M_QS_inf_H}? We want to write Eq.~\eqref{eq:M_exact} as
\begin{align}
\bm M\eff
=\frac{1}{\mu\eff k}
\begin{pmatrix}
1-\nu\eff & -\frac{i}{2}(1-2 \nu\eff)\\
\frac{i}{2}(1-2 \nu\eff)& 1-\nu\eff \\
\end{pmatrix}\ .
\label{eq:3}
\end{align}
One can show that $B(q)$ approaches unity and $C(q)$ vanishes in the limit $q\to\infty$. Thus, in the large $q$ limit (i.e. for wavelengths much smaller than $W$), Eq.~\eqref{eq:M_exact} coincides exactly with Eq.~\eqref{eq:3}, which means in this limit the response of the thicker (lower) block is described by 2D plane-strain elasticity (as expected physically for wavelengths much smaller than $W$).

Such a mapping does not emerge as cleanly for finite $q$'s. Clearly, for $\bm M$ of Eq.~\eqref{eq:M_exact} to have the same structure as $\bm M\eff$ of Eq.~\eqref{eq:3}, the $C(q)$ term in Eq.~\eqref{eq:M_exact} should be negligible with respect to the $B(q)$ term. As shown in Fig.~\ref{fig:bcq}, this is actually the case except at very small $q$. After neglecting the $C(q)$ term, a mapping between Eq.~\eqref{eq:M_exact} and Eq.~\eqref{eq:3} is obtained by equating the two independent terms in each matrix, i.e.~by solving the two equations
\begin{align}
\frac{(1-\nu ) B(q)}{\mu }&\simeq\frac{1-\nu\eff}{\mu\eff}\ , &
\frac{(1-2 \nu ) \left(1-e^{-\frac{q}{2}}\right)}{\mu}&\simeq\frac{1-2 \nu\eff}{\mu\eff}\ .
\end{align}
The $q$-dependent solution to these equations is
\begin{align}
\mu\eff(q)&\simeq\frac{\mu }{2 (1-\nu) B(q)-(1-2 \nu) \left(1-e^{-\frac{q}{2}}\right)}\ ,&
\nu\eff(q)&\simeq\frac{(1-\nu) B(q)-(1-2 \nu) \left(1-e^{-\frac{q}{2}}\right)}{2 (1-\nu) B(q)-(1-2 \nu) \left(1-e^{-\frac{q}{2}}\right)} \ ,
\label{eq:5}
\end{align}
which is identical to Eq. (7) in the main text. As stated above, in the limit $q\!\to\!\infty$ we have $B(q)\!\to\!1$ and clearly $\mu\eff\!\to\!\mu$ and $\nu\eff\!\to\!\nu$. In the opposite limit, $q\!\to\!0$, $C(q)$ is no longer negligible compared to $B(q)$ and a clean mapping to 2D does not emerge, i.e. the problem is fully 3D.

The effective constants for intermediate values of $q$ are plotted in Fig.~\ref{fig:analytical} of the main text. It is seen that for the chosen value of $\nu\=0.33$ we have $\mu\eff\!>\!\mu$ for all experimentally relevant values of $q$. For completeness, we note here that at very large $q$, when $\mu\eff(q)$ approaches $\mu$, $\mu\eff$ {\em minutely} deviates from $\mu$ and approaches it from below in the limit. That is, the effective material contrast $\mu\eff/\mu$ is practically unity, but slightly smaller. For realistic values of $\nu$, this effect is negligible and occurs at large $q$'s: For $\nu\=0.3$ the minimal value of $\mu\eff/\mu$ is $1-9.7\!\cdot\!10^{-7}$ and is obtained for $q\!\approx\!19.3$. The corresponding numbers for $\nu\=0.2$ and $\nu\=0.4$ are, respectively, $\min\{\mu\eff/\mu\}\=1-7\!\cdot\!10^{-4}, 1-8\!\cdot\!10^{-8}$ which are obtained at $q=8.6, 45$.

\section{Experiment}

\subsection{Sample construction}

The experiments reported on in the main text study the frictional motion of two poly(methyl methacrylate) (PMMA) blocks and compare two geometrically different experimental setups. Both experimental setups were conducted using same upper block of dimensions $200$ mm$ \times100$ mm$ \times5.5$ mm in the $x$, $y$ and $z$ direction, respectively (see Fig. 1a in the main text) while the lower block was of different geometry in the two setups. In the ``thin-on-thin'' (symmetric) experiment, a lower block of $250$ mm $\times100$ mm$\times 5.5$ mm dimensions was used~\cite{S-Svetlizky2016}. The ``thin-on-thick" experimental setup used a thicker lower block of $290$ mm $\times28$ mm $\times30$ mm dimensions~\cite{S-Svetlizky2014}. The two blocks were pressed together by an external normal force ($\sim 4.5$ MPa nominal pressure).

The shear and longitudinal wavespeeds, $c_s$ and $c_d$ respectively, were obtained by measuring the time of flight of ultrasonic pulses, yielding $c_s\=1345\!\pm\!10$ m/s and $c_d\=2700\!\pm\!10$ m/s. Due to the high frequency ($5$ MHz) of the ultrasonic pulses used, the measured $c_d$ corresponds to plane-strain conditions $(\varepsilon_{zz}\=0)$. Using these measured values, $c_d$ for plane stress ($\sigma_{zz}\=0$) was then calculated to be $c_d\=2333\pm10$ m/s. The corresponding Rayleigh wave speed is $c_R\!\approx\!1237$ m/s. This velocity is indeed consistent with the maximal measured front velocities for the ``thin-on-thin'' setup. The measured maximal velocities for the ``thin-on-thick'' setup are systematically larger, by about $2\%$, quite close to the values of $c_R$ for plain-strain conditions ($1255$ m/s). This value of $c_R$,  as well as the assumption of plane-strain conditions, were used in previous work~\cite{S-Svetlizky2014} where the ``thin-on-thick'' setup was utilized. The experimental loading system, strain and contact area measurements are described in detail in~\cite{S-Svetlizky2014}. We specify here only the main differences of the current study.

\subsection{Strain measurements}

\begin{figure} [h]
	\centerline{\includegraphics[width=0.4\textwidth]{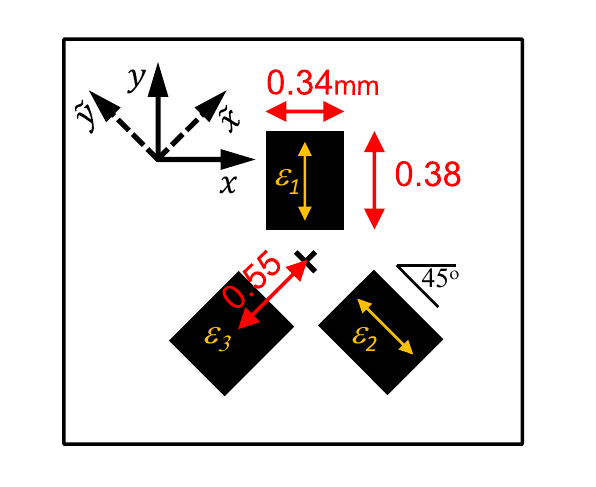}}
	\caption{Geometry and dimensions (in mm) of a single rosette strain gauge. The black rectangles represent the active area of the measuring components, $\varepsilon_1$, $\varepsilon_2$ and $\varepsilon_3$. Yellow arrows represent the direction of the measured strains.}
	\label{figS1}
\end{figure}

We used miniature Vishay 015RJ rosette strain gages for local strain measurements that were mounted $\simeq\!3.5$mm above the frictional interface (top block only). Each rosette strain gage  is composed of three active regions (each $0.34$mm $\times 0.38$mm size). Each active region provides a strain component, $\varepsilon_i$, along the directions denoted by the yellow arrows in Fig.~\ref{figS1}.

Electrical resistance strain gages can be calibrated to a high precision when are used on very stiff materials such as various metals. However, when these strain gages are embedded on less stiff materials such as plastics, their presence might locally alter the strain field in their surroundings (see~\cite{S-Ajovalasit2013} and references within). Analytical models and numerical efforts exist in the literature to estimate this effect and properly calibrate the measurement of strain.

For purposes of calibration, a rosette strain gage was glued at the center of $100$ mm diameter PMMA disk ($7.5$ mm width). The disk was subjected to radial compression at various angles with respect to the rosette axis ($y$ axis in Fig.~\ref{figS1}). We assumed that a transformation that relates the altered strain field due to the rosette presence (here denoted by $\varepsilon_i'$) to the ``actual" strain field in its absence ($\varepsilon_i$) could be found. We indeed found that the calibration measurements can be described by a phenomenological transformation of the following form
\begin{eqnarray}
\varepsilon_1'&=&a_1\cdot\varepsilon_1+k_1\cdot\varepsilon_{xx}+g_1\cdot\varepsilon_{xy}\ ,\\
\varepsilon_2'&=&a_2\cdot\varepsilon_2+k_2\cdot\varepsilon_3+g_2\cdot\varepsilon_{\widetilde{x}\widetilde{y}}\ ,\\
\varepsilon_3'&=&a_2\cdot\varepsilon_3+k_2\cdot\varepsilon_2+g_2\cdot\varepsilon_{\widetilde{x}\widetilde{y}}\ ,
\end{eqnarray}
where $a_i$ are corrections for the gage factors, $k_i$ represent the transverse sensitivity of the strain gages and $g_i$ represent shear sensitivities. ($\widetilde{x}$,\,$\widetilde{y}$) is the coordinate system rotated by $45^{\circ}$ relative to that of $\varepsilon_1$ (see Fig.~\ref{figS1}). $a_2\=1$ was chosen, as only the relative calibration of the components was of interest. Due to reflection symmetry with respect to the $y$ axis, coefficients of $\varepsilon_2$ and $\varepsilon_3$ are identical and $g_1\=0$. This reflection symmetry does not exist with respect to  $\varepsilon_2$ and $\varepsilon_3$, and hence shear sensitivity can not be excluded. As the effects of the elastic mismatch of the rosette configuration have not been previously considered, we note that shear sensitivity has not been discussed in the literature. Here, we find that shear sensitivity exists and is crucial for proper gage calibration. Our calibrations revealed that $a_1\!\approx\!0.95$, $k_1\!\approx\!-0.08$, $k_2\!\approx\!0$ and $g_2\!\approx\!0.1$ (details of the calibration procedure will be published elsewhere). Once $\varepsilon_i'$ are measured, $\varepsilon_i$ can be calculated using an inverse transformation.

\subsection{Experimental results}

\begin{figure}[h]
	\centerline{\includegraphics[width=1\textwidth]{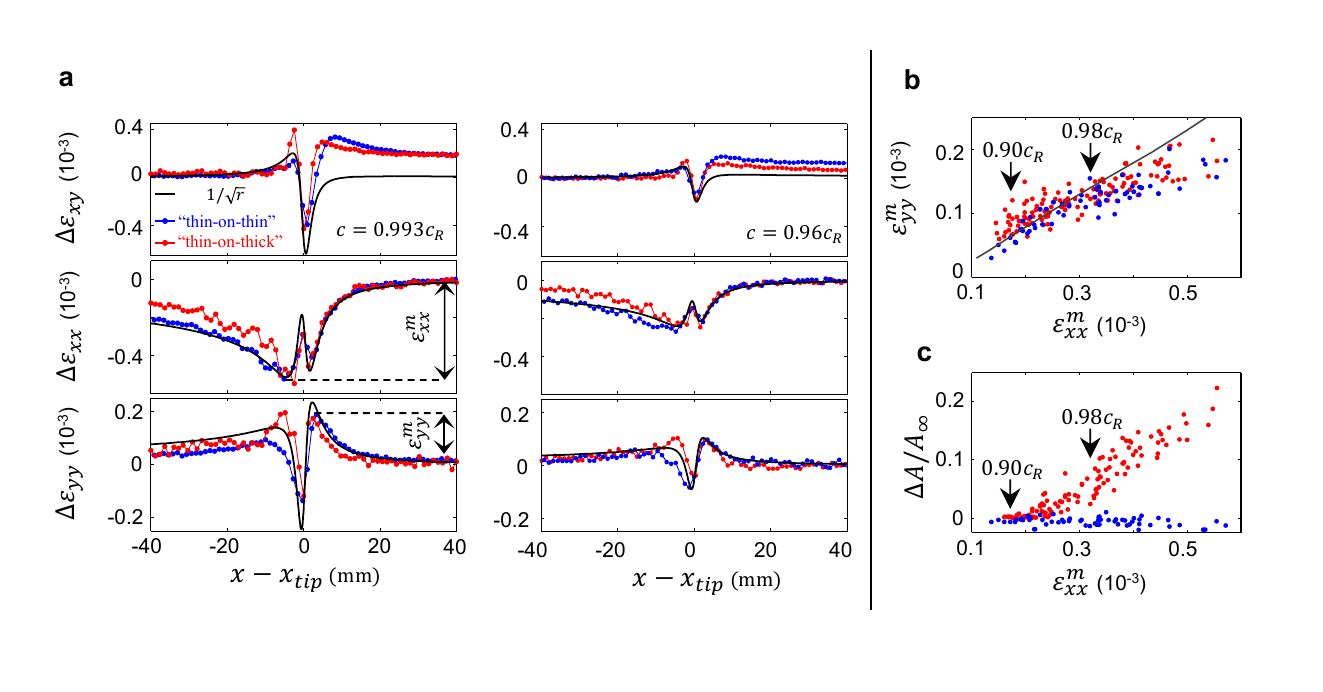}}
	\caption{Comparison of strain and contact area measurements for ``thin-on-thin" (blue symbols) and ``thin-on-thick" (red symbols) geometries. \textbf{a.} Strain tensor variations, $\Delta\varepsilon_{ij}$, after subtracting the initial strains from $\varepsilon_{xx}$ and $\varepsilon_{yy}$ and the residual strain from $\varepsilon_{xy}$. Strains were measured $3.5$ mm above the frictional interface and plotted with respect to the location of the rupture tip, $x_{tip}$. The singular term of the LEFM solution, that is plotted in black ($\Gamma\!=\!1.1$ J/m$^2 $, $c$ is noted in the panels), describes rather well both geometrical setups. The apparent discrepancy in the shear component, $\Delta\varepsilon_{xy}$, for $x\!-\!x_{tip}\!>\!0$ was shown to be related to nonsingular LEFM terms, as discussed in~\cite{S-Svetlizky2016}. These strain profiles correspond to two of the examples presented in Fig.~2b of the main text. \textbf{b.} Measured $\Delta\varepsilon_{xx}$ and $\Delta\varepsilon_{yy}$ were characterized by their peak values, $\varepsilon_{xx}^m$ and $\varepsilon_{yy}^m$, respectively, as denoted in \textbf{a(left)}. The prediction based on the singular LEFM solution, which corresponds to the black line, successfully captures the measurements with some systematic discrepancies at $c\!>\!0.98c_R$ \textbf{c.} The dependence of the undershoot $\Delta A /A_\infty$ on  $\varepsilon_{xx}^m$ for both geometries.
		Panel \textbf{c} here eventually transforms into Fig.~2c in the manuscript. This is done, as explained below (see text), in two steps. First, the measurements of $\Delta\varepsilon_{xx}$ at $y\!=\!3.5$ mm are extrapolated to the interface, i.e. $\Delta\varepsilon_{xx}$ at $y\!=\!0$. Then, $\Delta\varepsilon_{xx}(y\!=\!0)$ is related to the slip velocity, $\dot\epsilon_x$, according to $\dot\epsilon_x\!=\!-2c\!\cdot\!\Delta\varepsilon_{xx}(y\!=\!0)$.}
	\label{figS2}
\end{figure}

Typical examples of strain measurements, for both experimental setups, are presented in Fig.~\ref{figS2}. In previous work \cite{S-Svetlizky2014,S-Svetlizky2016}, it was found that the strains in vicinity of a rupture tip are well described by the singular Linear Elastic Fracture Mechanics (LEFM) solutions for ideal shear cracks with a single fitting parameter, the fracture energy $\Gamma$~\cite{S-Broberg1999Book}. It was found that for a wide range of rupture velocities, $c$, $\Gamma$ is approximately constant ($\Gamma\!\approx\!1.1$ J/m$^2$). Some systemic discrepancies (at most $30\%$) are observed at extreme rupture velocities (cf.~Fig.~\ref{figS2}b) between the measured strains and LEFM predictions. These discrepancies may result from either errors involved in the strain gage calibration (see previous section), or, possibly, violations of our implicit 2D assumption.

In this work we are especially interested in the rupture dynamics at high rupture velocities $0.9c_R\!<\!c\!<\!c_R$. While direct measurement of $c$ can be performed in our system to $\sim\!2\%$ precision, we can significantly decrease this experimental uncertainty by exploiting the significant growth of the strain amplitudes as $c\!\rightarrow\! c_R$~\cite{S-Svetlizky2014}. Using this, we improve our measurements of $c$ by fitting the measured $\Delta\varepsilon_{xx}$ amplitudes to the singular solution, while assuming that $\Gamma$ does not significantly change in the vicinity of $c_R$ and that the system obeys plane-stress boundary conditions. Using this method, $c$ is the only fitting parameter. Results of this procedure are demonstrated in Fig.~\ref{figS2}a and are employed to determine $c$ in Fig.~2 of the main text.

Note that the assumption of plane-stress conditions should be violated for the ``thin-on-thick'' setup. As mentioned above, the measured asymptotic velocities for the ``thin-on-thick'' setup are about $2\%$ above $c_R$ for plane-stress.  Nevertheless, for simplicity, we have used the plane-stress assumption in the above analysis. This, therefore, may lead to systematic errors in our estimated values of $c$ (for example, the directly measured velocities for the $3$ highest velocities in Fig.~2a are all $\simeq\!1255$ m/s). The use of $\Delta \varepsilon_{xx}$, however, enables us to quantitatively differentiate between the different high $c$ measurements, despite possible systematic errors in determining the absolute values of $c$. Consequently, the rupture propagation velocities stated in the legend of Fig.~2b in the manuscript should be understood in relative terms when normalized with respect to the relevant $c_R$.

\subsubsection{Slip velocity estimation}

Figure \ref{figS2}c demonstrates that $\Delta A /A_{\infty}$ is correlated with the amplitude of $\varepsilon_{xx}$, $\varepsilon^m_{xx}$, directly measured at $3.5$mm above the frictional interface. Relating $\Delta A /A_{\infty}$ to the slip velocity, defined at $y\=0$, is of great interest. The LEFM singular solution, which describes our measurements well at $y\=3.5$mm, predicts that the slip velocity (and actually all strain and stress components) should be singular at $y\=0$ and $x\=x_{tip}$. These singularities are naturally regularized at the crack tip. In this section we will explain the underlying assumptions that enable us to estimate the slip velocity and relate the direct measurements of $\Delta\varepsilon_{xx}$ presented in Fig.~\ref{figS2}c to the extrapolated slip velocities in Fig.~2b of the main text.
\begin{figure}[h]
	\centerline{\includegraphics[width=1\textwidth]{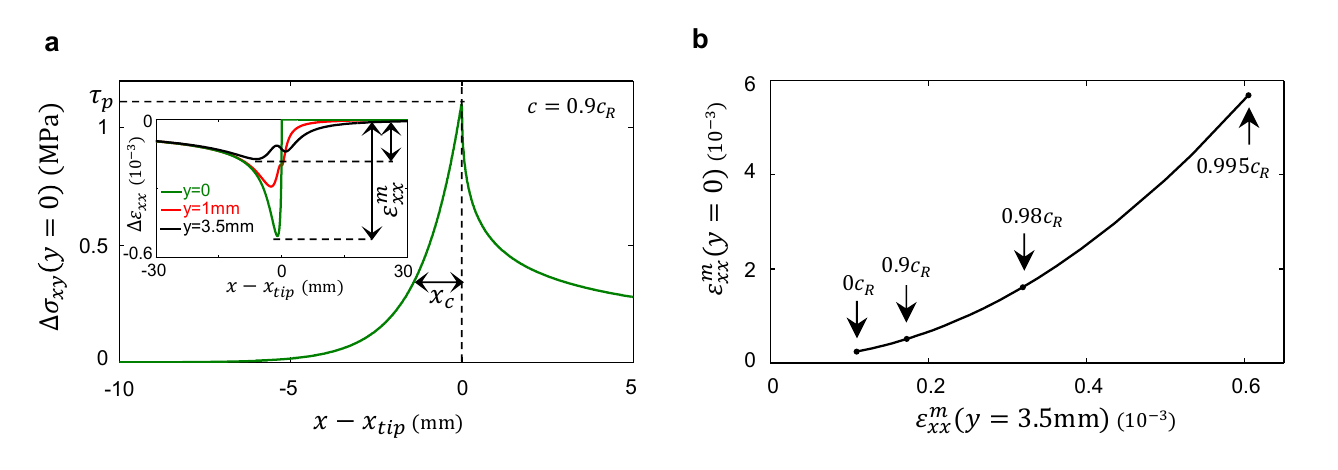}}
	\caption{Slip velocity estimation. \textbf{a.} The nonsingular cohesive zone model in which the shear stress is reduced exponentially behind the crack tip once the peak strength, $\tau_p$, is reached. The model is entirely defined by the measured values of $\Gamma$ and $X_c(c)$. The example provided is for $c\!=\!0.9c_R$. \textbf{(inset)} Snapshots of $\Delta\varepsilon_{xx}$ for various heights above the frictional interface ($c\!=\!0.9c_R$). The model is indistinguishable from the singular solution at the height of strain measurements ($y\!=\!3.5$mm). Once the frictional interface ($y\!=\!0$) is approached, amplitudes of $\Delta\varepsilon_{xx}$ significantly deviate from their value at $y\!=\!3.5$mm. \textbf{b.} The amplitudes of $\Delta\varepsilon_{xx}$ on the frictional interface are related to amplitudes of $\Delta\varepsilon_{xx}$ at $y\!=\!3.5$mm, by virtue of the model. The arrows indicate the crack velocities; as $c\!\rightarrow\!c_R$ the amplitudes diverge.}
	\label{figS3}
\end{figure}

We first note that even in the extreme case ($c\=0.993c_R$) presented in Fig.~\ref{figS2}, where $\Delta A /A_{\infty}\!\approx\!0.25$, strain measurements obtained in both experiments with different geometrical setups are surprisingly similar, where only some differences are observed at $x\!-\!x_{tip}\!<\!0$. These strain differences are only minor when compared to the large qualitative difference in strain measurements presented in~\cite{S-Shlomai2016}, where strong material contrast is considered. This observation enables us to adapt a perturbative approach in which we invoke the simplest cohesive zone model valid  for the ``thin-on-thin" case to estimate the slip velocity at the interface for \textit{both} geometrical setups. At this stage, however, we are not able to estimate the accuracy of this assumption for the ``thin-on-thick" setup.

In the non-singular cohesive zone model \cite{S-Poliakov2002,S-Samudrala2002}, weakening initiates once the shear stress has reached a finite peak strength, $\tau_p$, above the residual value, $\tau_r$ of the shear stress. The shear stress gradually decreases according to a prescribed shear stress profile, $\tau(x/X_c)\=\tau_p\cdot\widetilde{\tau}(x/X_c)$. $X_c$ is defined to be the cohesive zone size. Far ahead of the rupture tip the solution matches the square root singular form, i.e, $\sigma_{xy}(x\!\gg\!X_c, y\=0)\=K_{II}/\sqrt{2\pi x}$. Therefore, $\tau_p$, $X_c$ and $\Gamma\=K_{II}^2/E$, are related through~\cite{S-Samudrala2002}

\begin{equation}
K_{II}=\tau_p\cdot\sqrt{X_c}\cdot\sqrt{\frac{2}{\pi}}\cdot\int_{-\infty}^{0}\frac{\widetilde{\tau}(\xi)}{\sqrt{-\xi}}d\xi
\end{equation}
\par
In previous work \cite{S-Svetlizky2014} it was argued that the length scale over which $A$ is reduced provides an estimate of $X_c$. It was shown that $X_c$ contracts as $c \rightarrow c_R$. As would be expected from elastodynamic theory, these measurements were quantitatively described by $X_c(c)=X_c^0/f_{II}(c)$, where $X_c^0\=X_c(c\!\rightarrow\!0)\approx 2.5$ mm and $f_{II}(c)$ is a known function predicted by LEFM~\cite{S-Broberg1999Book}.

We use the experimentally measured variation of $X_c$, a constant value of $\Gamma$ and assume $\widetilde{\tau}(\xi)\=e^{\xi}$ (see example in Fig.~\ref{figS3}a). These constraints result in $\tau_p\!\approx\!1.1$ MPa, which is independent of $c$. Once the model is specified, all of the dynamic fields in the bulk can be calculated. For example, snapshots of $\Delta\varepsilon_{xx}$ for various heights above the frictional interface are plotted in Fig.~\ref{figS3}a-inset. This model was used to relate the measured amplitudes of $\Delta\varepsilon_{xx}$ at $y\=3.5$ mm to amplitudes of $\Delta\varepsilon_{xx}$ on the frictional interface for various rupture velocities (Fig.~\ref{figS3}c). Note that the systematic discrepancies at $c\!>\!0.98c_R$ shown in Fig.~\ref{figS2} may be further amplified by the non-linearity of the transformation. Finally, the slip velocity is calculated according to $v\=\dot\epsilon_x\=-2c\!\cdot\!\Delta\varepsilon_{xx}(y\=0)$ and the direct measurements presented in Fig.~\ref{figS2}c are converted and plotted in Fig.~2 of the main text.

\end{document}